%% file: paper.tex
\documentclass[11pt]{article}

\usepackage{amsmath}
\usepackage{amssymb}
\usepackage{graphicx}

\newcommand{\R}{\mathbb{R}}

\usepackage{ifthen}
\usepackage{epsfig}
\usepackage{array}
\usepackage{color}
\usepackage[table]{xcolor}
\usepackage{hyperref}
\usepackage{xspace}

\definecolor{darkgreen}{rgb}{0,0.44,0}
\definecolor{darkred}{rgb}{0.44,0,0}
\definecolor{darkblue}{rgb}{0,0,0.44}

\input{color_flatex}

\newcommand{\Rnn}{\R^{n \times n}}
\newcommand{\Rmn}{\R^{m \times n}}

\newcommand{\nref}[1]{(\ref{#1})}
\newcommand{\dbar}[1]{\bar{\bar{#1}}}

\newcommand{\Rjj}{\R^{j \times j}}

\newcommand{\bd}{w}
\newcommand{\bs}{b}

\begin{document}

\title{Look-Ahead in the Two-Sided Reduction to Compact Band Forms 
for Symmetric Eigenvalue Problems and the SVD}

\author{%
Rafael Rodr\'{\i}guez-S\'anchez\footnote{Depto. Ingenier\'{\i}a y Ciencia de Computadores,
Universidad Jaume I, Castell\'on, Spain. {\tt \{rarodrig,catalans,quintana,tomasan\}@icc.uji.es}}
\and
Sandra Catal\'an$^*$
\and
Jos\'e R. Herrero\footnote{Dept. d'Arquitectura de Computadors, Universitat Polit\`ecnica de Catalunya, Spain.
{\tt josepr@ac.upc.edu}}
\and
Enrique S. Quintana-Ort\'{\i}$^*$
\and
Andr\'es E. Tom\'as$^*$
}



\date{\today}

\maketitle

\begin{abstract}
\input{s0-abstract}
\end{abstract}

\input{body}

\subsection*{Acknowledgements}
This research  was partially sponsored by projects TIN2014-53495-R and 
TIN2015-65316-P  of  the Spanish  {\em  Ministerio  de Econom\'{\i}a  y 
Competitividad}, 
project 2014-SGR-1051 from the Generalitat  de Catalunya, 
and the EU H2020 project 732631 OPRECOMP.


\end{document}

%% file: color_flatex.tex
\usepackage{arydshln}

\newcolumntype{I}{!{\vrule width 1.5pt}}

\newlength\savedwidth 

\newcommand\whline{\noalign{\global\savedwidth\arrayrulewidth 
                            \global\arrayrulewidth 1.5pt}%
           \hline 
           \noalign{\global\arrayrulewidth\savedwidth}}

\newcommand{\routinename}{}

\newcommand{\precondition}{~}

\newcommand{\postcondition}{~}

\newcommand{\invariant}{~}

\newcommand{\guard}{~}

\newcommand{\partitionings}{~}

\newcommand{\partitionsizes}{~}

\newcommand{\blocksize}{blank}

\newcommand{\repartitionings}{~}

\newcommand{\repartitionsizes}{~}

\newcommand{\moveboundaries}{~}

\newcommand{\beforeupdate}{~}

\newcommand{\afterupdate}{~}

\newcommand{\update}{~}

\newcommand{\NoShow}[1]{}

\newcommand{\FlaAlgorithm}{
\begin{tabular}{| p{0.92\textwidth}|} \hline
$\mbox{\color{blue}Algorithm:~}\routinename$
\\ \whline
\partitionings \\
$\mbox{\color{blue} ~~~where~}$ \partitionsizes 
\\ 
$\mbox{\color{blue}while~} \guard \mbox{~\color{blue} do}$
\\
\ifthenelse{\equal{\blocksize}{1}}{\\}%
{%
\ifthenelse{ \equal{\blocksize}{blank} }{}%
{~~~~{\bf Determine block size $ \blocksize $}\\}%
}
~~~~ 
\repartitionings \\
~~~$\mbox{\color{blue} ~~~where~}$ \repartitionsizes
\\ \hline
~~~~  \update 
\\ \hline
~~~~ 
\moveboundaries 
\\
$\mbox{\color{blue} endwhile} $
\\ \hline 
\end{tabular}
}

\newcommand{\FlaWorksheet}{
\begin{tabular}{| c | p{0.98\textwidth} |}\hline
Step & $\mbox{\color{blue}Algorithm:~}\routinename$
\\ \hline
\rowcolor{yellow!75}
1a & $ \precondition $ 
\\ \whline
4 & 
\begin{minipage}[t]{0.9\textwidth}%
\partitionings~ \\
$\mbox{\color{blue} ~~~where~}$ \partitionsizes
\end{minipage}
\\ \hline
\rowcolor{yellow!75}
2 & $ \invariant $ 
\\ \hline
3 &$\mbox{\color{blue}while~} \guard \mbox{~\color{blue} do}$
\\ \hline 
\rowcolor{yellow!75}
2,3 & ~~~~ $ \invariant \wedge \guard$ 
\\ \hline
5a & ~~~~ \begin{minipage}[t]{0.85\textwidth}%
\ifthenelse{\equal{\blocksize}{1}}{}%
{%
\ifthenelse{ \equal{\blocksize}{blank} }{}%
{{\bf Determine block size $ \blocksize $}\\}%
}
\repartitionings~ \\
$\mbox{\color{blue} ~~~where~}$ \repartitionsizes
\end{minipage}
\\ \hline
\rowcolor{yellow!75}
6 & ~~~~ $\beforeupdate $
\\ \hline
8 & ~~~~  \update 
\\ \hline
5b & ~~~~ \begin{minipage}[t]{0.85\textwidth}%
\moveboundaries~
\end{minipage}
\\ \hline
\rowcolor{yellow!75}
7 & ~~~~ $\afterupdate $
\\ \hline
\rowcolor{yellow!75}
2 & ~~~~ $ \invariant  $ 
\\ \hline
 &$\mbox{\color{blue} endwhile} $
\\ \hline \whline
\rowcolor{yellow!75}
2,3 & $ \invariant \wedge \neg( \guard )$ 
\\ \hline
\rowcolor{yellow!75}
1b & $ \postcondition $ 
\\ \hline
\end{tabular}
}

\newcommand{\FlaWorksheetNine}{
\begin{tabular}{| c | p{0.98\textwidth} |}\hline
Step & $\mbox{\color{blue}Algorithm:~}\routinename$
\\ \hline
\rowcolor{yellow!75}
\phantom{1a} & $ \phantom\precondition $ 
\\ \whline
\phantom{4} & 
\begin{minipage}[t]{0.9\textwidth}%
\partitionings~ \\
$\mbox{\color{blue} ~~~where~}$ \partitionsizes
\end{minipage}
\\ \hline
\rowcolor{yellow!75}
\phantom{2} & $ \phantom\invariant $ 
\\ \hline
\phantom{3} &$\mbox{\color{blue}while~} \guard \mbox{~\color{blue} do}$
\\ \hline 
\rowcolor{yellow!75}
\phantom{2,3} & ~~~~ $ \phantom\invariant \phantom \wedge \phantom
                \guard $ 
\\ \hline
\phantom{5a} & ~~~~ \begin{minipage}[t]{0.85\textwidth}%
\ifthenelse{\equal{\blocksize}{1}}{}%
{%
\ifthenelse{ \equal{\blocksize}{blank} }{}%
{{\bf Determine block size $ \blocksize $}\\}%
}
\repartitionings~ \\
$\mbox{\color{blue} ~~~where~}$ \repartitionsizes
\end{minipage}
\\ \hline
\rowcolor{yellow!75}
\phantom{6} & ~~~~ $\phantom\beforeupdate $
\\ \hline
\phantom{8} & ~~~~  \update 
\\ \hline
\phantom{5b} & ~~~~ \begin{minipage}[t]{0.85\textwidth}%
\moveboundaries~
\end{minipage}
\\ \hline
\rowcolor{yellow!75}
\phantom{7} & ~~~~ $\phantom\afterupdate $
\\ \hline
\rowcolor{yellow!75}
\phantom{2} & ~~~~ $ \phantom\invariant  $ 
\\ \hline
 &$\mbox{\color{blue} endwhile} $
\\ \hline \whline
\rowcolor{yellow!75}
\phantom{2,3} & $ \phantom\invariant \wedge \neg( \phantom\guard )$ 
\\ \hline
\rowcolor{yellow!75}
\phantom{1b} & $ \phantom\postcondition $ 
\\ \hline
\end{tabular}
}

\newcommand{\FlaWorksheetEight}{
\begin{tabular}{| c | p{0.98\textwidth} |}\hline
Step & $\mbox{\color{blue}Algorithm:~}\routinename$
\\ \hline
\rowcolor{yellow!75}
1a & $ \precondition $ 
\\ \whline
4 & 
\begin{minipage}[t]{0.9\textwidth}%
\partitionings~ \\
$\mbox{\color{blue} ~~~where~}$ \partitionsizes
\end{minipage}
\\ \hline
\rowcolor{yellow!75}
2 & $ \invariant $ 
\\ \hline
3 &$\mbox{\color{blue}while~} \guard \mbox{~\color{blue} do}$
\\ \hline 
\rowcolor{yellow!75}
2,3 & ~~~~ $ \invariant \wedge \guard $ 
\\ \hline
5a & ~~~~ \begin{minipage}[t]{0.85\textwidth}%
\ifthenelse{\equal{\blocksize}{1}}{}%
{%
\ifthenelse{ \equal{\blocksize}{blank} }{}%
{{\bf Determine block size $ \blocksize $}\\}%
}
\repartitionings~ \\
$\mbox{\color{blue} ~~~where~}$ \repartitionsizes
\end{minipage}
\\ \hline
\rowcolor{yellow!75}
6 & ~~~~ $\beforeupdate $
\\ \hline
\rowcolor{orange!50}    
8 & ~~~~  \update 
\\ \hline
5b & ~~~~ \begin{minipage}[t]{0.85\textwidth}%
\moveboundaries~
\end{minipage}
\\ \hline
\rowcolor{yellow!75}
7 & ~~~~ $\afterupdate $
\\ \hline
\rowcolor{yellow!75}
2 & ~~~~ $ \invariant  $ 
\\ \hline
 &$\mbox{\color{blue} endwhile} $
\\ \hline \whline
\rowcolor{yellow!75}
2,3 & $ \invariant \wedge \neg( \guard )$ 
\\ \hline
\rowcolor{yellow!75}
1b & $ \postcondition $ 
\\ \hline
\end{tabular}
}

\newcommand{\FlaWorksheetSeven}{
\begin{tabular}{| c | p{0.98\textwidth} |}\hline
Step & $\mbox{\color{blue}Algorithm:~}\routinename$
\\ \hline
\rowcolor{yellow!75}
1a & $ \precondition $ 
\\ \whline
4 & 
\begin{minipage}[t]{0.9\textwidth}%
\partitionings~ \\
$\mbox{\color{blue} ~~~where~}$ \partitionsizes
\end{minipage}
\\ \hline
\rowcolor{yellow!75}
2 & $ \invariant $ 
\\ \hline
3 &$\mbox{\color{blue}while~} \guard \mbox{~\color{blue} do}$
\\ \hline 
\rowcolor{yellow!75}
2,3 & ~~~~ $ \invariant \wedge \guard$ 
\\ \hline
5a & ~~~~ \begin{minipage}[t]{0.85\textwidth}%
\ifthenelse{\equal{\blocksize}{1}}{}%
{%
\ifthenelse{ \equal{\blocksize}{blank} }{}%
{{\bf Determine block size $ \blocksize $}\\}%
}
\repartitionings~ \\
$\mbox{\color{blue} ~~~where~}$ \repartitionsizes
\end{minipage}
\\ \hline
\rowcolor{yellow!75}
6 & ~~~~ $\beforeupdate $
\\ \hline
8 & ~~~~  \phantom\update 
\\ \hline
5b & ~~~~ \begin{minipage}[t]{0.85\textwidth}%
\moveboundaries~
\end{minipage}
\\ \hline
\rowcolor{orange!50}    
7 & ~~~~ $\afterupdate $
\\ \hline
\rowcolor{yellow!75}
2 & ~~~~ $ \invariant  $ 
\\ \hline
 &$\mbox{\color{blue} endwhile} $
\\ \hline \whline
\rowcolor{yellow!75}
2 & $ \invariant \wedge \neg( \guard )$ 
\\ \hline
\rowcolor{yellow!75}
1b & $ \postcondition $ 
\\ \hline
\end{tabular}
}

\newcommand{\FlaWorksheetSix}{
\begin{tabular}{| c | p{0.98\textwidth} |}\hline
Step & $\mbox{\color{blue}Algorithm:~}\routinename$
\\ \hline
\rowcolor{yellow!75}
1a & $ \precondition $ 
\\ \whline
4 & 
\begin{minipage}[t]{0.9\textwidth}%
\partitionings~ \\
$\mbox{\color{blue} ~~~where~}$ \partitionsizes
\end{minipage}
\\ \hline
\rowcolor{yellow!75}
2 & $ \invariant $ 
\\ \hline
3 &$\mbox{\color{blue}while~} \guard \mbox{~\color{blue} do}$
\\ \hline 
\rowcolor{yellow!75}
2,3 & ~~~~ $ \invariant \wedge \guard $ 
\\ \hline
5a & ~~~~ \begin{minipage}[t]{0.85\textwidth}%
\ifthenelse{\equal{\blocksize}{1}}{}%
{%
\ifthenelse{ \equal{\blocksize}{blank} }{}%
{{\bf Determine block size $ \blocksize $}\\}%
}
\repartitionings~ \\
$\mbox{\color{blue} ~~~where~}$ \repartitionsizes
\end{minipage}
\\ \hline
\rowcolor{orange!50}   
6 & ~~~~ $\beforeupdate $
\\ \hline
8 & ~~~~  \phantom\update 
\\ \hline
5b & ~~~~ \begin{minipage}[t]{0.85\textwidth}%
\moveboundaries~
\end{minipage}
\\ \hline
\rowcolor{yellow!75}
7 & ~~~~ $\phantom\afterupdate $
\\ \hline
\rowcolor{yellow!75}
2 & ~~~~ $ \invariant  $ 
\\ \hline
 &$\mbox{\color{blue} endwhile} $
\\ \hline \whline
\rowcolor{yellow!75}
2,3 & $ \invariant \wedge \neg( \guard )$ 
\\ \hline
\rowcolor{yellow!75}
1b & $ \postcondition $ 
\\ \hline
\end{tabular}
}

\newcommand{\FlaWorksheetFive}{
\begin{tabular}{| c | p{0.98\textwidth} |}\hline
Step & $\mbox{\color{blue}Algorithm:~}\routinename$
\\ \hline
\rowcolor{yellow!75}
1a & $ \precondition $ 
\\ \whline
4 & 
\begin{minipage}[t]{0.9\textwidth}%
\partitionings~ \\
$\mbox{\color{blue} ~~~where~}$ \partitionsizes
\end{minipage}
\\ \hline
\rowcolor{yellow!75}
2 & $ \invariant $ 
\\ \hline
3 &$\mbox{\color{blue}while~} \guard \mbox{~\color{blue} do}$
\\ \hline 
\rowcolor{yellow!75}
2,3 & ~~~~ $ \invariant \wedge \guard $ 
\\ \hline
\rowcolor{orange!50}   
5a & ~~~~ \begin{minipage}[t]{0.85\textwidth}%
\ifthenelse{\equal{\blocksize}{1}}{}%
{%
\ifthenelse{ \equal{\blocksize}{blank} }{}%
{{\bf Determine block size $ \blocksize $}\\}%
}
\repartitionings~ \\
$\mbox{\color{blue} ~~~where~}$ \repartitionsizes
\end{minipage}
\\ \hline
\rowcolor{yellow!75}
6 & ~~~~ $\phantom\beforeupdate $
\\ \hline
8 & ~~~~  \phantom\update 
\\ \hline
\rowcolor{orange!50}   
5b & ~~~~ \begin{minipage}[t]{0.85\textwidth}%
\moveboundaries~
\end{minipage}
\\ \hline
\rowcolor{yellow!75}
7 & ~~~~ $\phantom\afterupdate $
\\ \hline
\rowcolor{yellow!75}
2 & ~~~~ $ \invariant  $ 
\\ \hline
 &$\mbox{\color{blue} endwhile} $
\\ \hline \whline
\rowcolor{yellow!75}
2,3 & $ \invariant \wedge \neg( \guard )$ 
\\ \hline
\rowcolor{yellow!75}
1b & $ \postcondition $ 
\\ \hline
\end{tabular}
}

\newcommand{\FlaWorksheetFour}{
\begin{tabular}{| c | p{0.98\textwidth} |}\hline
Step & $\mbox{\color{blue}Algorithm:~}\routinename$
\\ \hline
\rowcolor{yellow!75}
1a & $ \precondition $ 
\\ \whline
\rowcolor{orange!50}   
4 & 
\begin{minipage}[t]{0.9\textwidth}%
\partitionings~ \\
$\mbox{\color{blue} ~~~where~}$ \partitionsizes
\end{minipage}
\\ \hline
\rowcolor{yellow!75}
2 & $ \invariant $ 
\\ \hline
3 &$\mbox{\color{blue}while~} \guard \mbox{~\color{blue} do}$
\\ \hline 
\rowcolor{yellow!75}
2,3 & ~~~~ $ \invariant \wedge \guard $ 
\\ \hline
5a & ~~~~ \begin{minipage}[t]{0.85\textwidth}%
\ifthenelse{\equal{\blocksize}{1}}{}%
{%
\ifthenelse{ \equal{\blocksize}{blank} }{}%
{{\bf Determine block size $ \phantom\blocksize $}\\}%
}
$\mbox{\phantom\repartitionings}$~ \\
$\mbox{\color{blue} ~~~where~}$ \phantom\repartitionsizes
\end{minipage}
\\ \hline
\rowcolor{yellow!75}
6 & ~~~~ $\phantom\beforeupdate $
\\ \hline
8 & ~~~~  \phantom\update 
\\ \hline
5b & ~~~~ \begin{minipage}[t]{0.85\textwidth}%
\phantom\moveboundaries~
\end{minipage}
\\ \hline
\rowcolor{yellow!75}
7 & ~~~~ $\phantom\afterupdate $
\\ \hline
\rowcolor{yellow!75}
2 & ~~~~ $ \invariant  $ 
\\ \hline
 &$\mbox{\color{blue} endwhile} $
\\ \hline \whline
\rowcolor{yellow!75}
2,3 & $ \invariant \wedge \neg( \guard )$ 
\\ \hline
\rowcolor{yellow!75}
1b & $ \postcondition $ 
\\ \hline
\end{tabular}
}

\newcommand{\FlaWorksheetThree}{
\begin{tabular}{| c | p{0.98\textwidth} |}\hline
Step & $\mbox{\color{blue}Algorithm:~}\routinename$
\\ \hline
\rowcolor{yellow!75}
1a & $ \precondition $ 
\\ \whline
4 & 
\begin{minipage}[t]{0.9\textwidth}%
$\mbox{\phantom{\partitionings}}$~ \\
$\mbox{\color{blue} ~~~where~}$\phantom{\partitionsizes}  
\end{minipage}
\\ \hline
\rowcolor{yellow!75}
2 & $ \invariant $ 
\\ \hline
\rowcolor{orange!50}  
3 &$\mbox{\color{blue}while~} \guard \mbox{~\color{blue} do}$
\\ \hline 
\rowcolor{orange!50}   
2,3 & ~~~~ $ \invariant \wedge \guard $ 
\\ \hline
5a & ~~~~ \begin{minipage}[t]{0.85\textwidth}%
\ifthenelse{\equal{\blocksize}{1}}{}%
{%
\ifthenelse{ \equal{\blocksize}{blank} }{}%
{{\bf Determine block size $ \phantom\blocksize $}\\}%
}
$\mbox{\phantom\repartitionings}$~ \\
$\mbox{\color{blue} ~~~where~}$ \phantom\repartitionsizes
\end{minipage}
\\ \hline
\rowcolor{yellow!75}
6 & ~~~~ $\phantom\beforeupdate $
\\ \hline
8 & ~~~~  \phantom\update 
\\ \hline
5b & ~~~~ \begin{minipage}[t]{0.85\textwidth}%
\phantom\moveboundaries~
\end{minipage}
\\ \hline
\rowcolor{yellow!75}
7 & ~~~~ $\phantom\afterupdate $
\\ \hline
\rowcolor{yellow!75}
2 & ~~~~ $ \invariant  $ 
\\ \hline
 &$\mbox{\color{blue} endwhile} $
\\ \hline \whline
\rowcolor{orange!50}   
2,3 & $ \invariant \wedge \neg( \guard )$ 
\\ \hline
\rowcolor{yellow!75}
1b & $ \postcondition $ 
\\ \hline
\end{tabular}
}

\newcommand{\FlaWorksheetTwo}{
\begin{tabular}{| c | p{0.98\textwidth} |}\hline
Step & $\mbox{\color{blue}Algorithm:~}\routinename$
\\ \hline
\rowcolor{yellow!75}
1a & $ \precondition $ 
\\ \whline
4 & 
\begin{minipage}[t]{0.9\textwidth}%
$\mbox{\phantom{\partitionings}}$~ \\
$\mbox{\color{blue} ~~~where~} $ \phantom{\partitionsizes} 
\end{minipage}
\\ \hline
\rowcolor{orange!50} 
2 & $ \invariant $ 
\\ \hline
3 &$\mbox{\color{blue}while~} \phantom\guard \mbox{~\color{blue} do}$
\\ \hline 
\rowcolor{orange!50} 
2,3 & ~~~~ $ \invariant \wedge \phantom \guard $ 
\\ \hline
5a & ~~~~ \begin{minipage}[t]{0.85\textwidth}%
\ifthenelse{\equal{\blocksize}{1}}{}%
{%
\ifthenelse{ \equal{\blocksize}{blank} }{}%
{{\bf Determine block size $ \phantom\blocksize $}\\}%
}
$\mbox{\phantom\repartitionings}$~ \\
$\mbox{\color{blue} ~~~where~}$ \phantom\repartitionsizes
\end{minipage}
\\ \hline
\rowcolor{yellow!75}
6 & ~~~~ $\phantom\beforeupdate $
\\ \hline
8 & ~~~~  \phantom\update 
\\ \hline
5b & ~~~~ \begin{minipage}[t]{0.85\textwidth}%
\phantom\moveboundaries~
\end{minipage}
\\ \hline
\rowcolor{yellow!75}
7 & ~~~~ $\phantom\afterupdate $
\\ \hline
\rowcolor{orange!50} 
2 & ~~~~ $ \invariant  $ 
\\ \hline
 &$\mbox{\color{blue} endwhile} $
\\ \hline \whline
\rowcolor{orange!50} 
2 & $ \invariant \wedge \neg( \phantom\guard )$ 
\\ \hline
\rowcolor{yellow!75}
1b & $ \postcondition $ 
\\ \hline
\end{tabular}
}

\newcommand{\FlaWorksheetOne}{
\begin{tabular}{| c | p{0.98\textwidth} |}\hline
Step & $\mbox{\color{blue}Algorithm:~}\routinename$
\\ \hline
\rowcolor{orange!50}
1a & $ \precondition $ 
\\ \whline
4 & 
\begin{minipage}[t]{0.9\textwidth}%
$\mbox{\phantom{\partitionings}}$ ~ \\
$\mbox{\color{blue} ~~~where~}$ \phantom{\partitionsizes} 
\end{minipage}
\\ \hline
\rowcolor{yellow!75}
2 & $ \phantom\invariant $ 
\\ \hline
3 &$\mbox{\color{blue}while~} \phantom\guard \mbox{~\color{blue} do}$
\\ \hline 
\rowcolor{yellow!75}
2,3 & ~~~~ $ \phantom\invariant \wedge \phantom \guard$ 
\\ \hline
5a & ~~~~ \begin{minipage}[t]{0.85\textwidth}%
\ifthenelse{\equal{\blocksize}{1}}{}%
{%
\ifthenelse{ \equal{\blocksize}{blank} }{}%
{{\bf Determine block size $ \phantom\blocksize $}\\}%
}
$\mbox{\phantom\repartitionings}$ ~ \\
$\mbox{\color{blue} ~~~where~}$ \phantom\repartitionsizes
\end{minipage}
\\ \hline
\rowcolor{yellow!75}
6 & ~~~~ $\phantom\beforeupdate $
\\ \hline
8 & ~~~~  \phantom\update 
\\ \hline
5b & ~~~~ \begin{minipage}[t]{0.85\textwidth}%
\phantom\moveboundaries~
\end{minipage}
\\ \hline
\rowcolor{yellow!75}
7 & ~~~~ $\phantom\afterupdate $
\\ \hline
\rowcolor{yellow!75}
2 & ~~~~ $ \phantom\invariant  $ 
\\ \hline
 &$\mbox{\color{blue} endwhile} $
\\ \hline \whline
\rowcolor{yellow!75}
2,3 & $ \phantom\invariant \wedge \neg( \phantom\guard )$ 
\\ \hline
\rowcolor{orange!50}
1b & $ \postcondition $ 
\\ \hline
\end{tabular}
}

\newcommand{\FlaWorksheetZero}{
\begin{tabular}{| c | p{0.98\textwidth} |}\hline
Step & $\mbox{\color{blue}Algorithm:~}\routinename$
\\ \hline
\rowcolor{yellow!75}
1a & $ \phantom\precondition $ 
\\ \whline
4 & 
\begin{minipage}[t]{0.9\textwidth}%
$\mbox{\phantom{\partitionings}}$ ~ \\
$\mbox{\color{blue} ~~~where~}$ \phantom{\partitionsizes} 
\end{minipage}
\\ \hline
\rowcolor{yellow!75}
2 & $ \phantom\invariant $ 
\\ \hline
3 &$\mbox{\color{blue}while~} \phantom\guard \mbox{~\color{blue} do}$
\\ \hline 
\rowcolor{yellow!75}
2,3 & ~~~~ $ \phantom\invariant \wedge \phantom \guard$ 
\\ \hline
5a & ~~~~ \begin{minipage}[t]{0.85\textwidth}%
\ifthenelse{\equal{\blocksize}{1}}{}%
{%
\ifthenelse{ \equal{\blocksize}{blank} }{}%
{{\bf Determine block size $ \phantom\blocksize $}\\}%
}
$\mbox{\phantom\repartitionings}$~ \\
$\mbox{\color{blue} ~~~where~}$ \phantom\repartitionsizes
\end{minipage}
\\ \hline
\rowcolor{yellow!75}
6 & ~~~~ $\phantom\beforeupdate $
\\ \hline
8 & ~~~~  \phantom\update 
\\ \hline
5b & ~~~~ \begin{minipage}[t]{0.85\textwidth}%
\phantom\moveboundaries~
\end{minipage}
\\ \hline
\rowcolor{yellow!75}
7 & ~~~~ $\phantom\afterupdate $
\\ \hline
\rowcolor{yellow!75}
2 & ~~~~ $ \phantom\invariant  $ 
\\ \hline
 &$\mbox{\color{blue} endwhile} $
\\ \hline \whline
\rowcolor{yellow!75}
2,3 & $ \phantom\invariant \wedge \neg( \phantom\guard )$ 
\\ \hline
\rowcolor{yellow!75}
1b & $ \postcondition $ 
\\ \hline
\end{tabular}
}

\newcommand{\TBTinitialize}{}
\newcommand{\FlaAlgorithmTBT}{
\begin{tabular}{|l|} \hline
$\mbox{\color{blue}Algorithm:~}\routinename$
\\ \whline
\partitionings \\
$\mbox{\color{blue} ~~~where~}$ \partitionsizes 
\\ 
\TBTinitialize\\
$\mbox{\color{blue}while~} \guard \mbox{~\color{blue} do}$
\\
\ifthenelse{\equal{\blocksize}{1}}{}%
{%
\ifthenelse{ \equal{\blocksize}{blank} }{}%
{~~~~{\bf Determine block size $ \blocksize $}\\}%
}
~~~~ 
\repartitionings \\
~~~$\mbox{\color{blue} ~~~where~}$ \repartitionsizes
\\ \hline
~~~~  \update 
\\ \hline
~~~~ 
\moveboundaries 
\\
$\mbox{\color{blue} endwhile} $
\\ \hline 
\end{tabular}
}

%% file: s0-abstract.tex
We address the reduction to compact band forms, via unitary similarity transformations, for the solution of
symmetric eigenvalue problems and the computation of the singular value decomposition (SVD). 
Concretely, in the first case we revisit the reduction to
symmetric band form while, for the second case, we propose a similar alternative, which transforms the original matrix 
to (unsymmetric) band form, replacing the conventional reduction method that produces a triangular--band output.
In both cases, we describe algorithmic variants of the standard Level-3 BLAS-based procedures, enhanced with look-ahead,
to overcome the performance bottleneck imposed by the panel factorization. Furthermore, 
our solutions 
employ an algorithmic block size that differs from the target bandwidth, illustrating the important performance benefits of this
decision. Finally, we show that our alternative compact band form for the SVD is key to introduce an effective
look-ahead strategy into the corresponding reduction procedure.

%% file: body.tex
\input s1-intro
\input s2-eigen
\input s3-svd
\input s4-evaluation-SEVP
\input s4-evaluation-SVD
\input s5-remarks

%% file: s1-intro.tex
\section{Introduction}

The reduction to tridiagonal form is a crucial operation for the computation
of the eigenvalues of a dense symmetric matrix
when a significant part of the spectrum is required~\cite{GVL3}. 
Similarly, the reduction to bidiagonal form is the preferred option to obtain (all) the singular values
of a dense matrix
via the singular value decomposition (SVD)~\cite{GVL3}. The standard algorithms for these two reductions
in the legacy implementation of LAPACK ({\em Linear Algebra PACKage})~\cite{lapack}
compute these reduced forms via two-sided, fine-grained unitary transformations. 
Unfortunately, these routines are rich in Level-2 BLAS 
({\em Basic Linear Algebra Subroutines})~\cite{blas2}, which are memory-bounded kernels
and, therefore, deliver only a small fraction of the peak (computational) performance of recent computer architectures.

An alternative approach 
replaces these low-performance standard routines with
two-sided reduction (TSR) algorithms that consist of two stages~\cite{Bischof:2000:AST}. 
The idea is to initially transform the dense matrix into a {\em compact band} form (first stage) to next
operate on this by-product in order to yield the desired tridiagonal/bidiagonal form (second stage). 
For symmetric eigenvalue problems (SEVP), the compact by-product is a symmetric band matrix with upper and lower bandwidth $\bd$.
For the SVD, the compact representation corresponds, by convention, to an upper triangular--band 
matrix with upper bandwidth $\bd$.
The appealing property of the TSR algorithms is that the initial reduction mainly consists of 
high performance, compute-bounded Level-3 BLAS~\cite{blas3}, which explains the renewed interest in developing
two-stage TSR algorithms for multicore processors 
as well as manycore accelerators~\cite{CPE:CPE1680,PDP:Davor2011,6114396,Haidar:2013:IPS:2503210.2503292}. 
To conclude this brief review of two-stage TSR algorithms, we note that
in case $w$ is small compared with the problem dimension,
the computational cost of performing the reduction
of the band matrix to tridiagonal/bidiagonal form
is comparable with that of the initial reduction from the original dense matrix to the compact
representation~\cite{GROER1999969,PIROBAND}.
Furthermore, except for a few special problems, computing the eigenvalues/singular values of the tridiagonal/bidiagonal matrices
contributes a minor factor to the global cost of the procedure~\cite{Dhillon:2006:DIM:1186785.1186788,Petschow2013:208,Fernando1994,doi:10.1137/S0895479892242232}.
In contrast, when the associated eigenvectors/singular vectors are to be computed, the cost of accumulating
the unitary transformations during the second stage
can be significant, even if the bandwidth is small compared with the problem dimension~\cite{CPE:CPE1680}.

High-performance routines for the solution of linear systems (e.g., via the LU and QR factorizations~\cite{GVL3})
as well as the initial stage of TSR algorithms
usually implement a right-looking (RL) procedure that, 
at each iteration, factorizes
the {\em current panel} of the matrix, 
and then applies the transformations that realized this reduction to  update the {\em trailing submatrix}.
On today's multicore platforms, the 
factorization of the panel is a performance bottleneck because this operation is mostly memory-bounded 
and, moreover, exhibits a complex set of fine-grain data dependencies.
Fortunately, there exist three  
(to a certain extent complementary) 
techniques to tackle the constraint imposed by the panel factorization:
\begin{list}{}{}
\itemsep=0pt
\item {\sf T1}) exploit the fine-grain parallelism within the panel itself~\cite{Castaldo:2013:SLP:2491491.2491492}; 
\item {\sf T2}) divide the factorization of the current panel into multiple operations whose execution 
      can then be overlapped with certain parts of the trailing update, yielding the so-called algorithms-by-blocks or 
      tile algorithms~\cite{GROER1999969,Buttari200938,Quintana-Orti:2009:PMA:1527286.1527288}; and
\item {\sf T3}) overlap the trailing update with the factorization of the ``next'' panel(s)~\cite{Str98}. 
\end{list}
Note the distinction between 
{\sf T2}), which aims to exploit the parallelism among operations (tasks) in the same iteration of the RL algorithm; and
{\sf T3}), which exploits the parallelism among operations belonging to two (or more) consecutive iterations of the RL algorithm.
Here it is worth pointing out that recent developments on the semi-automatic task-parallelization of dense
linear algebra operations
with the support of a ``runtime'' (such as SuperMatrix, Quark, OmpSs, StarPU, OpenMP, etc.) have partially 
blurred the frontier between
{\sf T2}) and {\sf T3}). 
In particular, when this type of task-parallelization is applied to an algorithm-by-blocks for the solution of linear
systems in order to realize {\sf T2}), the result is that {\sf T3}) is often obtained for free.
For some TSR algorithms though, as we will discuss in the paper, this may be more difficult or even impossible.

In this paper we focus on {\sf T3}), which is usually known as
{\em look-ahead}~\cite{Str98}.
Here, as we do not rely on a runtime to exploit ``inter-iteration'' parallelism, 
we can refer more precisely to this strategy as ``static'' look-ahead.
While this technique has been long known and exploited for the solution of linear systems via 
the LU and QR factorizations\footnote{Static look-ahead is for example the technique embedded in
the implementation of these factorizations in Intel MKL.}, its application to TSR algorithms 
has not been fully discussed explicitly. 
In this paper we show that look-ahead can be introduced in the sophisticated TSR algorithms for SEVP and the SVD, delivering
remarkable performance benefits. In particular, our paper makes the following contributions:
\begin{itemize}
\itemsep=0pt
\item We explore the integration of look-ahead into the reduction of symmetric matrices to band form for SEVP 
      via two-sided unitary transformations. In this line, we propose two variants of the reduction algorithm, 
      enhanced with look-ahead, with distinct performance behaviour depending on the ratio between the algorithmic 
      block size $\bs$ (which dictates the number of columns in the panel,) and the target matrix bandwidth $\bd$.
      While LAPACK (version 3.7.1) and MAGMA (version 2.1.0) both include routines for SEVP to reduce the symmetric input
      matrix to band form, those implementations impose the restriction that
      the algorithmic block size must equal the bandwidth, limiting performance.
      Furthermore, the LAPACK routine for this reduction does not integrate look-ahead.
      The SBR (Successive Band Reduction) package~\cite{Bischof:2000:AST} 
      was a pioneer work that decoupled the bandwidth from the algorithmic block size in this type of reduction,
      but did not integrate look-ahead either.
\item We extend our analysis of look-ahead to the reduction of general matrices 
      to band form for the SVD via two-sided unitary transformations.
      Here we depart from the conventional TSR to band--triangular form, which imposes certain restrictions on
      the application of look-ahead, to advocate for the reduction to band form with equal lower
      and upper bandwidths. This change, in turn, yields 
      two variants for the reduction algorithm for the SVD which are analogous to those identified for SEVP.
\item We demonstrate the performance benefits of static-look ahead, using the reduction to band forms for SEVP
      and the SVD,
      on an Intel-based platform equipped with 8~Haswell cores.
      Our experimental analysis of the optimal block size clearly shows the importance of decoupling the algorithmic
      block size from the bandwidth, and the advantages of each variant.
\end{itemize}
The introduction of look-ahead paves the road to overlapping the panel factorization on a CPU with
the execution of the (rich in Level-3 BLAS) trailing update on an accelerator (e.g., a GPU).
Furthermore, on a multicore architecture, an algorithm that explicitly decomposes the TSR to
expose look-ahead can apply this technique with variable depth, using the support of a runtime such as OpenMP, OmpSs or StarPU.
In both cases, we can expect a notable increase of performance, as 
{\em i)} the panel factorization is potentially removed from the critical path of the algorithm; and 
{\em ii)} the algorithmic block size is decoupled from the bandwidth.


The rest of the paper is structured as follows.
In Sections~\ref{sec:eigen} and~\ref{sec:svd}
we describe the introduction of look-ahead in the first stage of 
the TSR algorithms  for SEVP and the SVD, respectively.
In Section~\ref{sec:evaluation} we assess the benefits of a flexible implementation 
of this technique by experimentally demonstrating its effects
for the TSR of dense matrices to the selected band forms for SEVP and the SVD. 
Finally, in Section~\ref{sec:remarks} we close our paper with a few concluding remarks and a discussion of future work.

To close this introduction, we note that the mathematical equations, algorithms, and the evaluation in the remainder
of the paper are all formulated for problems with real data entries, using orthogonal transformations,
but their extension to the Hermitian case, involving unitary transformations, is straight-forward.

%% file: s2-eigen.tex
\section{TSR for SEVP}
\label{sec:eigen}

\newcommand{\Tp}{{\sf T_S}\xspace}
\newcommand{\Tr}{{\sf T_P}\xspace}
\newcommand{\spi}[1]{^{\mbox{{\sf \footnotesize #1}}}}

\subsection{Basic algorithm}
\label{subsec:basic}

\begin{figure}[t!]
\centerline{\includegraphics[width=0.4\linewidth]{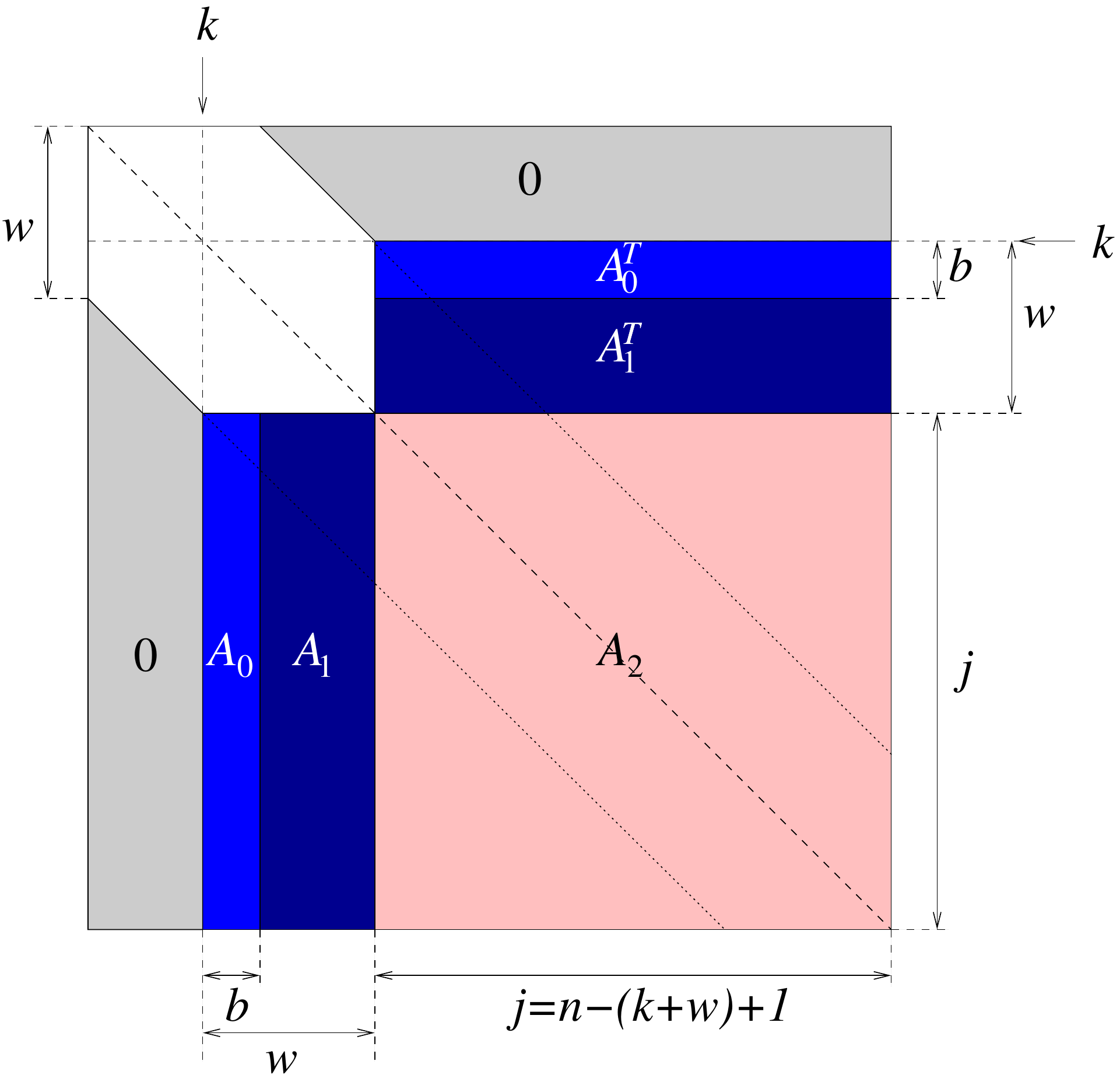}}
\caption{Partitioning of the matrix during one iteration of the reduction to symmetric band form for the solution of SEVP.}
\label{fig:syrdb}
\end{figure}

Let us first describe the algorithm that reduces a dense symmetric matrix $A \in \Rnn$ to symmetric band form,
with bandwidth $\bd$, via orthogonal similarity transformations. This procedure is numerically stable and, moreover, 
preserves the eigenvalues of the matrix~\cite{GVL3}.  
Suppose that the first $k-1$ rows/columns of $A$ have been already reduced to band form; 
the algorithmic block size satisfies $\bs \leq \bd$;
and assume 
for simplicity that $k+\bd+\bs-1 \leq n$; 
see Figure~\ref{fig:syrdb}. 
At this point we note the key roles of the bandwidth $\bd$ and the block size $\bs$, and their interaction. 
The optimal bandwidth itself depends on the efficiency of the second stage of the reduction
and, therefore, it cannot be chosen independently. To complicate things a bit further, the optimal bandwidth also depends
on the problem dimensions and the selected block size. The take-away lesson of this short discussion is that
the best combination of bandwidth and block size depends on several factors, 
some of which are external to the implementation of the first stage.

During the current iteration of the reduction procedure, 
$\bs$ new rows/columns of the band matrix are computed as follows:
\begin{enumerate}
\item {\sc Panel Factorization}. Compute the QR factorization 
\begin{equation}
\label{eqn:updateA0}
   A_0 = Q R,
\end{equation}
where $A_0 \in \R^{j \times \bs}$, with $j=n-(k+\bd)+1$;
$R \in \R^{j \times \bs}$ is upper triangular; and 
the orthogonal matrix $Q$ 
is implicitly assembled using the WY representation~\cite{GVL3}
      as $ Q = I_j + WY^T$, where $W,~Y \in \R^{j \times \bs}$ and $I_j$ denotes the square identity matrix of order $j$.
\item {\sc Trailing Update}. Apply the orthogonal matrix $Q$ to $A_1 \in \R^{j \times \bd-\bs}$ from the left:
\begin{equation}
\label{eqn:updateA1}
   A_1 := Q^T A_1  = (I_j+WY^T)^T A_1 = A_1 + Y (W^T A_1);
\end{equation}
and to $A_2 \in \Rjj$ from both left and right:
\begin{equation}
\label{eqn:updateA2}
\begin{array}{rcl}
   A_2 & := & Q^T A_2 Q  = (I_j+WY^T)^T A_2 (I+WY^T) \\
       & =          & A_2 + Y W^T A_2 + A_2WY^T + YW^T A_2 WY^T.
\end{array}
\end{equation}
During this last operation (only the lower or upper
triangular part of) $A_2$ is updated, via the following
sequence of Level-3 BLAS operations:
\begin{eqnarray}
\label{eqn:updateA2symm} 
X_1 &:=& A_2 W, \\
\label{eqn:updateA2gemm1} 
X_2 &:=& \frac{1}{2} X_1^T W, \\
\label{eqn:updateA2gemm2} 
X_3 &:=& X_1 + Y X_2, \\
\label{eqn:updateA2syr2k} 
A_2 &:=& A_2 + X_3 Y^T + Y X_3^T.
\end{eqnarray}
\end{enumerate}

Provided $\bs$ and $\bd$ are both small compared with $n$, the global cost of the reduction of a full matrix
to band form is $4n^3/3$ floating-point arithmetic operations (flops).\footnote{Hereafter, lower order terms are neglected 
in the theoretical costs.} Furthermore,
the bulk of the computation is performed in terms of the
Level-3 BLAS operations 
in~\nref{eqn:updateA2symm} and~\nref{eqn:updateA2syr2k}. 

The problem with this basic algorithm is that the panel factorization in~\nref{eqn:updateA1}
is mainly memory-bounded (at least, for the usual values of $\bs$) 
as well as features some complex dependencies so that, as the 
number of cores performing the factorization is increased, the panel operation 
rapidly becomes a performance bottleneck. 
We next describe how to solve this problem via two algorithmic variants that implement a static look-ahead
in order to overlap in time (i.e., run concurrently) 
the execution of the trailing update for the current iteration with the factorization of the next panel.

\subsection{Introducing look-ahead}
\label{subsec:look-ahead}

Consider the blocks $A_0$, $A_1$, $A_2$ involved in iteration $k$ of the basic algorithm 
(see Figure~\ref{fig:syrdb});
and let us refer to the panel that will be factorized 
in the subsequent iteration $\bar{k}=k+\bs-1$ as $\bar{A}_0$. 
The key to formulate a variant of the basic algorithm enhanced with look-ahead lies in:
\begin{enumerate}
\itemsep=0pt
\item identifying the parts of the trailing submatrix $\left[A_1,~A_2\right]$ that will become $\bar{A}_0$ 
during the next iteration;  
\item isolating the updates corresponding to
application of the orthogonal transformations in~\nref{eqn:updateA2} that
affect $\bar{A}_0$ from those which modify those parts of $\left[A_1,~A_2\right]$ that do not overlap with
$\bar{A}_0$; and 
\item during iteration $k$, overlapping
      the factorization of the subsequent panel $\bar{A}_0$ (look-ahead factorization) with the updates corresponding to this iteration.
\end{enumerate}

\begin{figure}[t!]
\centerline{\includegraphics[width=0.4\linewidth]{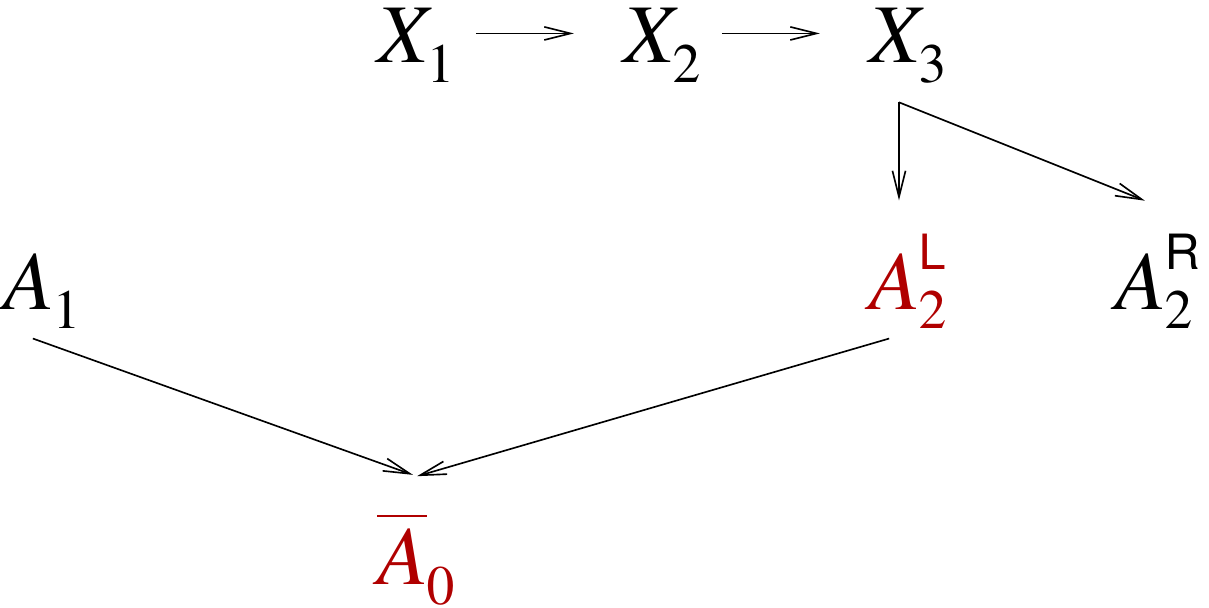}}
\caption{Dependencies among operations appearing in Variant~V2 of the initial TSR to symmetric band form. For simplicity, each operation
         is identified by its output operand.} 
\label{fig:dependencies_variant2}
\end{figure}

At this point we distinguish two cases, leading to two variants of the TSR algorithm with look-ahead, depending on the
relation between $\bs$ and $\bd$:
\begin{itemize}
\item Variant~V1: $2\bs \leq \bd$.
      In this case, $\bar{A}_0$ lies entirely within $A_1$, as the number of columns in the latter satisfies $\bd-\bs\geq \bs$. 
      We then define the following partitioning of the trailing submatrix:
\begin{equation}
\label{eqn:partitionV1}
\renewcommand{\arraystretch}{1.2}
\left[\begin{array}{cIc}\color{black}{A_1} & \color{black}{A_2} \end{array} \right] = \left[\begin{array}{c|cIc} \color{darkred}{A_1\spi{L}} & \color{black}{A_1\spi{R}} & \color{black}{A_2} \end{array}\right]   
    = \left[\begin{array}{c|cIc} \begin{array}{c}\color{darkred}{A_1\spi{TL}} \\\hline \color{darkred}{\bar{A}_0}\end{array} & \color{black}{A_1\spi{R}} & \color{black}{A_2} \end{array}\right],
\end{equation}
where $A_1\spi{L}$ consists of $\bs$ columns, $A_1\spi{TL}$ is $\bs \times \bs$, and we (use the red color to) distinguish 
those blocks that overlap in the column range of
$\bar{A}_0$.

During iteration $k$, we can then perform the following three groups of operations (\textcolor{darkred}{left},
\textcolor{black}{middle} and \textcolor{black}{right}) concurrently:
\[
  \begin{array}{@{}l}
    \left. \begin{array}{@{}l}
    \mbox{{\color{darkred}{Sequential panel factorization}}}\\ [0.05in]
    {\color{darkred}{A_{1}\spi{L}} := Q^T A_1\spi{L}} \\ [0.05in]
    {\color{darkred}{\bar{A}_{0}} = \bar{Q} \bar{R}}
\end{array} \right. 
\hspace{0.5in}
    \left. \begin{array}{@{}l}
    ~\\ [0.05in]
    {\color{black}{A_{1}\spi{R}} := Q^T A_1\spi{R}} \\ [0.05in]
    ~
\end{array} \right. 
\hspace{0.5in}
    \left. \begin{array}{@{}l}
    \mbox{{\color{black}{Parallel remainder update}}}\\ [0.05in]
    {\color{black}{A_{2}} := Q^T A_2 Q} \\ [0.05in]
    ~
\end{array} \right. 
  \end{array}
\]
Note that, with this partitioning, $\bar{A}_0$ is generally small compared with
$A_2$. Therefore, we can expect that
the factorization of the subsequent panel $\bar{A}_0$ can be overlapped with the update of $A_2$ on the right, eliminating
the former from the critical path of the reduction. 
On a multicore architecture, we can achieve this by dedicating a few threads/cores to the panel factorization while
the remaining ones compute the trailing update. On a CPU-GPU system, the CPU can take care of the panel factorization
while the GPU updates the trailing submatrix.
Hereafter, we will refer to these (two groups of) computational resources, few threads/CPU and many threads/GPU, 
as $\Tp$ (for sequential) and $\Tr$ (for parallel) respectively.

There exists a direct dependency between the two operations on the left-hand side group, that we can denote as
$A_{1}\spi{L} \rightarrow \bar{A}_0$. Here,
the update of $A_{1}\spi{L}$ is a Level-3 BLAS operation that in general will offer
low performance as the width of the panel is usually small. 
Due to the dependency and this low performance, the update of $A_{1}\spi{L}$ will be performed by
$\Tp$.
As for the update of $A_{1}\spi{R}$, in the middle ``group'',
in case this is also a narrow column panel (i.e., $\bd-2\bs$ is small), 
we can expect low performance from it, so that it should join the group of ``sequential'' 
operations on the left (red group), to be performed by $\Tp$.
Otherwise, it can be merged with the ``parallel'' group on the right,
to be computed by $\Tr$.

\item Variant~V2: $2\bs > \bd$.
      In this case, $\bar{A}_0$ expands beyond the columns of $A_1$ to partially overlap with $A_2$.
      Let us consider the following partitioning:
\begin{equation}
\renewcommand{\arraystretch}{1.2}
\left[\begin{array}{cIc}\color{black}{A_1} & \color{black}{A_2} \end{array} \right] = \left[\begin{array}{cIc|c} \color{darkred}{A_1} & \color{darkred}{A_2\spi{L}} & \color{black}{A_2\spi{R}} \end{array}\right]   
    = \left[\begin{array}{c|c} 
\begin{array}{c}
\begin{array}{cIc}\color{darkred}{A_1\spi{T}} & \color{darkred}{A_2\spi{TL}}\end{array} \\\hline
\color{darkred}{\bar{A}_0} \end{array}
& 
\color{black}{A_2\spi{R}}  
\end{array}\right],
\end{equation}
where $\left[A_1,~A_2\spi{L}\right]$ consists of $\bs$ columns and 
$\left[A_1\spi{T},~A_2\spi{TL}\right]$ is $\bs \times \bs$.

During iteration $k$, 
we initially compute the following operations: 
\begin{eqnarray}
    A_{1} &:=& Q^T A_1, \\ 
    X_1 &:=& A_2 W, \\ 
    X_2 &:=& \frac{1}{2}X_1^T W,  \\ 
    X_3 &:=& X_1+YX_2,
\end{eqnarray}
which correspond to the update of $A_1$ and part of the computations necessary for the update of
$A_2$; see~\nref{eqn:updateA2symm}--\nref{eqn:updateA2syr2k}.
After this is completed, we can concurrently perform 
the following two groups of operations:
\[
  \begin{array}{@{}l}
    \left. \begin{array}{@{}l}
    \mbox{{\color{darkred}{Sequential panel factorization}}}\\ [0.05in]
    {\color{darkred}{A_{2}\spi{L}} := A_2\spi{L} + X_3(Y\spi{L})^T + Y(X_3\spi{L})^T} \\ [0.05in]
    {\color{darkred}{\bar{A}_0} = \bar{Q}\bar{R}} 
\end{array} \right. 
\hspace{0.5in}
    \left. \begin{array}{@{}l}
    \mbox{{\color{black}{Parallel remainder update}}}\\ [0.05in]
    {\color{black}{A_{2}\spi{R}} := A_2\spi{R} + X_3(Y\spi{R})^T + Y(X_3\spi{R})^T} \\ [0.05in]
~\\
\end{array} \right.
\end{array}
\]
Here, $Y=\left[Y\spi{L},~Y\spi{R}\right]$ and
      $X_3=\left[X_3\spi{L},~X_3\spi{R}\right]$ are partitionings conformal with $A_2=\left[A_2\spi{L},~A_2\spi{R}\right]$.
As in the previous variant (case), this pursues the goal of overlapping the factorization of the next
panel $\bar{A}_0$ with a sufficiently-large Level-3 BLAS. In general, $\bar{A}_0$ is small compared with the trailing
submatrix $A_2\spi{R}$ so that we can expect this is the case.

To close the discussion of Variant~V2, we note the collection of dependencies appearing among the operations identified
in this case; see Figure~\ref{fig:dependencies_variant2}.
For these operations, the panel width
determines whether they
involve narrow panels and, therefore, can be considered memory-bounded 
low-performance kernels.
Thus, together with the dependencies, 
this property will 
ultimately decide whether they are moved to the groups of either sequential or parallel kernels, to be tackled 
by $\Tp$ or $\Tr$, respectively.
For example, one possibility is to update $A_1$ on $\Tp$, while $X_1,~X_2,~X_3$ are being computed by $\Tr$; when all these
operations are completed,
we can continue with the update of
    $A_{2}\spi{L}$ and the factorization of
    $\bar{A}_0$ on $\Tp$, while
    $A_{2}\spi{R}$ is being updated by $\Tr$.
\end{itemize}

At this point, it is fair to ask what is the value of explicitly exposing static look-ahead if the same effect 
could be obtained, in principle, 
with the combination of an algorithm-by-blocks and the support of a task-parallelizing runtime. 
Armed with the previous discussion of look-ahead, we can now offer several
arguments in response to this question:
\begin{enumerate}
\itemsep=0pt
\item Exposing the look-ahead variant provides a better understanding of the algorithms.
\item Static look-ahead can be as efficient as or even outperform a runtime-assisted dynamic solution~\cite{catalan17}. 
      The reason is that, for regular dense linear
      algebra operations such as those in the Level-3 BLAS, dividing these kernels into fine-grain operations incurs into 
      some packing/unpacking overheads. In addition, the use of a runtime promotes the exploitation of
      task-parallelism at the cost of a suboptimal use of the cache hierarchy.
\item As exposed in the next section, 
      for the tile algorithm proposed for the reduction to triangular--band form, the application of a runtime may not allow 
      {\em per se}
      the exploitation of task-parallelism among operations belonging to different iterations.
\item The look-ahead variants do not require the implementation of tuned kernels to factorize blocks with special structures as those
      that appear in the algorithm-by-blocks, and apply the corresponding transformations.\footnote{We recognize that 
      this problem can be overcome by a careful reconstruction of the orthogonal factor, but this comes 
      at the cost of an increase in the computational cost of the panel factorization~\cite{Ballard20153}.}
      Moreover, they do not incur the overhead
      due to the operation with these kernels and do not have an internal block size that needs to be
      tuned~\cite{Buttari200938,Quintana-Orti:2009:PMA:1527286.1527288}.
\item For CPU-GPU platforms, static look-ahead can be the only practical means to eliminate the panel factorization from
      the critical path of the algorithm. 
\end{enumerate}

%% file: s3-svd.tex
\section{TSR for the SVD}
\label{sec:svd}

\newcommand{\ql}{U}
\newcommand{\qr}{V}
\newcommand{\wl}{W_U}
\renewcommand{\wr}{W_V}
\newcommand{\yl}{Y_U}
\newcommand{\yr}{Y_V}

\subsection{Triangular--band form}

\begin{figure}[t!]
\centerline{\includegraphics[width=0.4\linewidth]{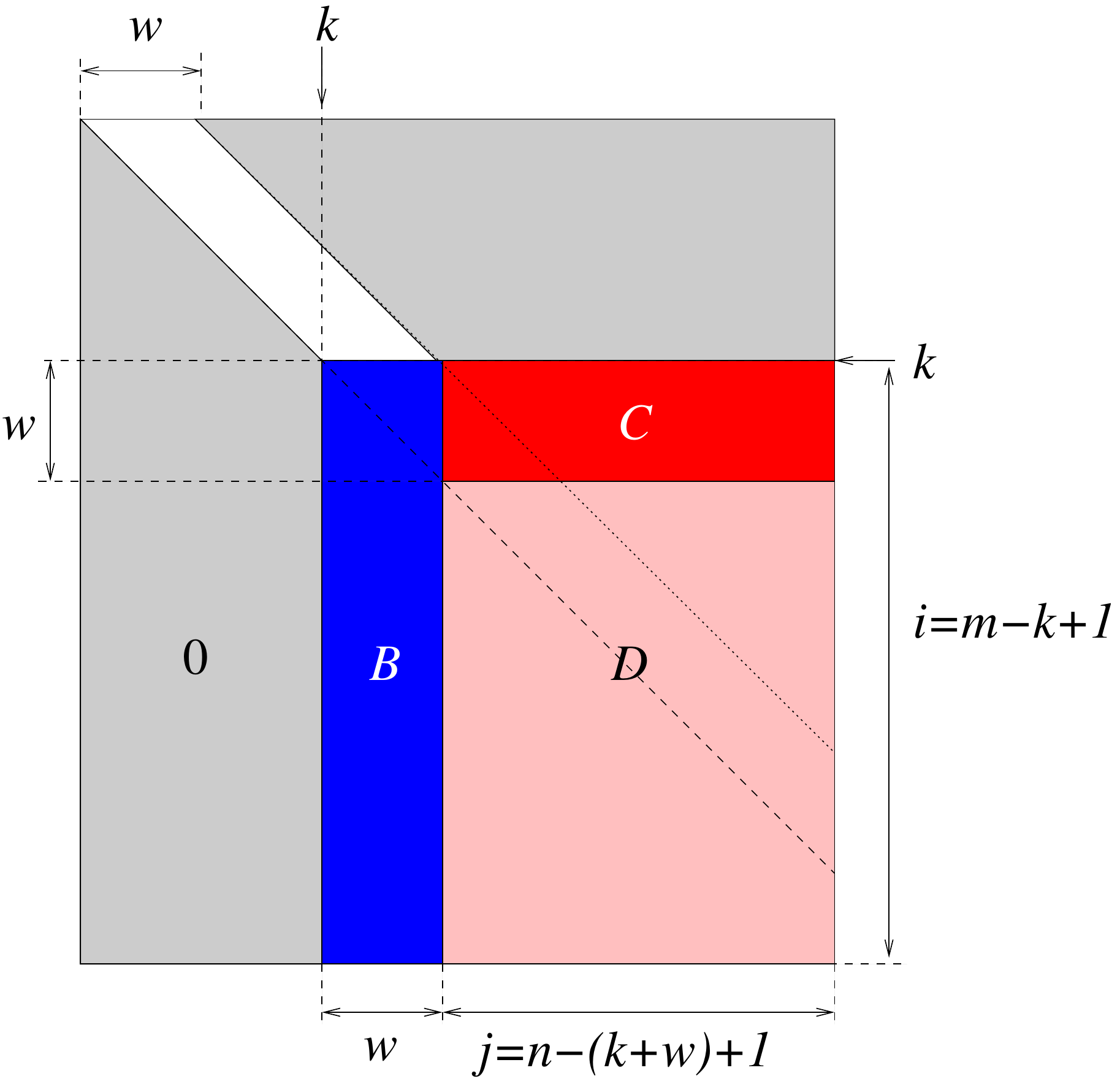}}
\caption{Partitioning of the matrix during one iteration of the reduction to triangular--band form for the SVD ($\bd=\bs$).}
\label{fig:gerdb}
\end{figure}

The conventional algorithm for the first stage of the TSR algorithm for the SVD computes an upper triangular matrix with
upper bandwidth $\bd$. To describe this procedure, consider a matrix $A\in\Rmn$, where the first $k$ rows/columns
have already been reduced to the desired triangular--band form; and assume that $k+w+b-1 \leq m,n$, with $m \geq n$.
Furthermore, {\em let us consider initially the simpler case} with $\bd=\bs$; see Figure~\ref{fig:gerdb}.
During the current iteration, the following computations thus advance the reduction by
$\bd$ additional rows/columns:
\begin{enumerate}
\item {\sc Left Panel Factorization}. Compute the QR factorization
\begin{equation}
\label{eqn:updateB}
   B = \ql R,
\end{equation}
where $B \in \R^{i \times \bd}$, with $i=m-k+1$;
$R \in \R^{i \times \bd}$ is upper triangular; and
$ \ql = I_i + \wl\yl^T$ is orthogonal, with $\wl,~\yl \in \R^{i \times \bd}$.
\item {\sc Left Trailing Update}. Apply $\ql$ to $E = \left[\begin{array}{c} C\\ D \end{array}\right] \in \R^{i \times j}$, with $j=n-(k+w)+1$, from the left:
\begin{equation}
\label{eqn:updateCD}
E := \ql^T
E = 
   (I_i+\wl\yl^T)^T 
E = 
E  + \yl (\wl^TE).
\end{equation}
\item {\sc Right Panel Factorization}. Compute the LQ factorization~\cite{GVL3}
\begin{equation}
\label{eqn:updateC}
   C = L \qr^T,
\end{equation}
where $C \in \R^{\bd \times j}$;
$L \in \R^{\bd \times j}$ is lower triangular; and
$ \qr = I_j + \wr\yr^T$ is orthogonal, with $\wr,~\yr \in \R^{j \times \bd}$.
\item {\sc Right Trailing Update}. Apply $\qr$ to $D \in \R^{(i-w) \times j}$, from the right:
\begin{equation}
\label{eqn:updateD}
D := D \qr = 
   D (I_j+\wr\yr^T) = 
D + (D\wr) \yr^T.
\end{equation}
\end{enumerate}
Assuming $\bd\ll m,n$, this algorithm requires $4(mn^2-n^3/3)$ flops and the major part of these operations are concentrated in 
the trailing updates~\nref{eqn:updateCD}, \nref{eqn:updateD}, which correspond to high performance Level-3 BLAS. 

\subsection{Triangular--band form and look-ahead}

Unfortunately, the two panel factorizations 
in~\nref{eqn:updateB}, \nref{eqn:updateC} impose the same bottleneck as that discussed for the reduction
to the symmetric band form.
Furthermore, in the reduction to triangular--band form
overcoming this problem via a look-ahead strategy enforces certain constraints on the relation
between $\bd$ and $\bs$ that may impair performance.
Let us explain this in detail via three cases, where the first one corresponds to the simple scenario with $\bd=\bs$, 
and the remaining two decouple the block size from the bandwidth so that $\bs \leq \bd$.


\begin{figure}[t!]
\begin{tabular}{cc}
\begin{minipage}{0.28\textwidth}
\centerline{\includegraphics[width=\textwidth]{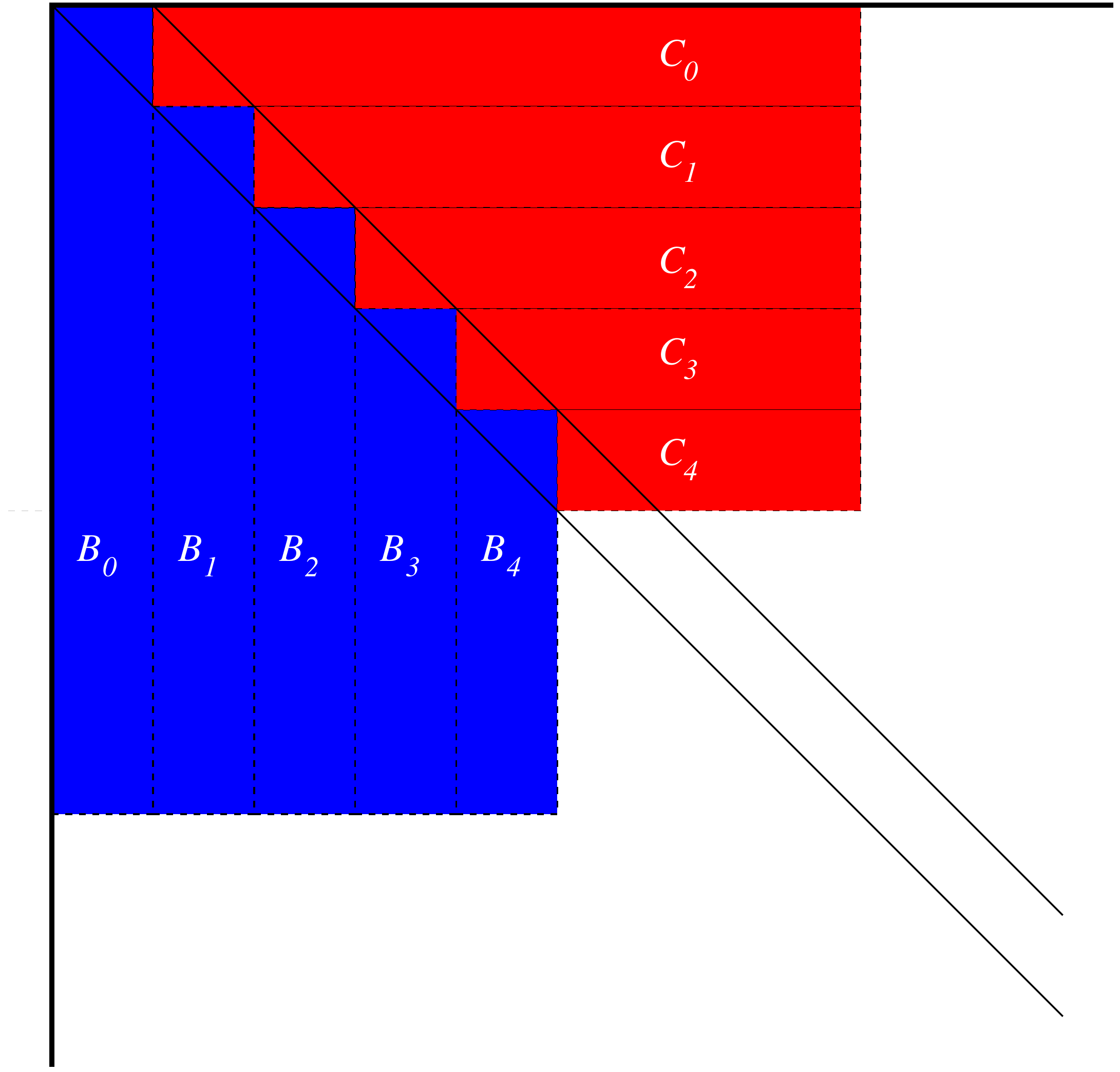}}
\end{minipage}
&
\begin{minipage}{0.7\textwidth}
\centerline{\includegraphics[width=\textwidth]{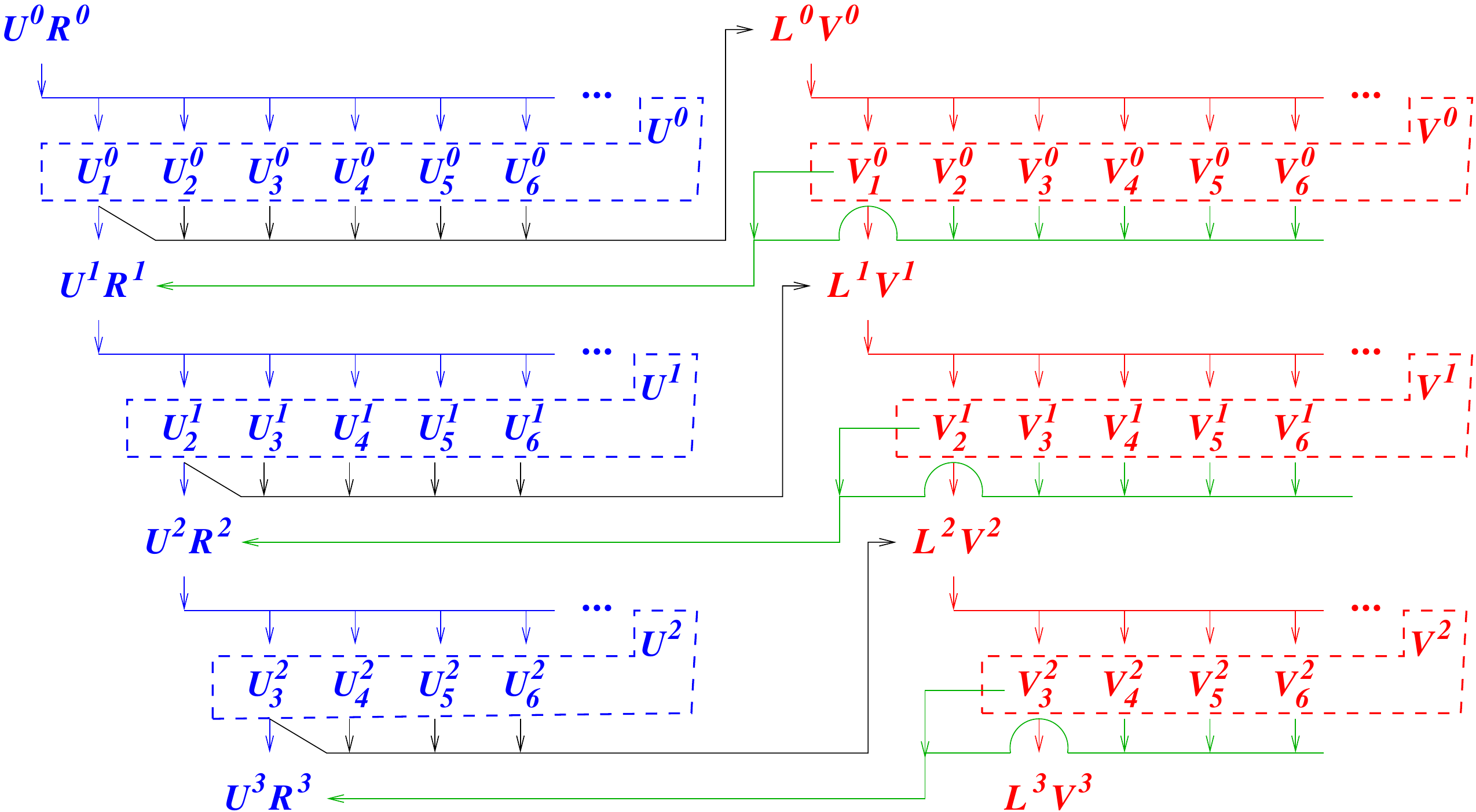}}
\end{minipage}
\end{tabular}
\caption{Matrix partitioning (left) and dependencies (right) for the reduction to triangular--band form for the SVD ($\bd=\bs$).}
\label{fig:gerdb_wb}
\end{figure}

\paragraph{First case: $\bd=\bs$.}
Consider the scenario illustrated in Figure~\ref{fig:gerdb_wb} where
the superindices (as, e.g., in the factorization $B^0=U^0 R^0$) indicate the iteration count (starting at 0), and the subindices 
specify the index of the block being updated by the corresponding transformations (either from the left, as in
$U^0_1$, or from the right, as in $V^0_1$). The arrows correspond to data dependencies and thus define a partial ordering
for the execution of the operations.
For simplicity, let us aggregate the updates
$U^{k}_{k+1},U^{k}_{k+2},U^{k}_{k+3},\ldots$ into a single macro-update $U^k$ and, similarly,
$V^{k}_{k+1},V^{k}_{k+2},V^{k}_{k+3},\ldots$ into the macro-update $V^k$.
Then, it is easy to verify that the existing dependencies enforce
the strict ordering
\textcolor{black}{%
$U^0R^0 ~\textcolor{black}{\rightarrow}~ 
 U^0    ~\textcolor{black}{\rightarrow}~ 
 L^0V^0 ~\textcolor{black}{\rightarrow}~ 
 V^0    ~\textcolor{black}{\rightarrow}~ 
 U^1R^1 ~\textcolor{black}{\rightarrow}~ \ldots
$},
revealing that it is not possible to exploit look-ahead in this case.

\begin{figure}[t!]
\begin{tabular}{cc}
\begin{minipage}{0.28\textwidth}
\centerline{\includegraphics[width=\textwidth]{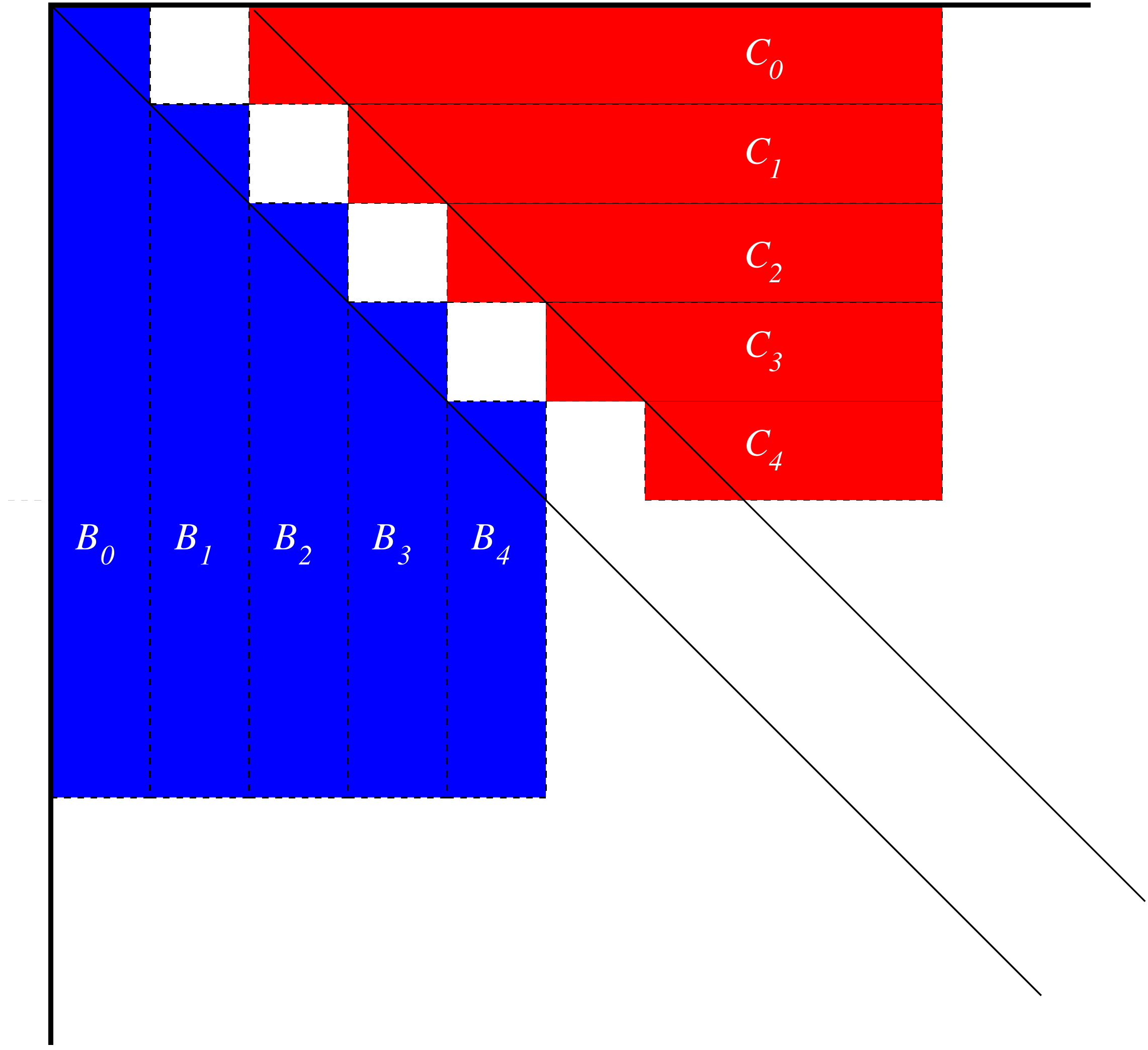}}
\end{minipage}
&
\begin{minipage}{0.7\textwidth}
\centerline{\includegraphics[width=\textwidth]{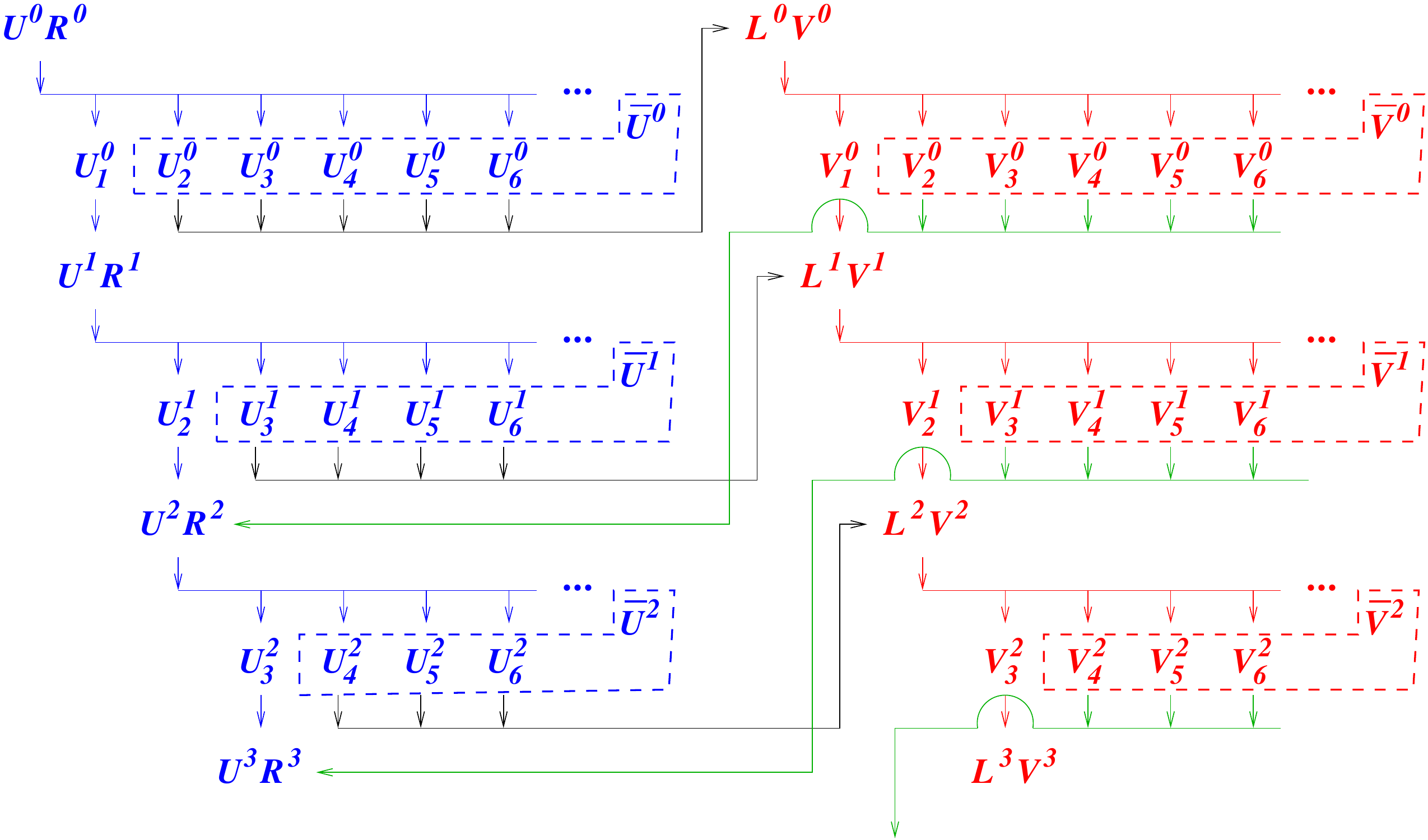}}
\end{minipage}
\end{tabular}
\caption{Matrix partitioning (left) and dependencies (right) for the reduction to triangular--band form for the SVD ($\bd=2\bs$).}
\label{fig:gerdb_w2b}
\end{figure}

\paragraph{Second case: $\bd=2\bs$.}
The new situation is displayed in Figure~\ref{fig:gerdb_w2b}, which will be leveraged to expose that the dependency problem identified
in the previous case partially remains. 
In particular, let us now aggregate the updates 
$U^{k}_{k+2},U^{k}_{k+3},U^{k}_{k+4},\ldots$ into a single macro-update $\bar{U}^k$ and
$V^{k}_{k+2},V^{k}_{k+3},V^{k}_{k+4},\ldots$ into $\bar{V}^k$.
Moreover, for simplicity let us consider that the updates of the form $U^{k}_{k+1}$ and $V^{k}_{k+1}$ 
respectively occur inside the factorizations
$U^{k+1}R^{k+1}$ and $L^{k+1}V^{k+1}$.
Then, we can initially compute $U^0R^0$;
followed by the overlapped execution of
the factorization $U^1R^1$ with the macro-update $\bar{U}^0$; and the factorization
$L^1V^1$ next. 
At this point, we would like to 
overlap 
$U^2R^2$ with $\bar{U}^1$ {\em and}
$L^1V^1$ with $\bar{V}^0$.
{\em However, because of the dependencies, we can exploit one of the overlappings, but not both}.
To see this, assume our goal is to encode the first overlapping. Then, 
$\bar{V}^0$ must be available (green lines), but $L^1V^1$ cannot be computed yet (black lines). 
Therefore, the second overlapping is not possible.
Conversely, assume that we intend to encode the second overlapping.
Then, $\bar{U}^1$ must be available (black lines), but $L^1V^1$ cannot be computed yet (green lines), 
and the first overlapping is not possible.

\begin{figure}[t!]
\begin{tabular}{cc}
\begin{minipage}{0.28\textwidth}
\centerline{\includegraphics[width=\textwidth]{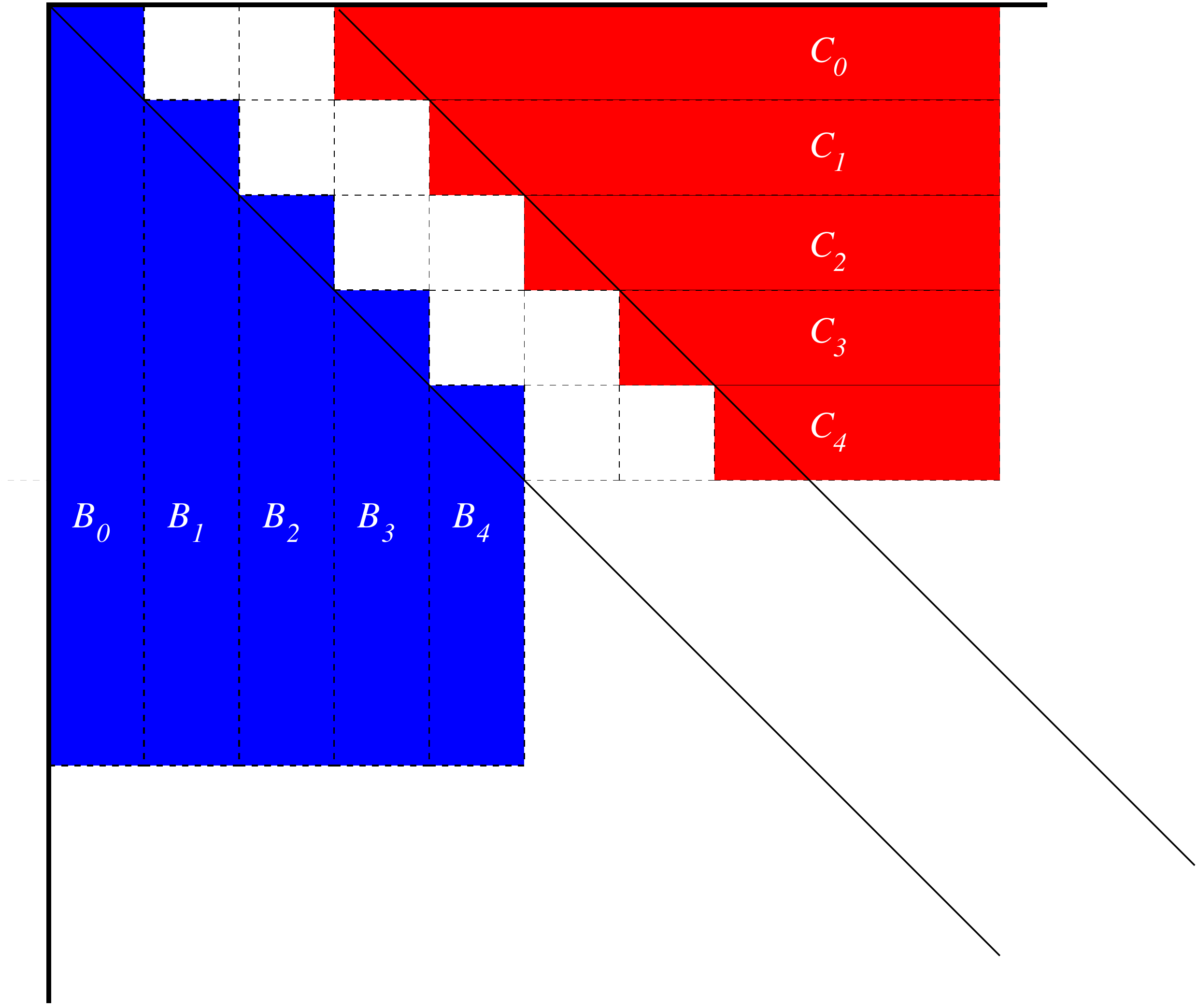}}
\end{minipage}
&
\begin{minipage}{0.7\textwidth}
\centerline{\includegraphics[width=\textwidth]{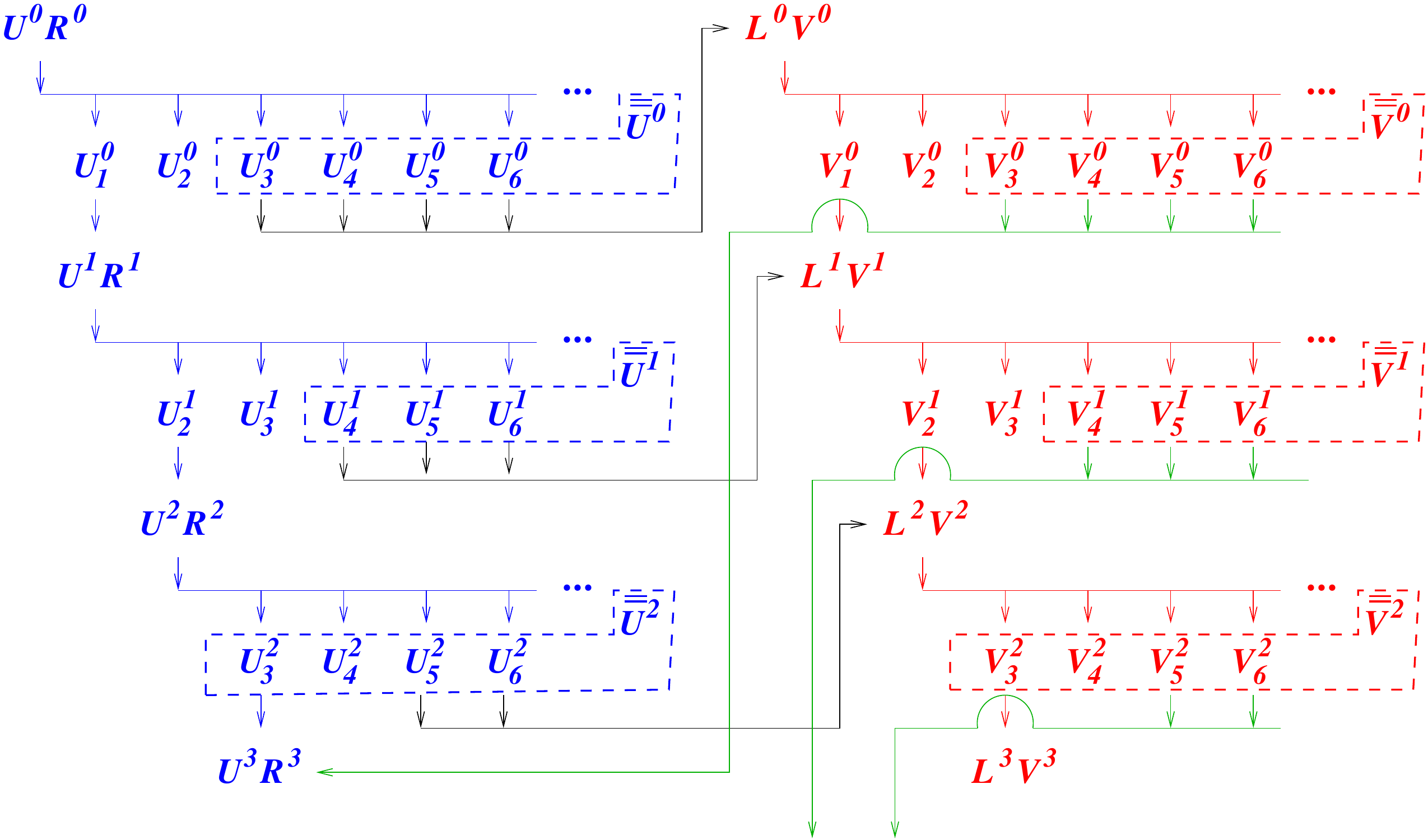}}
\end{minipage}
\end{tabular}
\caption{Matrix partitioning (left) and dependencies (right) for the reduction to triangular--band form for the SVD ($\bd=3\bs$).}
\label{fig:gerdb_w3b}
\end{figure}

\paragraph{Third case: $\bd=3\bs$.} Consider next the scenario in Figure~\ref{fig:gerdb_w3b}.
Let us use 
the macro-update $\dbar{U}^k$ to stand now for 
$U^{k}_{k+3},U^{k}_{k+4},U^{k}_{k+5},\ldots$; and 
$\dbar{V}^k$ for
$V^{k}_{k+3},V^{k}_{k+4},V^{k}_{k+5},\ldots$.
Also, assume for simplicity that the updates of the form 
$U^{k}_{k+1},~U^{k}_{k+2}$ and
$V^{k}_{k+1},~V^{k}_{k+2}$ 
respectively occur as part of the factorizations
$U^{k+1}R^{k+1}$ and $L^{k+1}V^{k+1}$.
As in the previous case, we can initially compute $U^0R^0$;
followed by the overlapped execution of
the factorization $U^1R^1$ with the macro-update $\bar{U}^0$; and the factorization
$L^1V^1$ next. 
However, because of the distinct dependencies that are present in this third case, nothing prevents us 
in the following steps from overlapping 
$U^2R^2$ with $\bar{U}^1$; 
$L^1V^1$ with $\bar{V}^0$;  
$U^3R^3$ with $\bar{U}^2$; and so on.

\paragraph{}
The conclusion from this study is that, in the reduction to triangular--band form, 
applying look-ahead for both the left and right panel factorizations requires $\bd\geq 3\bs$.
While this is doable, it has some practical implications on the relation between the practical
values of $\bd$ and $\bs$. On one hand, $\bd$ should be kept small to moderate because the selection of a large value
delays much of the computational cost into the second stage (reduction from triangular--band to bidiagonal form), which 
is realized via slower Level-2 BLAS. 
On the other hand, $\bs$ needs to be set to a large value as otherwise the updates will not fully benefit
from the performance of Level-3 BLAS. 
The practical consequence is that the constraint that $w\geq 3\bs$ in this type of decomposition can exert a strong negative 
impact on performance.


\paragraph{Pipelining the factorizations.}
Consider again the simple case $\bd=\bs$ and 
the operations~(\ref{eqn:updateB})--(\ref{eqn:updateD}) to be computed in a given iteration.
A different (but related) possibility to attain an overlapped execution
is to decompose the left panel factorization in this iteration into several
column micro-panels, of width $\bs_l < \bd$, and then overlap the factorization of the micro-panels with their application
to the remaining part(s) of the matrix (i.e., within the same micro-panel within $B$ as well as to $E$). 
When this is completed, the algorithm proceeds to the right panel factorization 
in the same iteration, and basically applies the same idea using row micro-panels of height $\bs_r$. 
With this strategy we can choose a value for the micro-panels that simply satisfies
$\bd=2\bs_l=2\bs_r$. However, note that with this approach, 
the first micro-panel for both the left and right panel factorizations (at each iteration) cannot
be overlapped, with a strong negative impact on performance. This will enforce us to select
$\bd\geq 3\bs_l, 3\bs_r$, with the same consequences as those discussed in the previous paragraph.
Even worse, with this approach, the first left and right panel factorizations of each iteration cannot be overlapped.

\paragraph{Other implementations.}
The discussion of the reduction  to triangular--band reveals a strong limitation 
when aiming to exploit task-parallelism among operations belonging to different iterations.
We note that this algorithm is precisely the selection that was made for the 
message-passing implementation of TSR to triangular--band form in~\cite{GROER1999969}. 
It is also the choice for the tile algorithms in PLASMA that perform 
this reduction on multicore platforms in~\cite{6114396,Haidar:2013:IPS:2503210.2503292}.
In addition, all of these implementations couple the algorithmic block size to the bandwidth, so that $\bs=\bd$, with a potential
negative effect on performance.


\subsection{Band form and look-ahead}

In~\cite{GROER1999969} the authors explored the triangular--band reduction as well as an 
alternative algorithm that reduces the dense
matrix to a band form with the same upper and lower bandwidth. 
However, the latter algorithm was abandoned in that work as it did
not offer any special advantage. In this subsection we show that this approach is actually the key to obtaining two variants for the 
reduction to band form for the SVD, enhanced with look-ahead, which are analogous to those already presented for SEVP
in subsection~\ref{subsec:look-ahead}.
Importantly, this approach does not enforce that $\bd\geq 3\bs$, as was the case of the reduction
to triangular--band form.
Before we review this algorithm, we note that the cost of applying this procedure to reduce a dense matrix to 
band form, with upper and lower bandwidth $\bd=\bd'/2$,
is about the same as that of reducing the matrix to triangular--band form with bandwidth $\bd'$. 

\begin{figure}[t!]
\centerline{\includegraphics[width=0.6\linewidth]{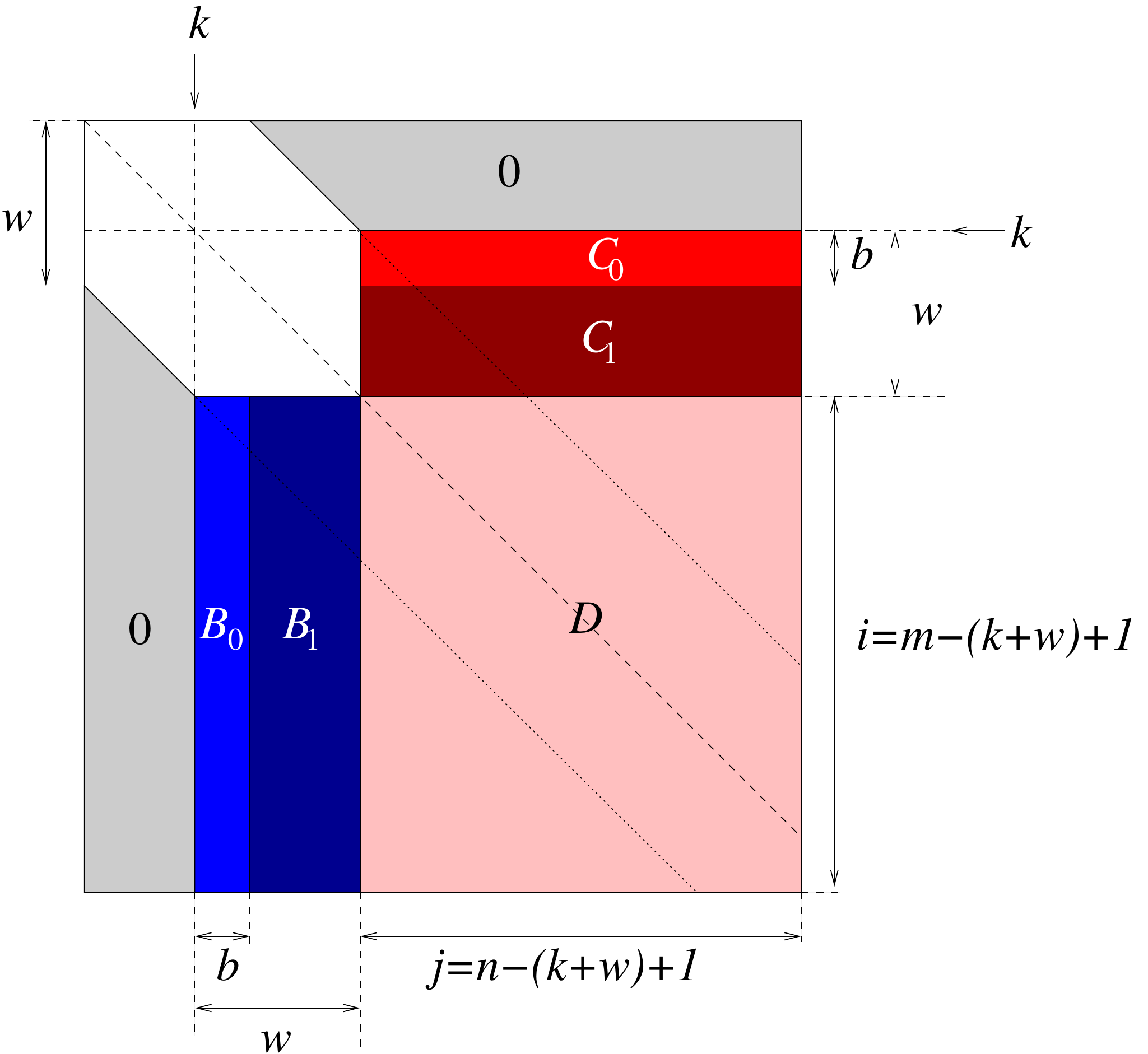}}
\caption{Partitioning of the matrix during one iteration of the reduction to band form for the SVD.}
\label{fig:gerdb_2k}
\end{figure}

The basic algorithm (i.e., without look-ahead) 
is very similar to the reduction to symmetric band form, with the differences stemming from the fact that
$A$ is now an unsymmetric matrix, which requires separate left and right factorizations.
As usual, consider that the first $k-1$ rows/columns of $A$ have been already reduced to band form; 
select $\bs \leq \bd$;
and assume for simplicity that $k+\bd+\bs-1 \leq m,n$; 
see Figure~\ref{fig:gerdb_2k}. 

During the current iteration of the reduction procedure, 
$\bs$ new rows/columns of the band matrix are computed as follows (for brevity, we do not state explicitly the dimensions and
properties of the matrix blocks/factors in the following, as they can be easily derived from the context and Figure~\ref{fig:gerdb_2k}):
\begin{enumerate}
\item {\sc Left Panel Factorization}. Compute the QR factorization 
\begin{equation}
\label{eqn:updateB0}
   B_0 = \ql R,~~\mbox{\rm with}~~\ql = I_i + \wl\yl^T.
\end{equation}
\item {\sc Left Trailing Update}. Apply $\ql$ to the trailing submatrix: 
\begin{eqnarray}
\label{eqn:updateBD}
   B_1 &:=& \ql^T B_1  = B_1 + \yl (\wl^T B_1);\\
   D &:=& \ql^T D  \,\,= D + \yl (\wl^T D);
\end{eqnarray}
\item {\sc Right Panel Factorization}. Compute the LQ factorization  
\begin{equation}
\label{eqn:updateC0}
   C_0 = L \qr^T,~~\mbox{\rm with}~~\qr = I_j + \wr\yr^T.
\end{equation}
\item {\sc Right Trailing Update}. Apply $\qr$ to the trailing submatrix: 
\begin{eqnarray}
\label{eqn:updateCD2}
   C_1 &:=& C_1 \qr = A_1 + (A_1\wr) \yr^T;\\
\label{eqn:updateCD22}   
   D   &:=& D\qr \,\,= D + (D\wr) \yr^T.
\end{eqnarray}
\end{enumerate}

From these expressions, let us now re-consider the two cases leading to Variants~V1 and~V2 of the look-ahead strategy:
\begin{itemize}
\item Variant~V1: $2\bs \leq \bd$.
      The next panels $\bar{B}_0$  and $\bar{C}_0$ 
      lie entirely within $B_1$ and $C_1$, respectively. Therefore, the update and factorization of these panels can
      be overlapped with the updates performed on~$D$ from the left and right, respectively.
\item Variant~V2: $2\bs > \bd$.
      Now both $\bar{B}_0$  and $\bar{C}_0$ extend to overlap with $D$.
      The key to introduce look-ahead 
      is that the left and right updates of $D$ can be performed ``simultaneously'' as follows~\cite{GROER1999969}:
\begin{eqnarray}
Z_L&:=&D^T\wl,\\
Z_R&:=&D\wr,\\
X&:=&Z_R+\yl(Z_L^T\wr),\\
D&:=&\ql^T D \qr \nonumber\\ &=& D + \yl\wl^TD + D\wr\yr^T + \yl\wl^T D \wr\yr^T \nonumber\\
  &=& D + [X,~\yl][\yr,~Z_L]^T.
\label{eqn:gebrd_2k}
\end{eqnarray}
\end{itemize}
Therefore, we can initially perform the updates of $B_1$, $C_1$ and compute $Z_L$, $Z_R$, $X$. Next,
we partition the update of $D$ to expose those parts of the result that overlap with
$\bar{B}_0$  and/or $\bar{C}_0$:
\begin{equation}
D = 
\left[
\begin{array}{c|c}
D_{11} & D_{12} \\ \hline
D_{21} & D_{22} 
\end{array}
\right], 
\label{eqn:partD}
\end{equation} 
where $D_{11} \in \R^{(2\bs-\bd)\times (2\bs-\bd)}$.
Finally, by partitioning the operands 
  $[X,~\yl]$, $[\yr,~Z_L]^T$
in~\nref{eqn:gebrd_2k} conformally with $D$ in~\nref{eqn:partD},
we can overlap the updates of 
$D_{11}$, 
$D_{21}$, 
$D_{12}$, and the
 small left and right panel factorizations
in $\Tp$ with the update of the larger $D_{22}$ in $\Tr$.

%% file: s4-evaluation-SEVP.tex
\section{Experimental Evaluation}
\label{sec:evaluation}

In this section, we analyze in detail the performance benefits obtained by introducing the look-ahead strategies
formulated in this paper as well as the decoupling of the algorithmic block size from the bandwidth
in the TSR algorithms for SEVP and the SVD.
All experiments were performed using IEEE double-precision arithmetic on an 
Intel Xeon E5-2630 v3 processor (8 cores running at a nominal frequency of 2.4 GHz). 
The implementations were linked with BLIS (version 0.1.8)%
~\cite{BLIS2}.

In the experiments, we employed square symmetric matrices \textcolor{black}{for SEVP, and both square and rectangular matrices for the SVD},
with random entries uniformly distributed in $(0,1)$, and dimensions of up to 10000 in steps of 500. 
We reiterate that 
the optimal bandwidth $\bd$ depends not only on the implementation of the first stage, but also on that of the second stage,
for which there exist multiple algorithms and tuned implementations, depending on the target 
architecture~\cite{PDP:Davor2011,6114396,Haidar:2013:IPS:2503210.2503292,PIROBAND},
the problem size, etc. For this reason, we decided to test the algorithms 
using six bandwidths: \textcolor{black}{$\bd=\{32, 64, 96, 128, 192, 256\}$}. 
For these cases, the block size $\bs$ was then tuned using values
ranging from \textcolor{black}{16} up to $\bd/2$ for Variant~V1 and up to $\bd$ for V2, in steps of \textcolor{black}{16}. 
We employed one thread per core in all executions.
For the look-ahead versions, we set $\Tp$ with 1~core and $\Tr$ with the remaining 7~cores; 
for the reference implementations without look-ahead,
there is no separation of the threads into groups so that all of them participate in the execution of each BLAS.

\textcolor{black}{In all cases, we use the nominal flop count to compute the GFLOPS (billions of flops/sec.). For example, for
the reduction in the SEVP, we employ $4n^3/3$ flops independently of the target bandwidth $\bd$. Since the comparison between 
algorithms/variants is performed in an scenario with fixed $\bd$, this is a reasonable approach to obtain a scaled version of the
execution time, with the differences being more visible for smaller problem sizes than those that could have been exposed
using the execution time itself.}

\subsection{Reference implementation}
\label{subsec:reference}

Our implementation of (the first stage of) the TSR algorithms for SEVP is based on the
codes in the SBR package~\cite{Bischof:2000:AST}.
On the original version of these codes, we performed two relevant optimizations:
\begin{itemize}
\itemsep=0pt
\item We replaced the routine for the panel factorization in the SBR package, based on Level-2 BLAS, for an
      alternative that factorizes this panel using a blocked left-looking (LL) algorithmic variant that, furthermore, relies
      on Level-3 BLAS. The inner block size for this routine was set to \textcolor{black}{$16$}, with this value being determined (i.e., tuned) in
      an independent experiment. The LL variant was selected because it offered higher performance than its
      RL counterpart in our experiments.
\item The routine that builds the matrices $W,Y$ in SBR, which define the orthogonal factors, was modified to assemble
      $W$ as the product of $Y$ and the triangular $\bs \times \bs$ matrix $T$ that is obtained from
      the alternative {\em compact WY representation}~\cite{GVL3}. This modification considerably reduced the cost of building
      $W$ as this can then be based entirely on Level-3 BLAS.
\end{itemize}
The implementations of the TSR algorithms for the SVD re-utilized as much as possible of these building blocks,
including the ideas underlying the previous two code optimizations.
The legacy LAPACK (version 3.7.1) comprises routine {\tt dsytrd\_sy2sb} 
for the reduction of a symmetric matrix to symmetric band form which also features
these optimizations (except for the use of the left-looking factorization). However,
the LAPACK routine does not include look-ahead and, furthermore, it imposes the restriction that
$\bs=\bd$. As our experimental evaluation of the optimal block size will show, this limitation severely impairs performance.

\subsection{The role of the block size}

In practice, the algorithmic block size $\bs$ has an important impact on performance. 
For the particular case (of the first stage) of both TSR algorithms, the block size should balance two criteria:
\begin{enumerate}
\itemsep=0pt
\item Deliver high performance for the BLAS-3 kernels that compose the trailing update. 
Small values of $\bs$ turn 
$W,Y, 
W_U,Y_U,
W_V,Y_V$ 
into narrow column panels, affecting the performance of the 
Level-3 BLAS, since the amount of data reuse is greatly reduced so that, eventually, the kernels become memory-bounded.
\item Reduce the amount of flops performed in the panel factorization. Small values of 
$\bs$ reduce the number of operations performed in these intrinsically-sequential operations.
\end{enumerate}

\begin{figure}[t!]
\begin{center}
\includegraphics[width=0.49\columnwidth]{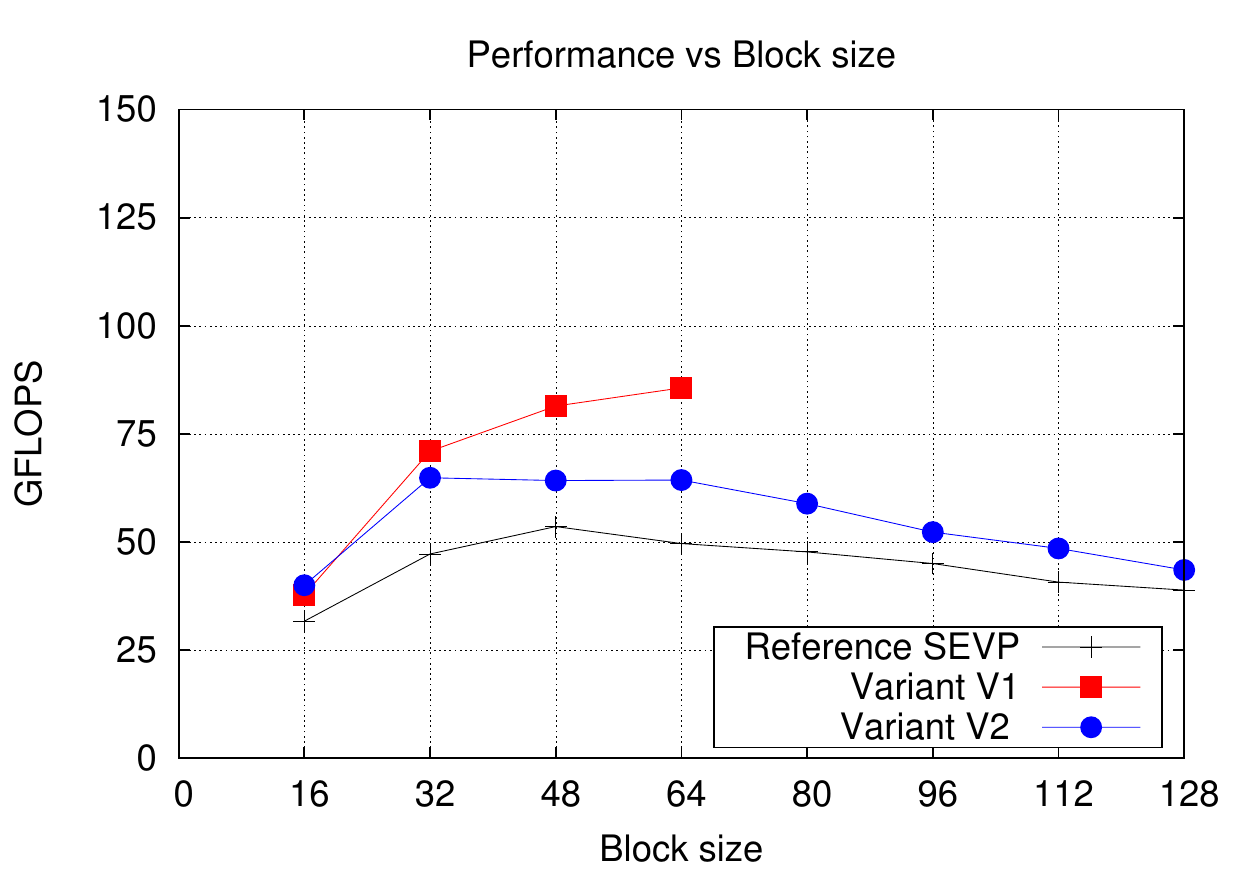}
\includegraphics[width=0.49\columnwidth]{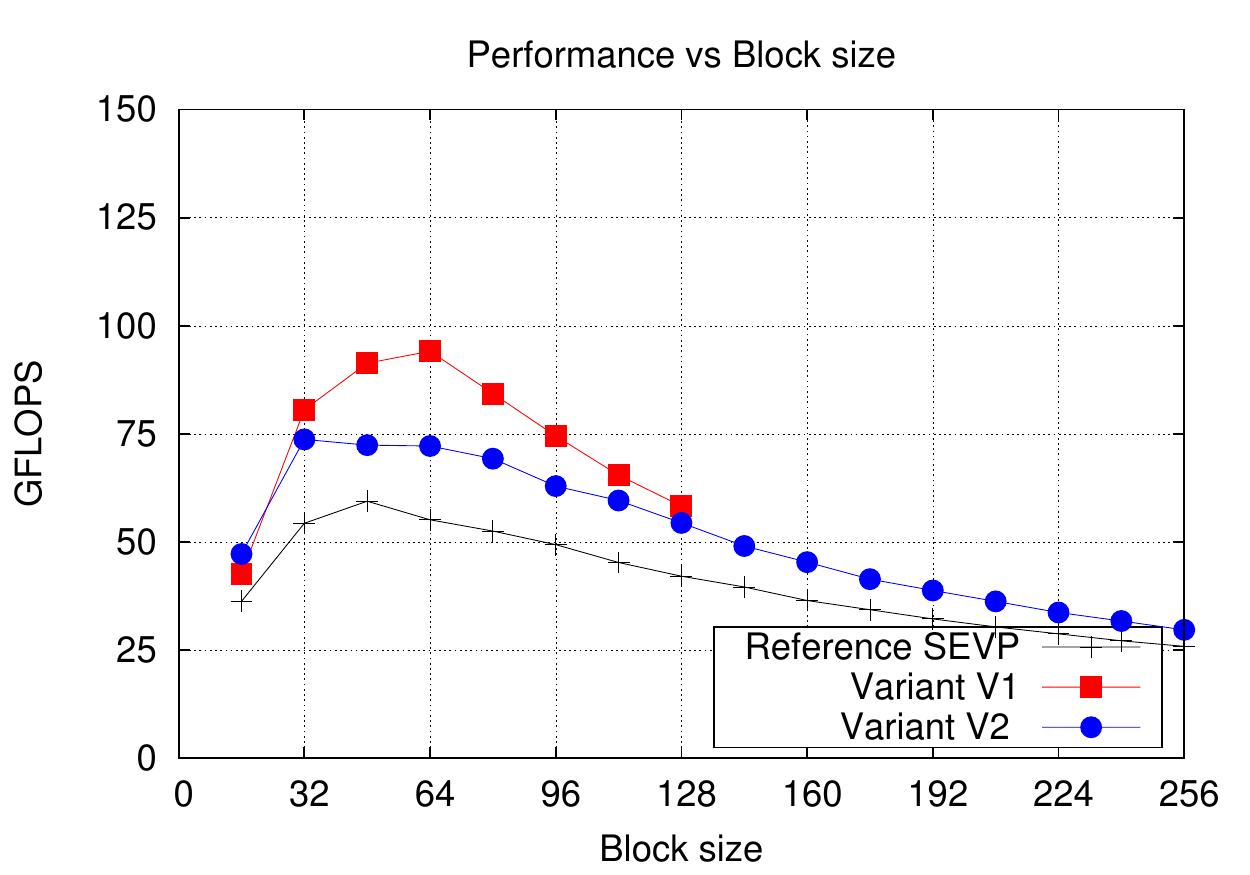}
\includegraphics[width=0.49\columnwidth]{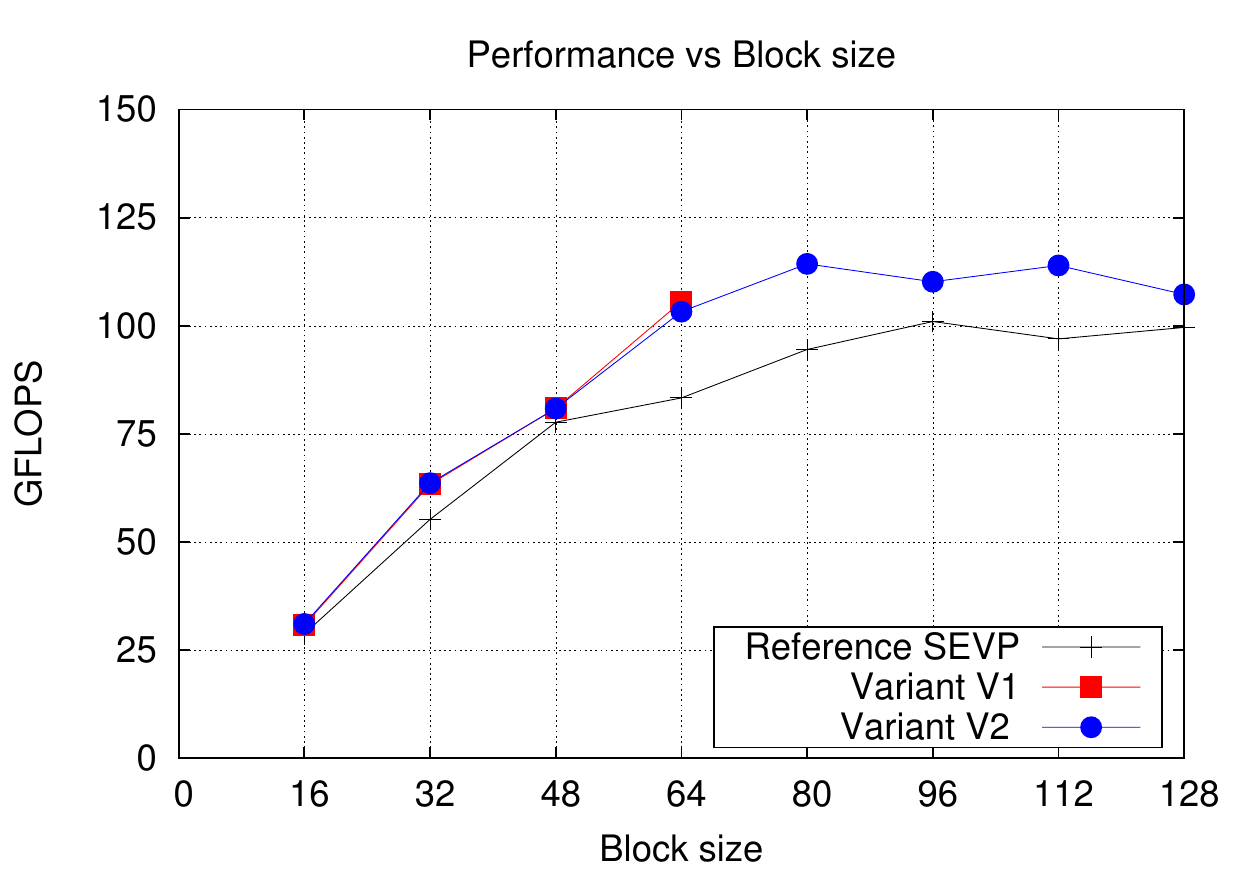}
\includegraphics[width=0.49\columnwidth]{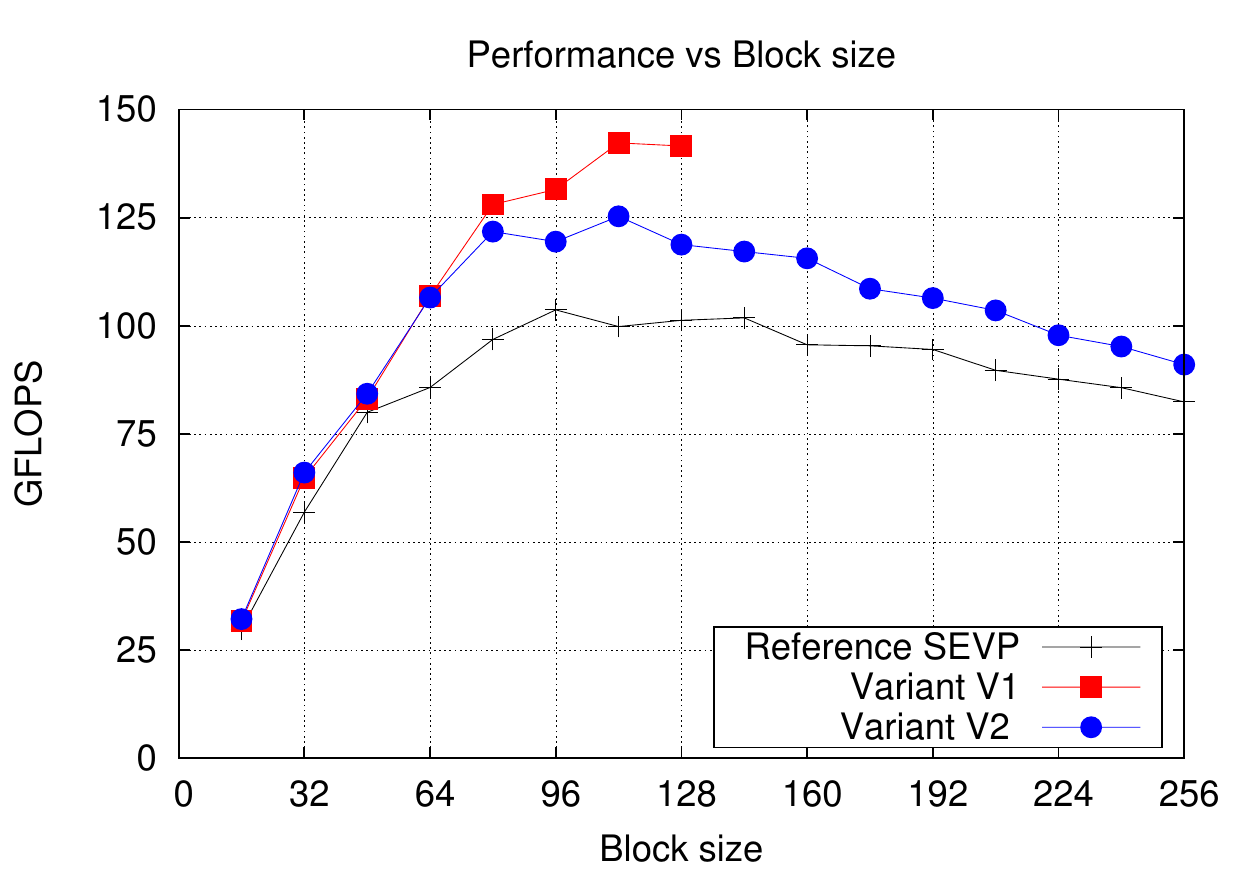}
\end{center}
\caption{Performance vs block size of the SEVP implementations; 
		The bandwidth is set to $w = 128$ (left) and $w = 256$ (right), 
         and the problem dimension to $n=2500$ (top) and $n=10000$ (bottom).}
\label{fig:block_performance}
\end{figure}

Figure~\ref{fig:block_performance} illustrates the interplay between 
the block size and the bandwidth, using 
the reduction of symmetric matrices to a banded form (first stage of SEVP). The figure includes
the reference implementation (hereafter, labeled as ``Reference SEVP'') 
as well as the two look-ahead variants introduced in this work. 

This experiment reveals that, for the small problem (two top plots in Figure~\ref{fig:block_performance}),
the optimal performance is attained for
small values of the block size, since they 
balance the execution times of the panel factorization and the trailing update. 
In rough detail, for the small problem dimensions,
if the block size is too large, the 
panel factorization will require more time to complete than the trailing update. 
As a consequence, many computing resources (i.e., threads/cores) will become idle during the iteration, degrading performance. 
At this point, we note that the degree of resource concurrency 
also exerts its impact on the workload balance, as this factor can change the 
problem size threshold from which the trailing update 
is more expensive than the panel factorization.

In addition, the plots show that, regardless of the implementation, 
the lowest performance 
for the large problem (two bottom plots in Figure~\ref{fig:block_performance})
is observed for the smallest and the largest block sizes, 
with the optimal choice residing 
in the middle range. For the smallest block size, the BLAS-3 kernels invoked from
the trailing update cannot efficiently utilize all the computational potential of the platform. 
For largest block sizes too many flops are devoted to the panel factorization. 

The previous experimental analysis conforms that in~\cite{catalan17} for the LU factorization,
and exposes that choosing the optimal block size is a non-trivial task since
this parameter depends on many other factors such the problem dimension $n$, the bandwidth $\bd$, and
 the degree of parallelism. To further complicate the selection of the optimal block size, 
as the reduction progresses, the operation is decomposed into sub-problems of dimension $n-\bs, n-2\bs, \ldots$. 
This implies that the optimal block size for the initial sub-problem(s) may not be the optimal for the subsequent ones. 
To partially compensate for this, the computational cost of the sub-problems rapidly decreases with their dimensions. 

These experiments clearly show that coupling the block size to the bandwidth, so that $\bs=\bd$, in general
results in suboptimal performance. Taking into consideration the elaboration and results in this subsection,
in the remainder of this paper we will perform an extensive tuning of the block size, 
for each problem dimension and bandwidth.

\subsection{Performance of TSR to symmetric band form for SEVP}

In this section, we analyze in detail the performance behavior of the multi-threaded variants with look-ahead
aimed to enhance the computational throughput of the algorithms for the first stage of SEVP.
Specifically, the following implementations are compared:
\begin{itemize}
\itemsep=0pt
\item \textbf{Reference SEVP: } Reference implementation from SBR with the optimizations described
 in subsection~\ref{subsec:reference}. 

\item \textbf{Variant V1: } Look-ahead variant for problems with $2\bs \le \bd$. Two different 
mappings of the update of $A_{1}^{R}$ to the threads/cores were considered,
depending on whether this operation is performed by 
either $\Tp$ or
$\Tr$.
In both cases, the factorization of $\bar{A}_0$ and the update of $A_{1}^{L}$ are performed by $\Tp$; 
and the update of $A_2$ is done by $\Tr$.
Our experiments with these mappings demonstrated that the first option, which updates 
the full $A_1$ using $\Tp$, always delivers equal or lower performance than the alternative mapping for this variant. 
(The reason is that, for small bandwidths, $A_{1}^{R}$ is consequently small, and its execution time 
does not affect the overall execution time of the reduction; 
for large bandwidths, the multi-threaded execution of $A_{1}^{R}$ is the preferred choice.)
Therefore, for clarity, we removed the first mapping from the following plots. 

\item \textbf{Variant V2: } Look-ahead variant for problems with $2\bs > \bd$. Two different 
mappings are possible,
depending on which threads update $A_1, X_1, X_2, X_3$. 
\begin{itemize}
\item $A_{1}$ on $\Tp$ (while $X_1, X_2$, and $X_3$ are computed concurrently on $\Tr$).
\item $A_{1}$ on $\Tp+\Tr$ (after which, $X_1, X_2, X_3$ are updated by the same $\Tp + \Tr$ threads).
\end{itemize}
\end{itemize}

\begin{figure}[t!]
\begin{center}
\includegraphics[width=0.49\columnwidth]{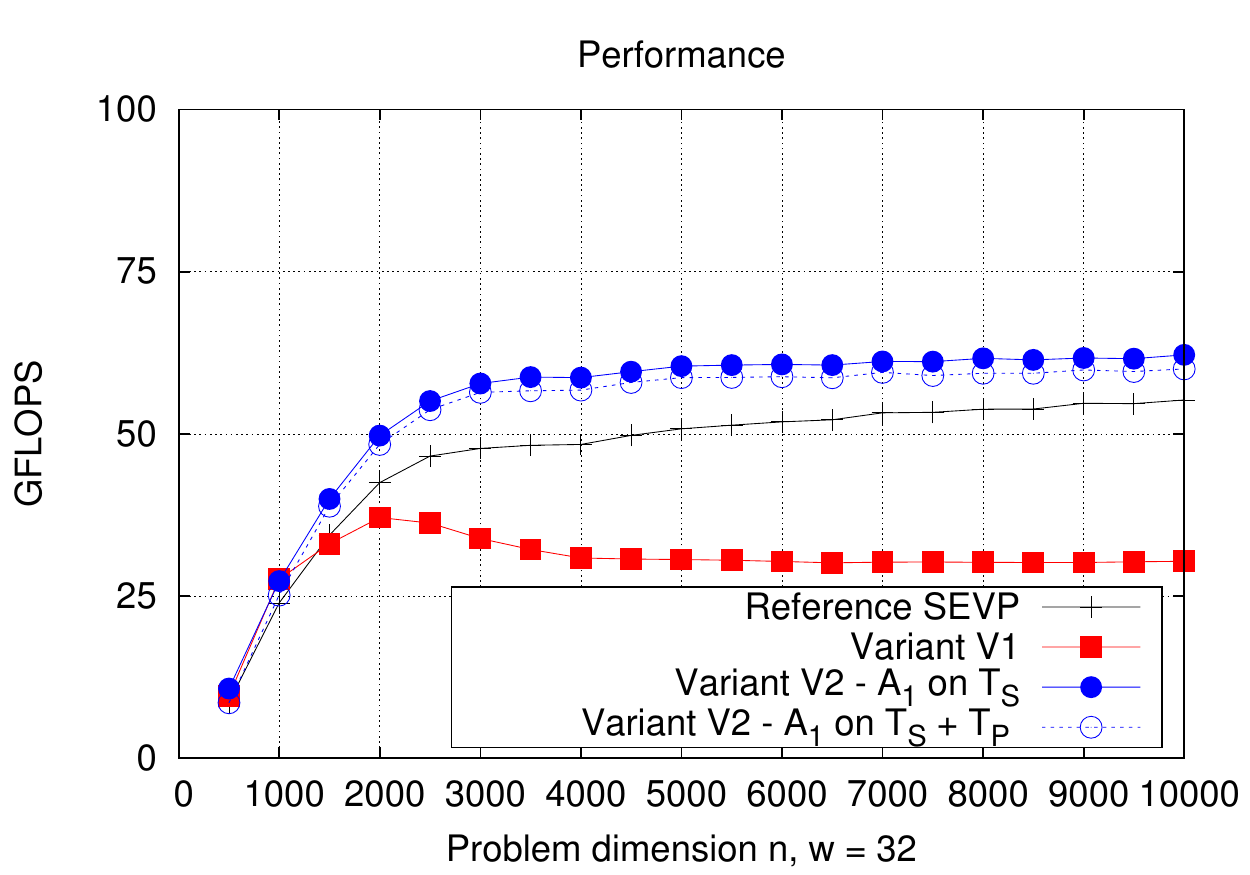}
\includegraphics[width=0.49\columnwidth]{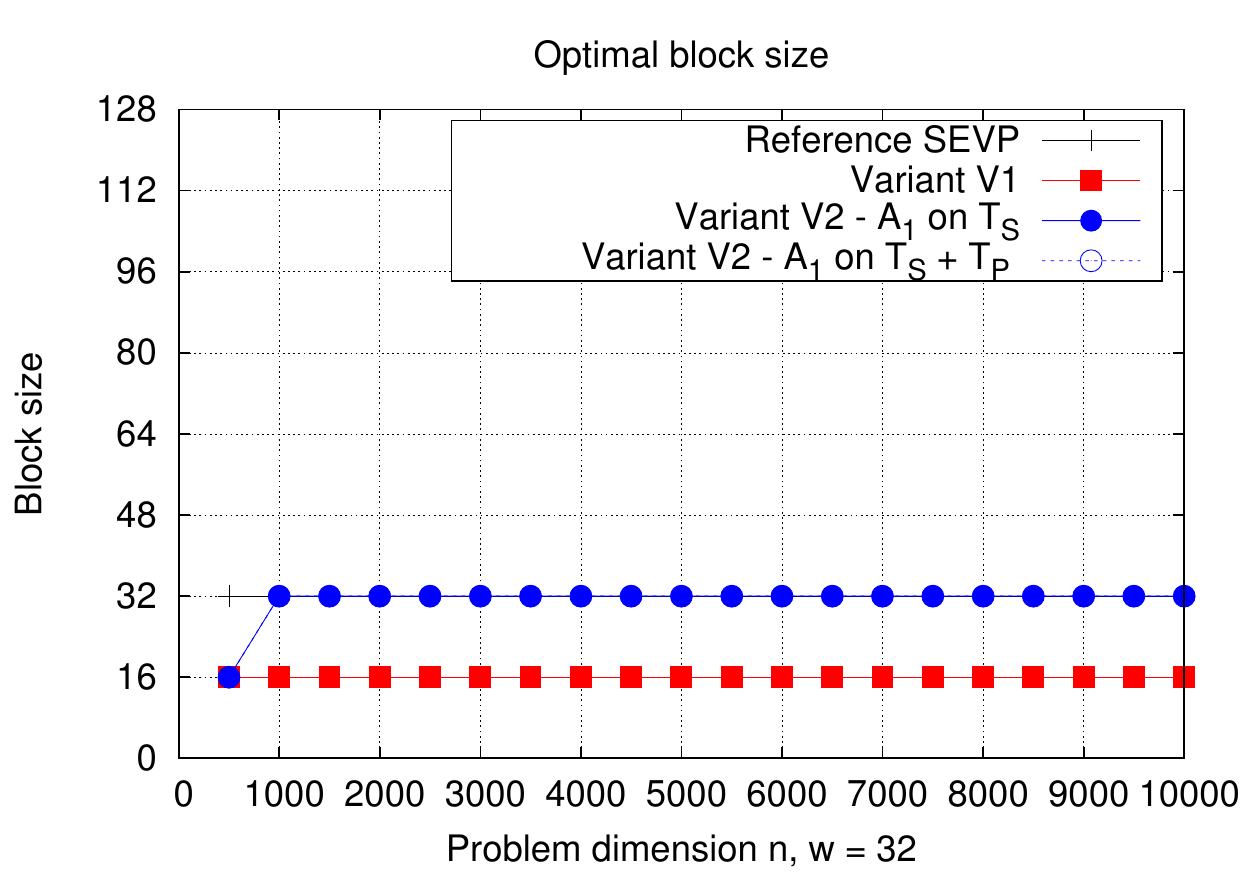}
\includegraphics[width=0.49\columnwidth]{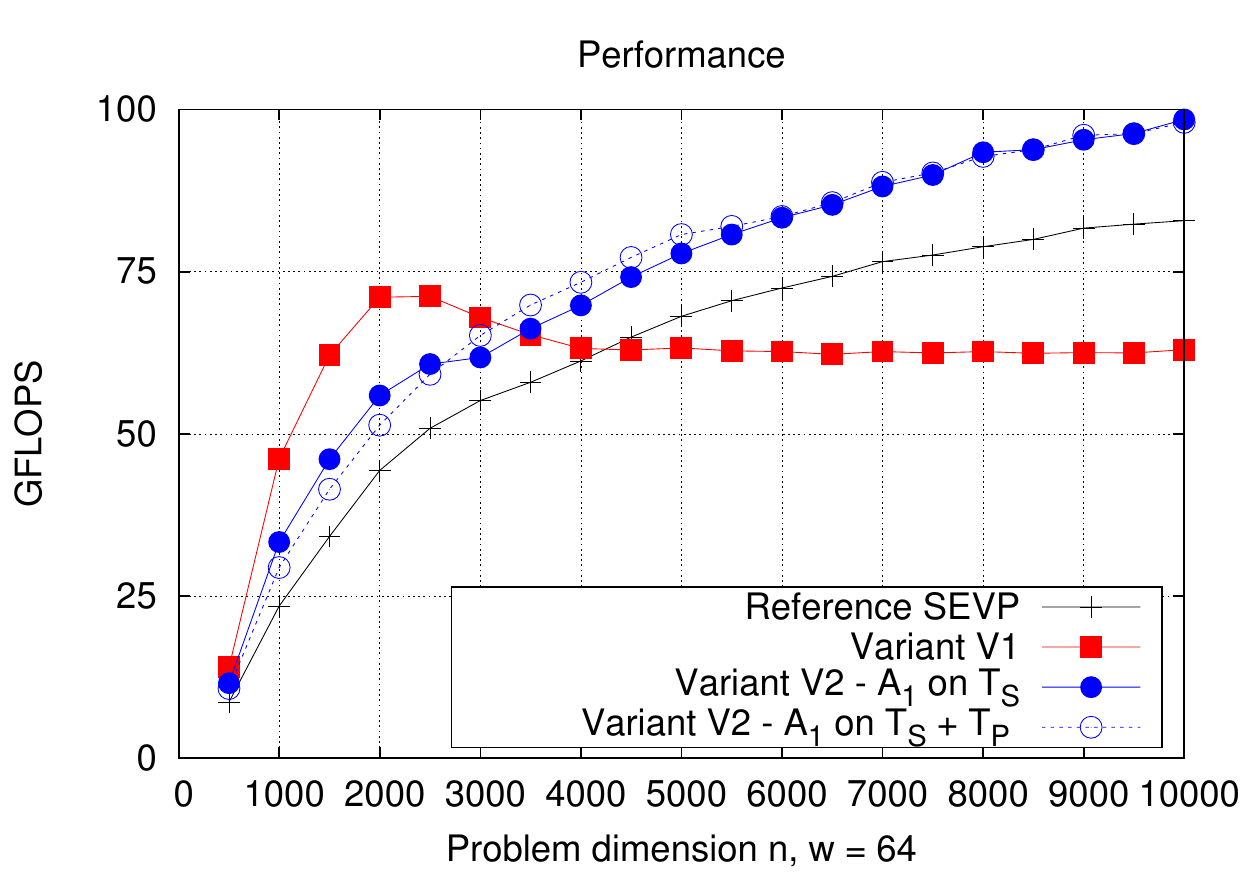}
\includegraphics[width=0.49\columnwidth]{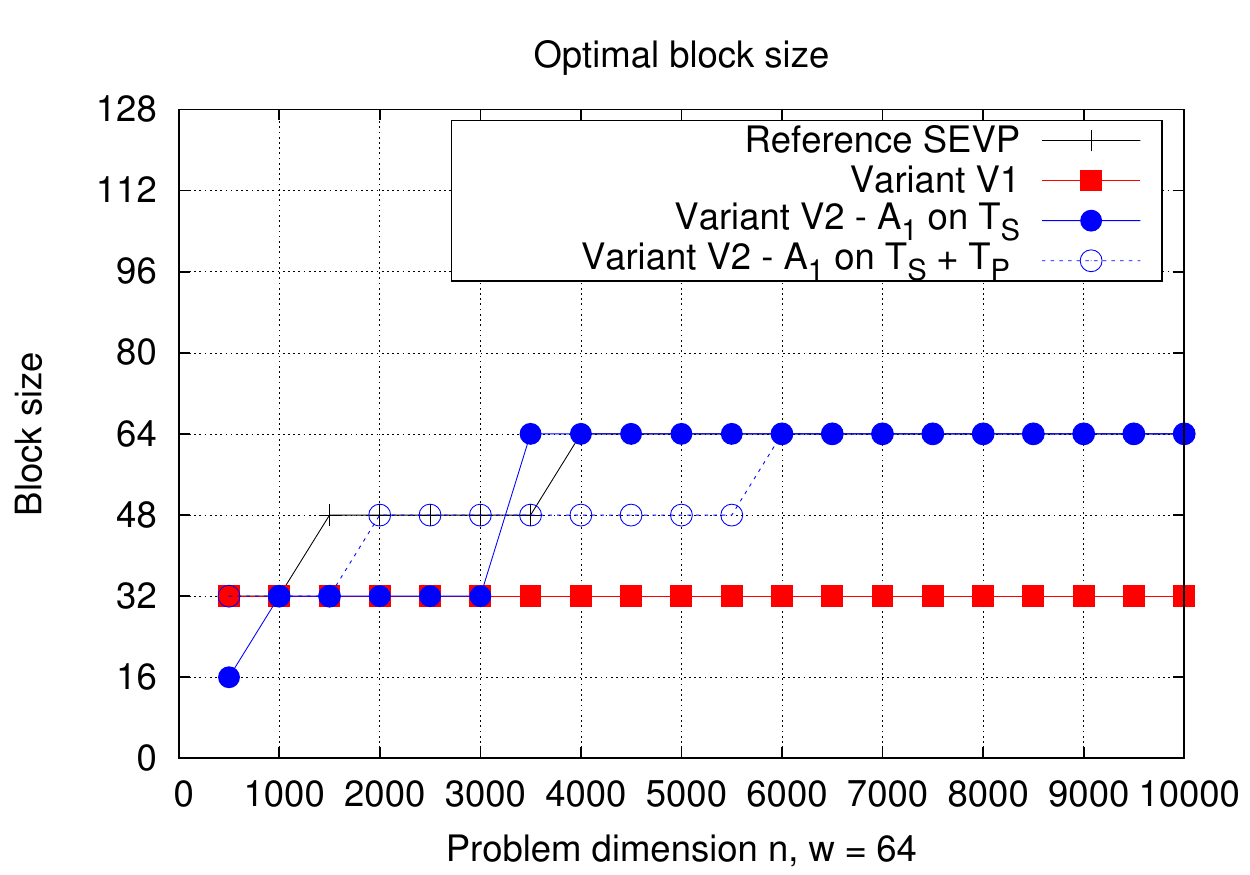}
\includegraphics[width=0.49\columnwidth]{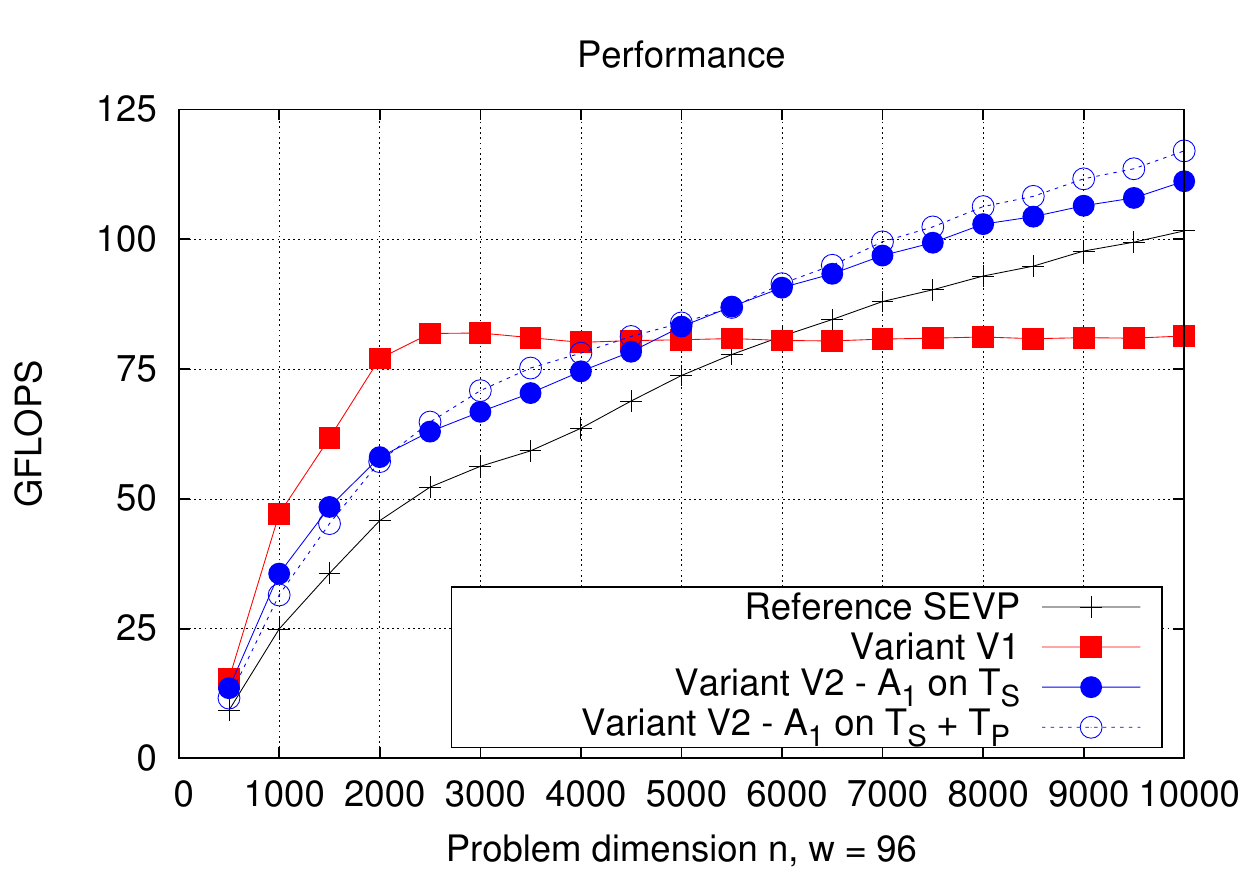}
\includegraphics[width=0.49\columnwidth]{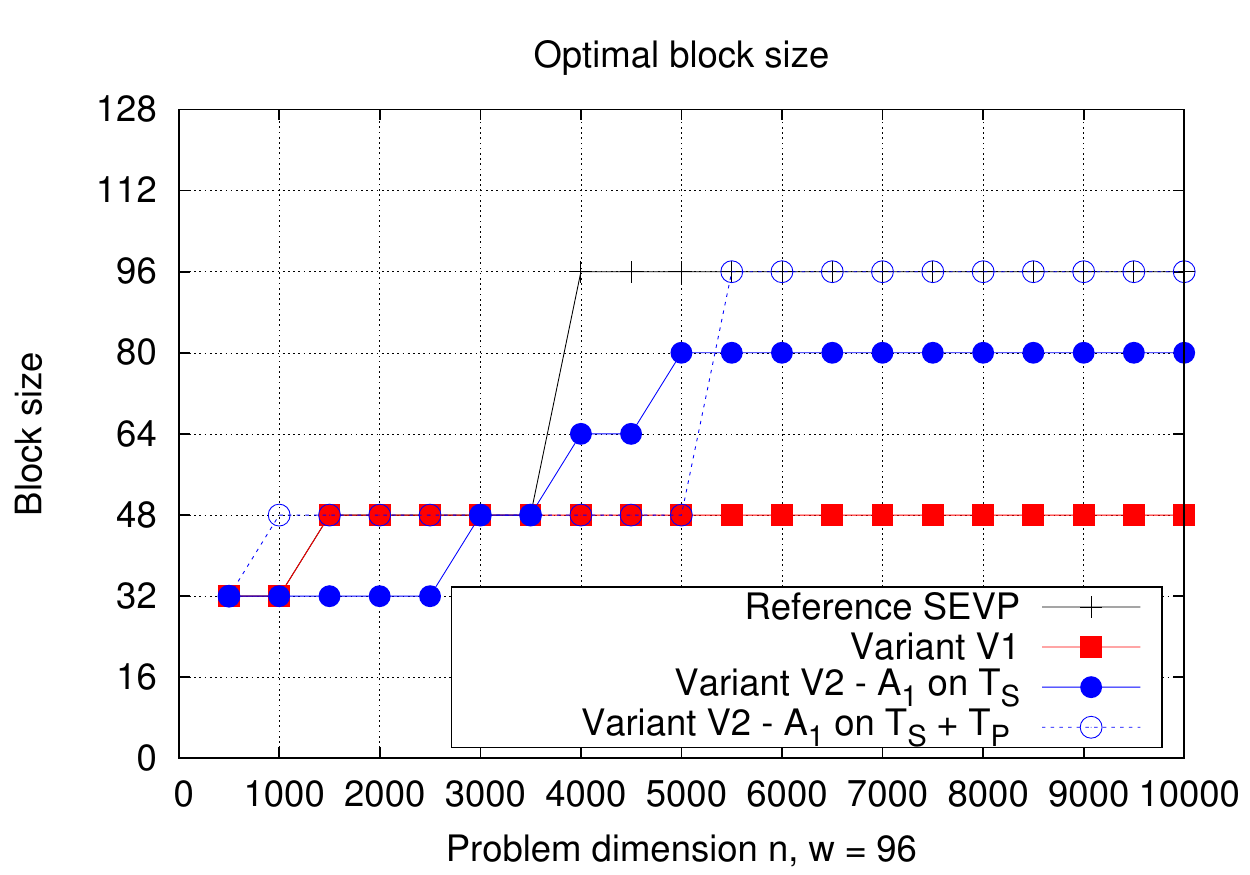}
\end{center}
\caption{Performance vs problem dimension (left) of the SEVP implementations with $w = 32, 64$ and $96$; 
and optimal block size vs problem dimension (right). }
\label{fig:Banda_small}
\end{figure}

\begin{figure}[t!]
\begin{center}
\includegraphics[width=0.49\columnwidth]{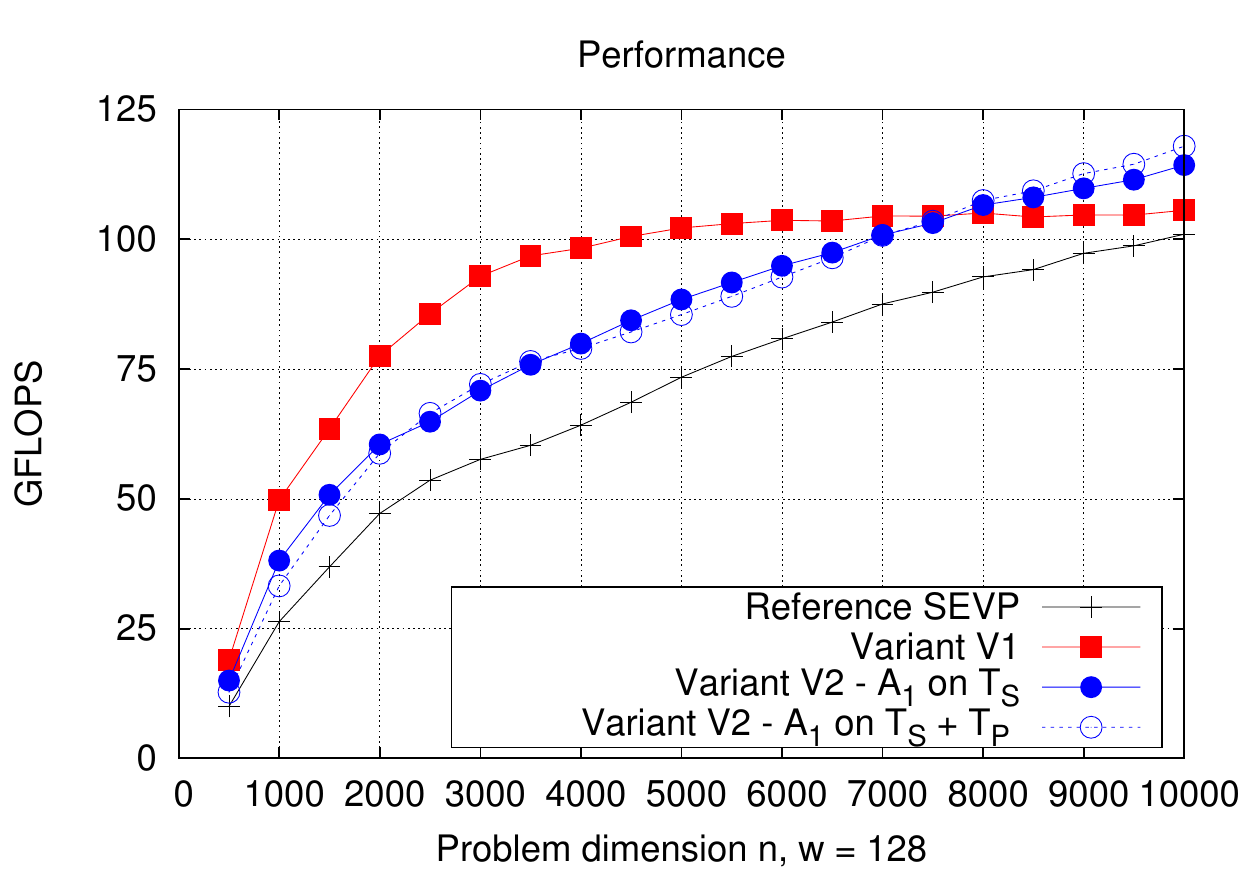}
\includegraphics[width=0.49\columnwidth]{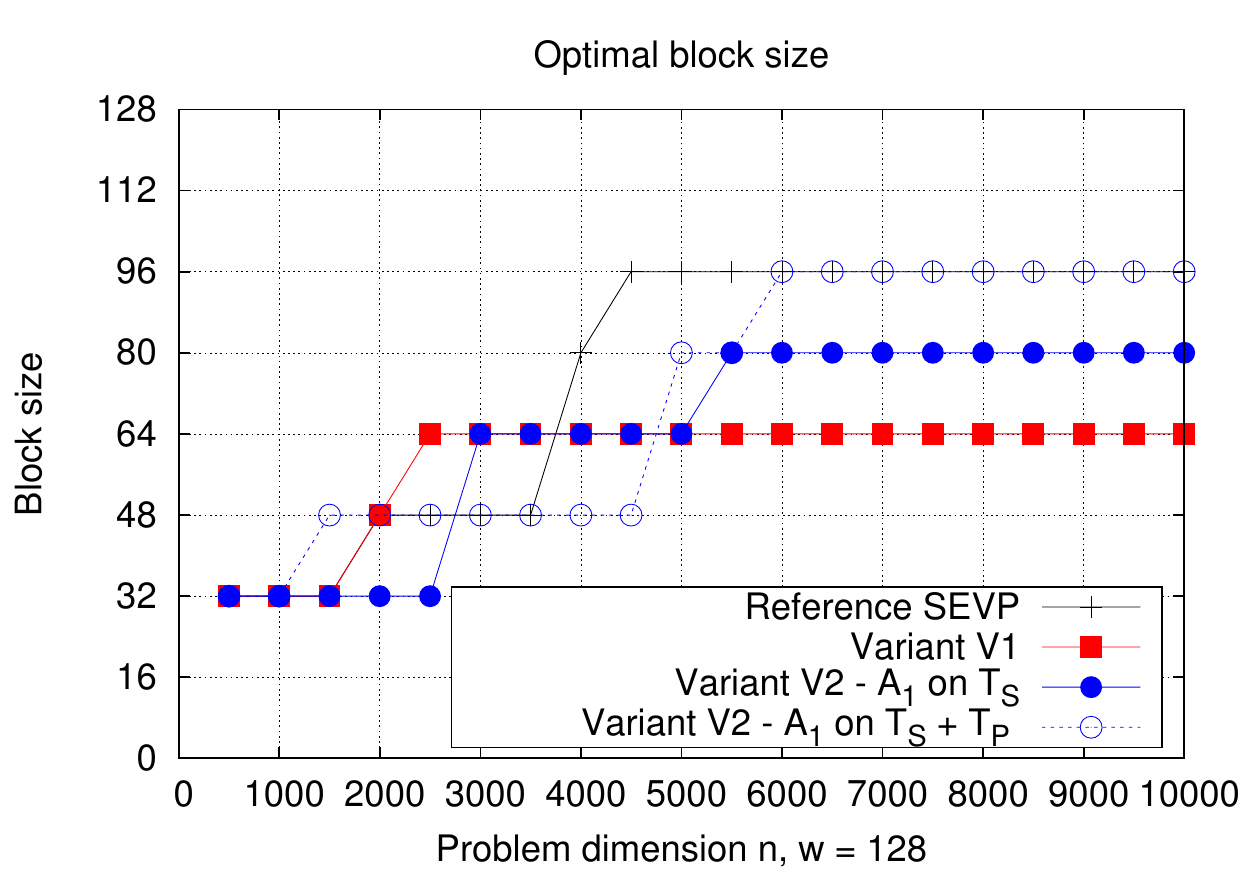}
\includegraphics[width=0.49\columnwidth]{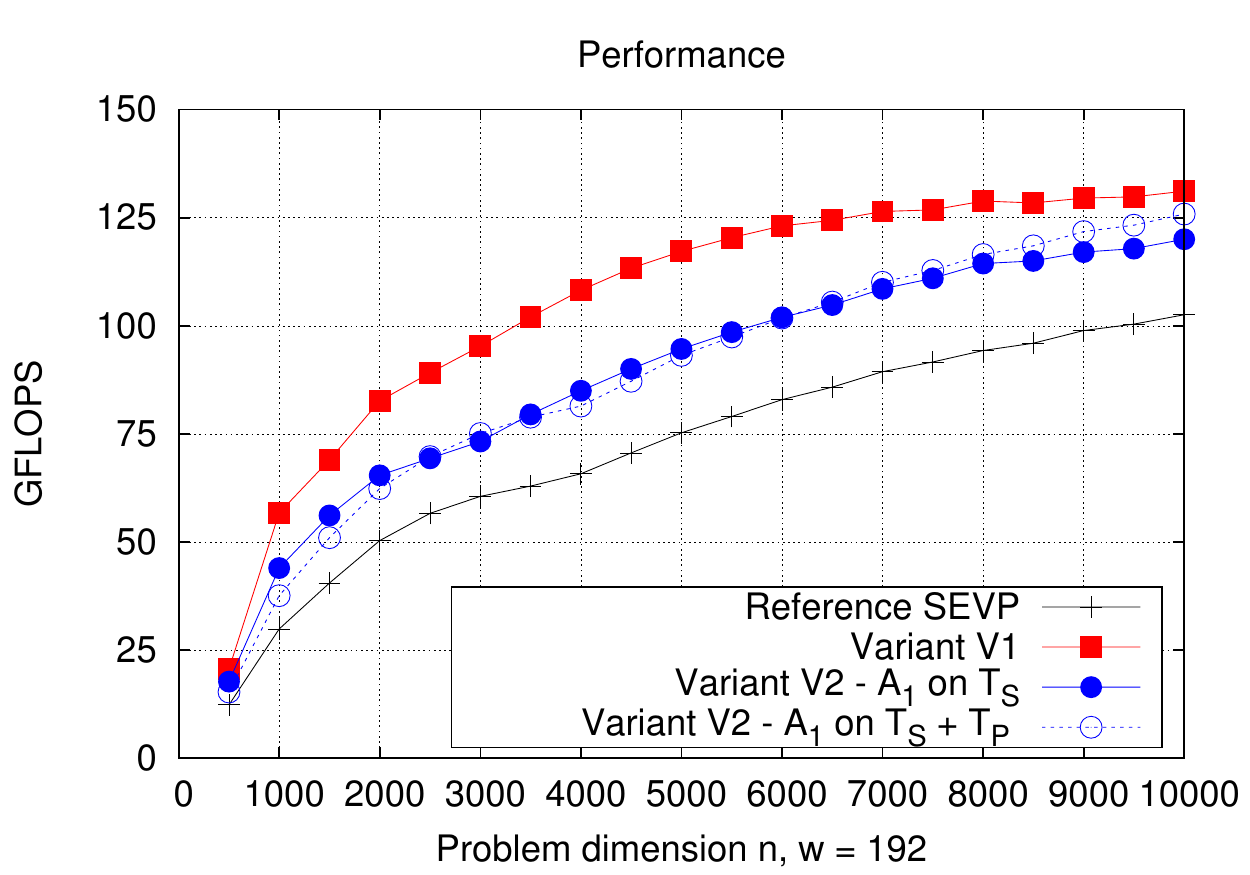}
\includegraphics[width=0.49\columnwidth]{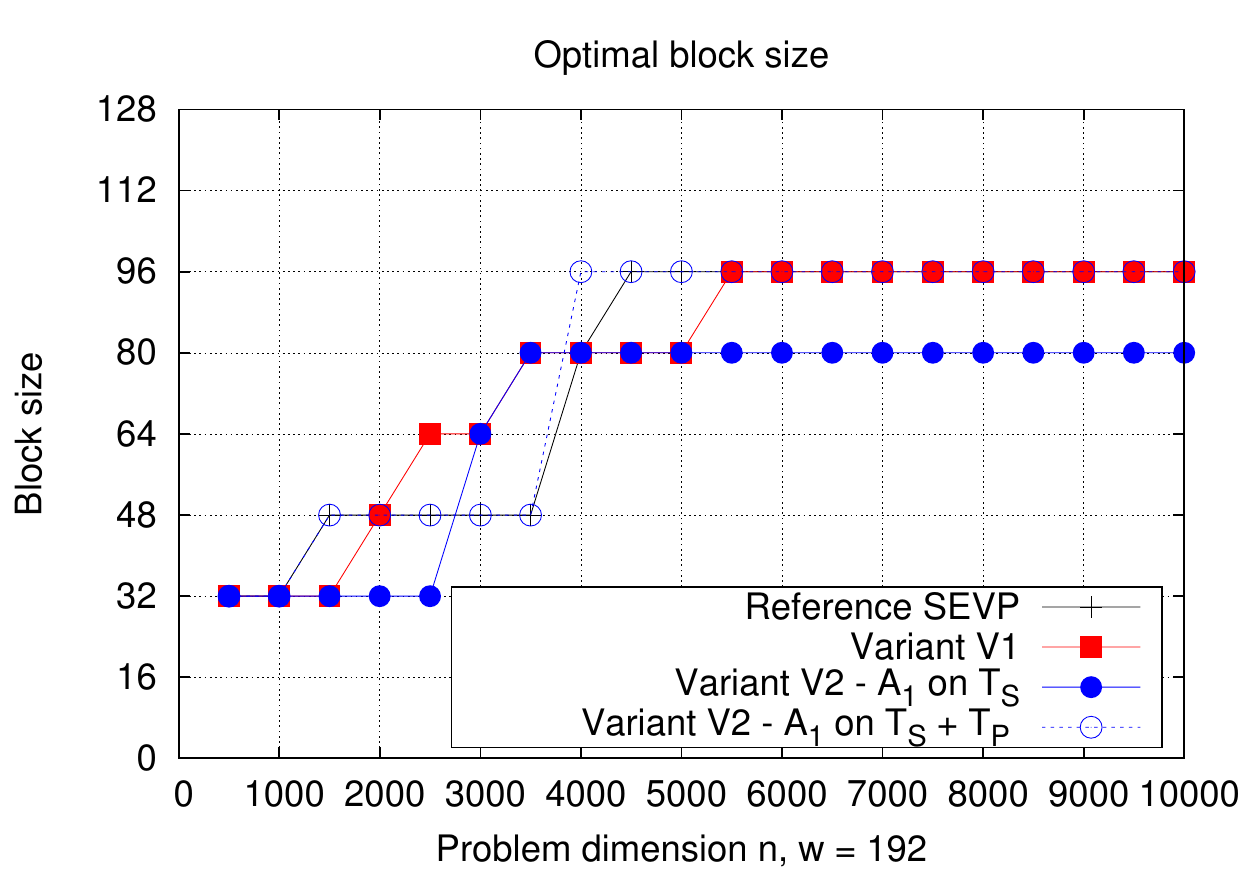}
\includegraphics[width=0.49\columnwidth]{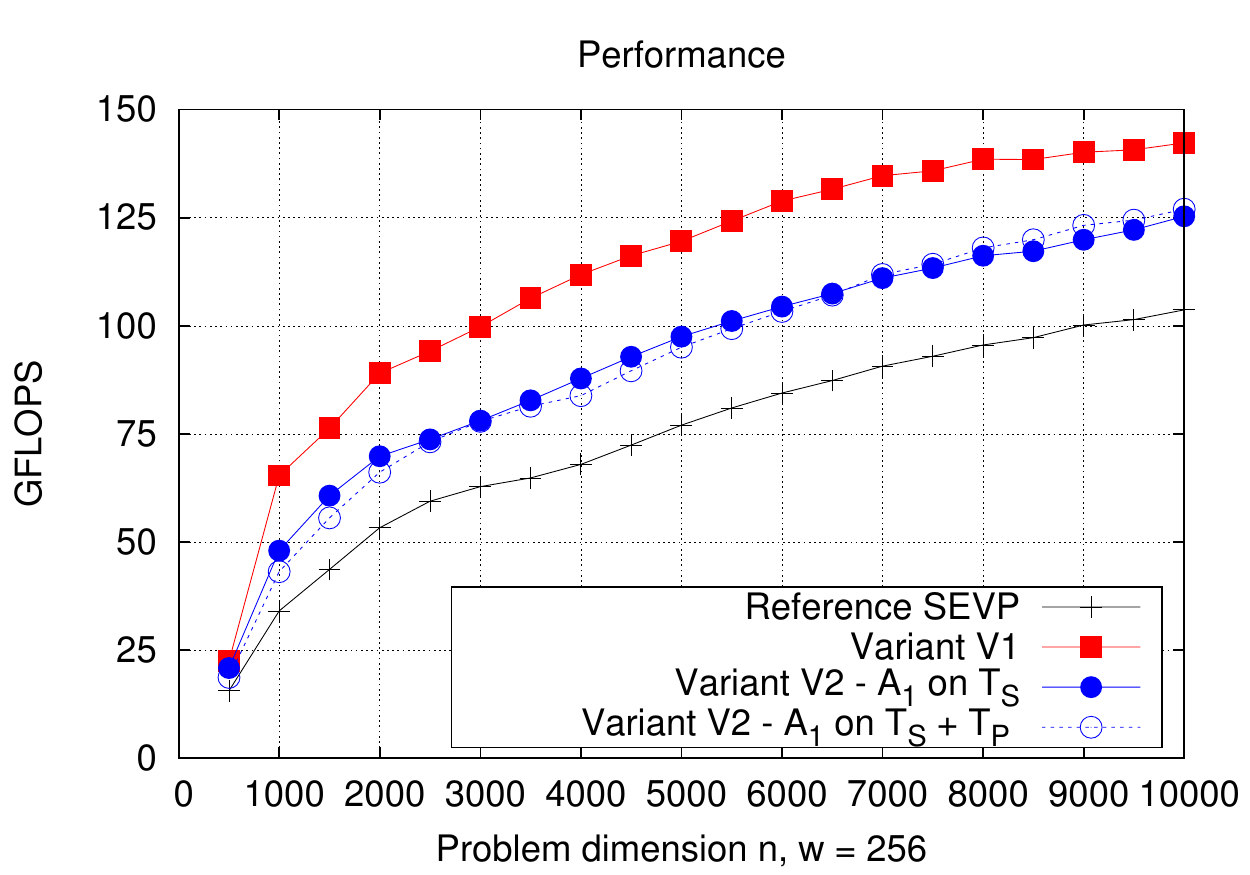}
\includegraphics[width=0.49\columnwidth]{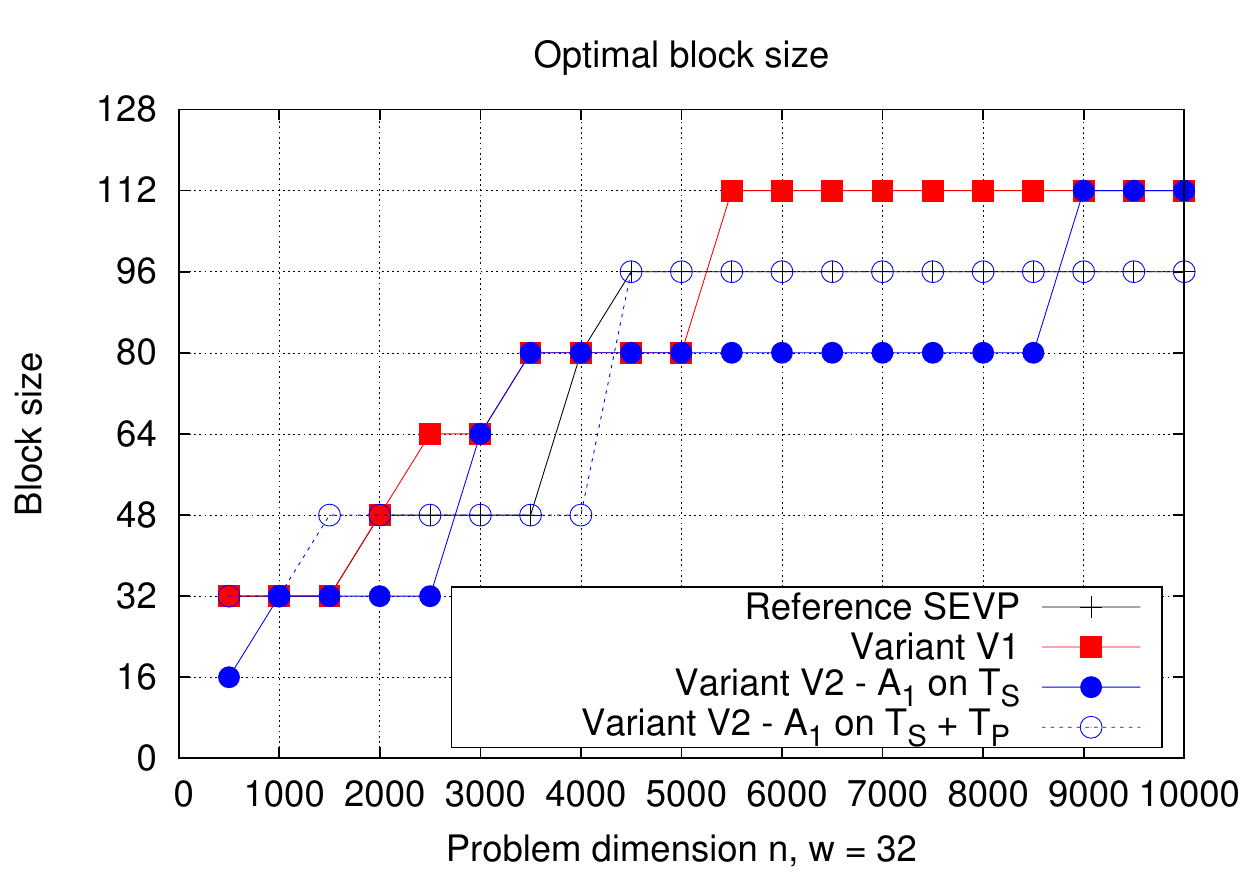}
\end{center}
\caption{Performance vs problem dimension (left) of the SEVP implementations with $w = 128, 192$ and $256$; 
and optimal block size vs problem dimension (right). }
\label{fig:Banda_large}
\end{figure}

The plots in the left-hand side of 
Figure~\ref{fig:Banda_small} and Figure~\ref{fig:Banda_large} report the GFLOPS rates 
attained by the configurations (namely algorithms/variants/mappings) for several bandwidths
in the range \textcolor{black}{32--256}. The right-hand side plots in both figures illustrate
the optimal block size for each problem dimension and configuration, showing the crucial role of this parameter.  

The first conclusion that can be extracted from the plots is that the performance of the two Variants~V1 and~V2
enhanced with look-ahead depends on the ratio between the algorithmic 
block size $\bs$ (equivalent to the number of columns in the panel), the target matrix bandwidth $\bd$, and 
the problem dimension $n$.

Focusing on Variant~V1, we identify a drawback due to 
the limitation imposed on the block size by the condition $2b \le w$. This implies that, for small bandwidths, 
Variant~V1 can only employ very reduced block sizes. In consequence, the invocation to the  Level-3 BLAS 
to perform the trailing update cannot efficiently exploit all resources of the processor. 
To illustrate this behavior, let us focus on \textcolor{black}{the experiments with $\bd = 64$ in~Figure~\ref{fig:Banda_small}}. 
The right-hand side plot there shows that, for Variant~V1,
the block size is always 32, which is smaller than 
those selected for Variant~V2 and the reference implementation.
In addition, the left-hand side plot shows that, despite Variant~V1 integrates look-ahead, 
its performance is inferior to that of the reference implementation for large problem dimensions
as the overall execution time in those cases is  dominated by the trailing update.
In contrast, in the plots \textcolor{black}{using larger bandwidths (i. e. $\bd = 128$ and $256$)} 
we observe that this fact enables the selection of larger block sizes,
considerably improving the throughput of Variant~V1, which now outperforms all other configurations.

Our implementations of Variant~V2 always improve the performance of the reference routine;
and also that of Variant~V1 for small bandwidths\textcolor{black}{, and for medium-size bandwidths combined with small problem dimensions}. 
Indeed, as there are no strict restrictions on the 
block size for Variant~V2 (other than $\bs\leq \bd$), large values can be selected for
$\bs$, and the Level-3 BLAS for the trailing update tend to 
deliver a good fraction of the peak performance even for small bandwidths. 
However, the drawback of Variant~V2 lies in the parts of the algorithm that may be overlapped as part of the look-ahead strategy.
In particular, as $2\bs$ can be larger than $\bd$, the factorization of $\bar{A}_0$ cannot start till the updates of $A_1$ 
and $X_3$ are completed (requiring a synchronization point after them). 
This implies that only the factorization of $\bar{A}_0$ and the update of $A_2^L$ can be overlapped with the update of $A_2^R$.
In contrast, for Variant~V1, the factorization of $\bar{A}_0$ only requires that the update of $A_1^L$ 
is completed, therefore removing this synchronization point; in consequence, the update of 
$A_1^L$ and the factorization of $\bar{A}_0$ can be overlapped with the update of~$A_2$.

\textcolor{black}{To close the experiments in this subsection, we evaluate 
the impact that the TSR algorithms for SEVP
make on the overall computation of the eigenvalue decomposition. We remind that, in the two-stage reduction to tridiagonal form, 
the matrix is reduced to a symmetric band form employing one of the TSR algorithms presented earlier in this subsection
(the SBR-based Reference, Variant V1 or Variant V2) and this banded matrix is next reduced to tridiagonal form.
For the second stage, in our evaluation we will employ routine 
{\sf SBRDT} from the SBR package. After that, the eigenvalues are obtained using routine {\sf DSTERF} from LAPACK.
Alternatively, when using the the traditional solver for SEVP in LAPACK, the input matrix is 
directly reduced to tridiagonal form, using routine {\sf DSYTRD} routine from LAPACK, after which routine {\sf DSTERF}
is applied to obtain the eigenvalues. Routines {\sf SBRDT}, {\sf DSTERF} and {\sf DSYTRD} are mostly composed of calls to kernels
in the Level-1 and Level-2 of BLAS. Due to the sequential implementation and lack of optimization of these kernels in BLIS, in our
experiments we linked these routines to Intel MKL. (The initial reduction to band form via the TSR algorithms was still performed
using the kernels from BLIS.)}

\textcolor{black}{Table~\ref{tab:Overall_mkl} reports the execution time of the different stages that are present in the solution 
of SEVP as well as the acceleration with respect to the single-stage reduction approach to tridiagonal form for
three problem sizes and several bandwidth dimensions. 
These results show that, for the smallest problem, as the symmetric matrix fits in the L3 cache on chip, the 
best option is to employ the conventional solver in LAPACK, based on routines {\sf DSYTRD+DSTERF}. 
On the other hand, for the larger two problems, the best option corresponds to
the two-stage reduction to tridiagonal form, using Variant~V2 with a narrow bandwidth $w=64$ in both cases.
In particular, the speed-ups with respect to LAPACK's solver with a single-stage reduction are 1.96 and 2.78 when the complete
process is considered. Focussing on the two-stage approach, the results also expose the need to limit the bandwidth of the
compact form as the cost of routine {\sf SBRDT} rapidly grows with $w$. Finally, the table reveals that the speed-ups observed
for Variant~V2 with respect to the reference implementation vary between 1.16 and 1.19 for the two largest problem sizes
and $w=64$. At this point, we note that the contribution of the new TSR 
to band form to the total cost of the eigenvalue computation
is largely dependent on the 
implementation and efficiency of the subsequent stages. 
(Thus, for example, the results can be significantly different
if one employs a solver that directly obtains the eigenvalues from the band
form, without requiring the reduction to tridiagonal form~\cite{Moldaschl:2014:CES:2748152.2748600}, or just applies an iterative solver on the band matrix
to obtain a few selected eigenvalues~\cite{ALIAGA2016314}.)
However, the acceleration factors observed for Variants~V1 and~V2
with respect to the reference implementation in the first stage will remain constant.
}

\begin{table}[!t]
{\small
\begin{center}
\begin{tabular}{|rr||rrr|r|r||rrr|}
\hline
 		& 		& \multicolumn{3}{c|}{Variants}	& 	&  & \multicolumn{3}{c|}{Speed-up vs}\\
 		& 		& \multicolumn{3}{c|}{TSR to band form}	& 	&  & \multicolumn{3}{c|}{DSYTRD + DSTERF}\\
 $n$		& $\bd$		& {\sf Ref}	& {\sf V1}	& {\sf V2} & SBRDT	& DSTERF & {\sf Ref}	& {\sf V1}	& {\sf V2}\\
 \hline \hline
2000  	& 32	& 0.25	& 0.30	& 0.22	& 0.20	& 0.09	& 0.44	& 0.41	& 0.47\\
  		& 64	& 0.24	& 0.16	& 0.19	& 0.33	& 0.09	& 0.36	& 0.41	& 0.39\\
  		& 96	& 0.24	& 0.14	& 0.19	& 0.43	& 0.09	& 0.32	& 0.36	& 0.34\\
  		& 128	& 0.23	& 0.14	& 0.18	& 0.93	& 0.11	& 0.19	& 0.20	& 0.20\\
  		& 192	& 0.22	& 0.13	& 0.17	& 1.21	& 0.11	& 0.15	& 0.16	& 0.16\\
  		& 256	& 0.20	& 0.12	& 0.16	& 1.27	& 0.11	& 0.15	& 0.16	& 0.16\\
\hline
6000  	& 32	& 5.55	& 9.50	& 4.76	& 1.51	& 0.74	& 1.69	& 1.12	& 1.89\\
  		& 64	& 3.97	& 4.62	& 3.42	& 2.58	& 0.74	& 1.81	& 1.66	& 1.96\\
  		& 96	& 3.54	& 3.57	& 3.18	& 3.71	& 0.74	& 1.64	& 1.65	& 1.73\\
  		& 128	& 3.55	& 2.78	& 3.05	& 8.98	& 0.78	& 0.99	& 1.05	& 1.03\\
  		& 192	& 3.48	& 2.34	& 2.82	& 12.08	& 0.78	& 0.81	& 0.87	& 0.84\\
  		& 256	& 3.41	& 2.23	& 2.77	& 14.40	& 0.78	& 0.71	& 0.76	& 0.74\\
\hline
10000 	& 32	& 24.15	& 43.85	& 21.52	& 4.18	& 2.02	& 2.13	& 1.29	& 2.33\\
  		& 64	& 16.11	& 21.26	& 13.50	& 7.78	& 2.02	& 2.50	& 2.08	& 2.78\\
  		& 96	& 13.14	& 16.41	& 12.01	& 10.68	& 2.02	& 2.50	& 2.22	& 2.62\\  		
  		& 128	& 13.15	& 12.59	& 11.70	& 25.72	& 2.05	& 1.58	& 1.60	& 1.64\\
  		& 192	& 12.99	& 10.15	& 11.09	& 35.14	& 2.05	& 1.29	& 1.37	& 1.34\\
  		& 256	& 12.86	& 9.28	& 10.65	& 41.88	& 2.06	& 1.14	& 1.22	& 1.19\\
\hline		
  	  	  	
\end{tabular}
\end{center}
}
\caption{Execution time (in sec.) of the different stages for the solution of the SEVP (computation of the eigenvalues only) via
         the TSR algorithms and speed-up with respect to the conventional approach in LAPACK ({\sf DSYTRD+DSTERF}).}
\label{tab:Overall_mkl}
\end{table}

%% file: s4-evaluation-SVD.tex
\subsection{Performance of TSR to band form for the SVD.}

We next analyze the performance behavior of the multi-threaded variants with look-ahead
aimed to enhance the computational throughput of the TSR algorithms for the first stage of the SVD.
Specifically, the following implementations are compared:
\begin{itemize}
\itemsep=0pt
\item \textbf{Reference implementations:} These routines depart from the one presented for SEVP
in that the matrix is not symmetric and, therefore, distinct left and right panel factorizations are required. 
In addition, two different implementations are possible for the SVD, depending on how the trailing update is performed:
\begin{itemize}
\item Reference SVD. This case adheres to the formulation in equations~\nref{eqn:updateB0}--\nref{eqn:updateCD22}. 
At each iteration this implementation first computes the QR factorization; then applies the resulting orthogonal matrix $\ql$
to the trailing submatrix; next computes the subsequent LQ factorization; and finally applies the resulting $\qr$ 
to the trailing submatrix.
\item Reference SVD Simultaneous. This implementation performs the update of the trailing submatrix as in~\nref{eqn:gebrd_2k}. 
That is, at each iteration of the algorithm, the QR and LQ factorizations are performed first; 
and then the updates of the trailing submatrix with $\ql$ and $\qr$ are fused into a single ``step''.
\end{itemize}

\item \textbf{Variant V1:} Look-ahead variant of the ``Reference SVD'' implementation for problems with $2\bs \le \bd$. 
Similarly to Variant~V1 for SEVP, two different 
mappings of the updates of $B_1$ and $C_1$ are possible, where these two blocks are 
split into two independent sub-blocks ($B_{1}^{L}$ and $B_{1}^{R}$ ; $C_{1}^{T}$ and $B_{1}^{B}$). 
However, for similar reasons, in the experiments we only consider the case with $B_{1}^{R}$ and $C_{1}^{B}$ updated by $\Tr$.

\item \textbf{Variant V2:} Look-ahead variant of the ``Reference SVD Simultaneous'' implementation for problems with $2\bs > \bd$. 
Two different mappings are possible, depending on which threads update $B_1, C_1, Z_L, Z_R, X$. 
\begin{itemize}
\item $B_{1}$ and $C_{1}$ on $\Tp$ (while $Z_L, Z_R$, and $X$ are computed concurrently on $\Tr$).
\item $B_{1}$ and $C_{1}$ on $\Tp+\Tr$ (after which, $Z_L, Z_R$, and $X$ are updated by the same $\Tp + \Tr$ threads).
\end{itemize}
\end{itemize}
Following the optimizations presented earlier for the SBR routine,
all routines for the SVD perform the factorization of the panels via Level-3 BLAS procedures, and compute the matrices
$W_U,W_V$ from the product of the corresponding compact WY factors $T_U,T_V$ and $Y_U,Y_V$.

\begin{figure}[t!]
\begin{center}
\includegraphics[width=0.49\columnwidth]{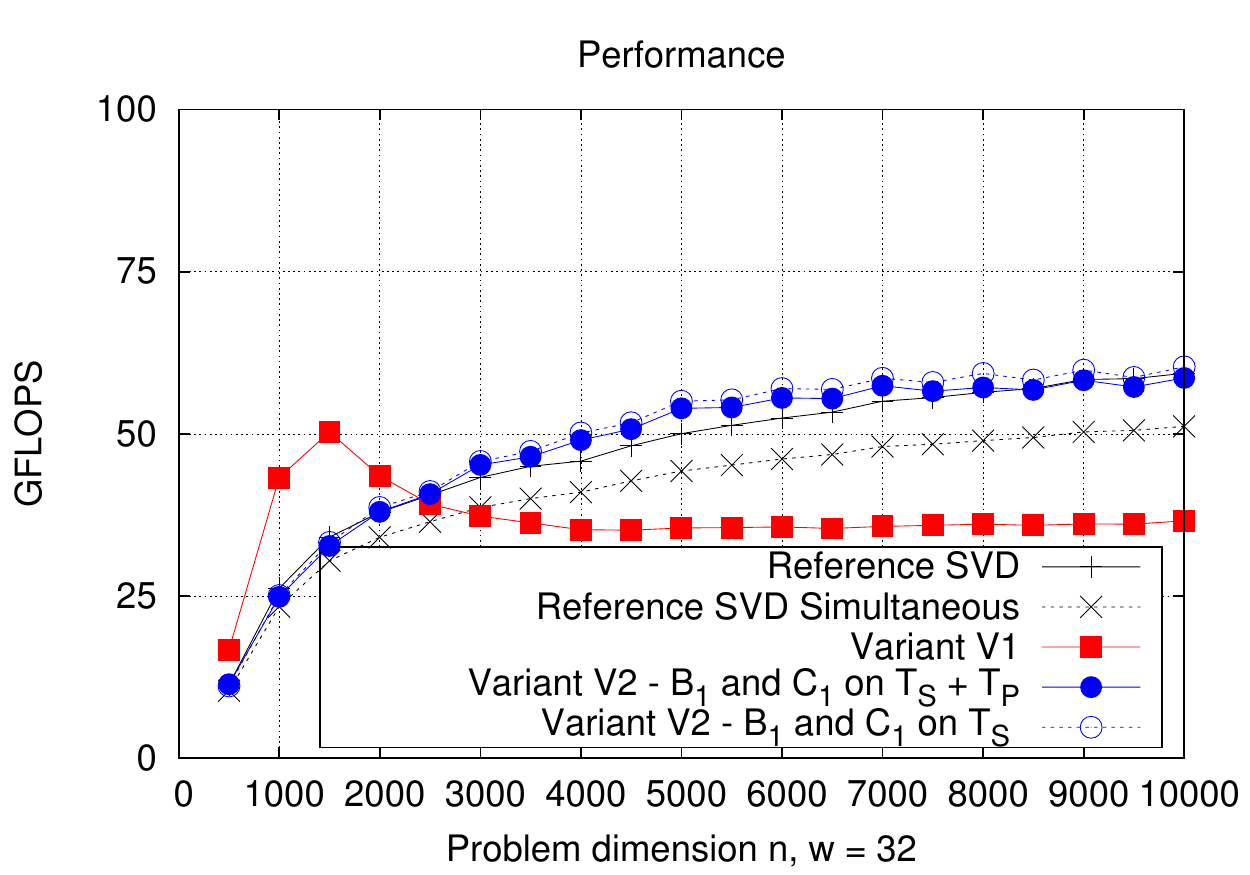}
\includegraphics[width=0.49\columnwidth]{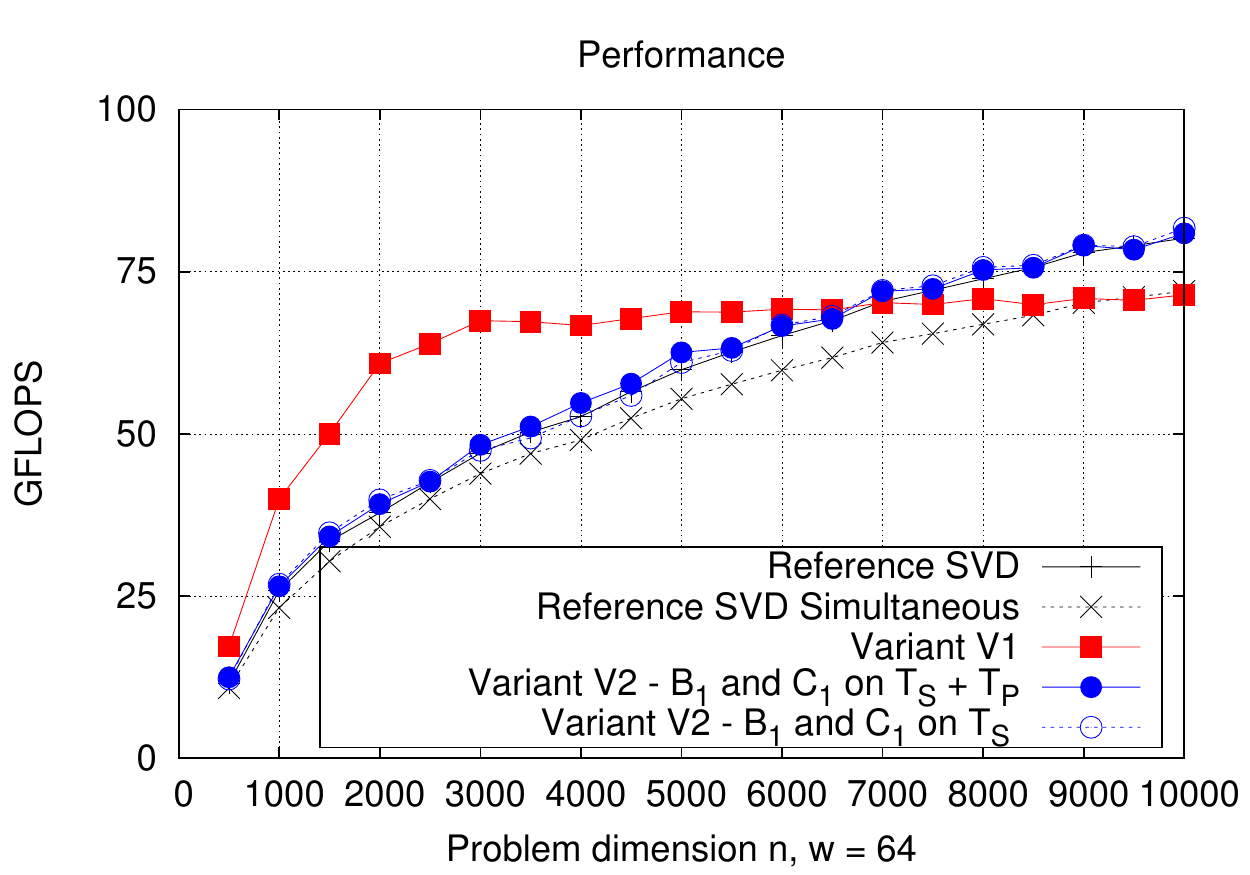}\\
\includegraphics[width=0.49\columnwidth]{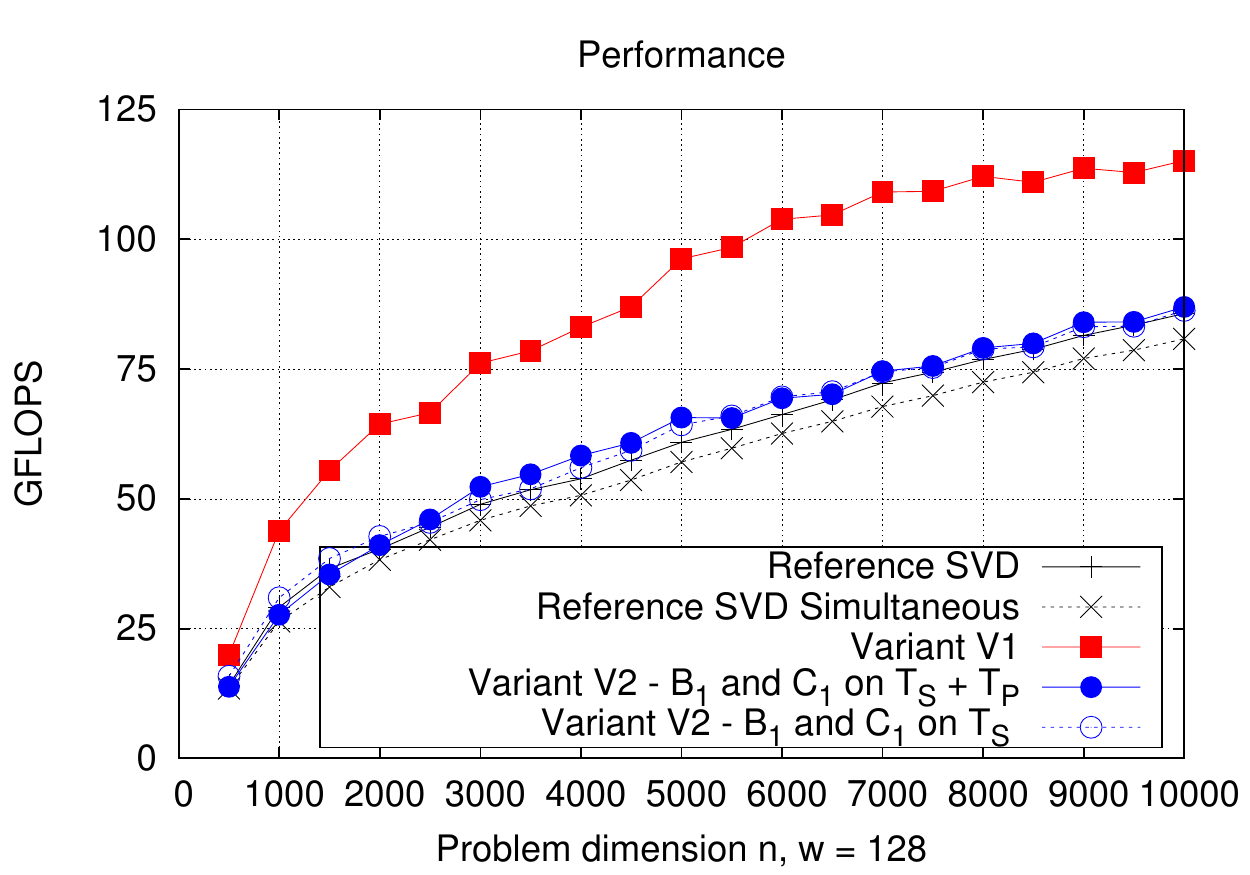}
\includegraphics[width=0.49\columnwidth]{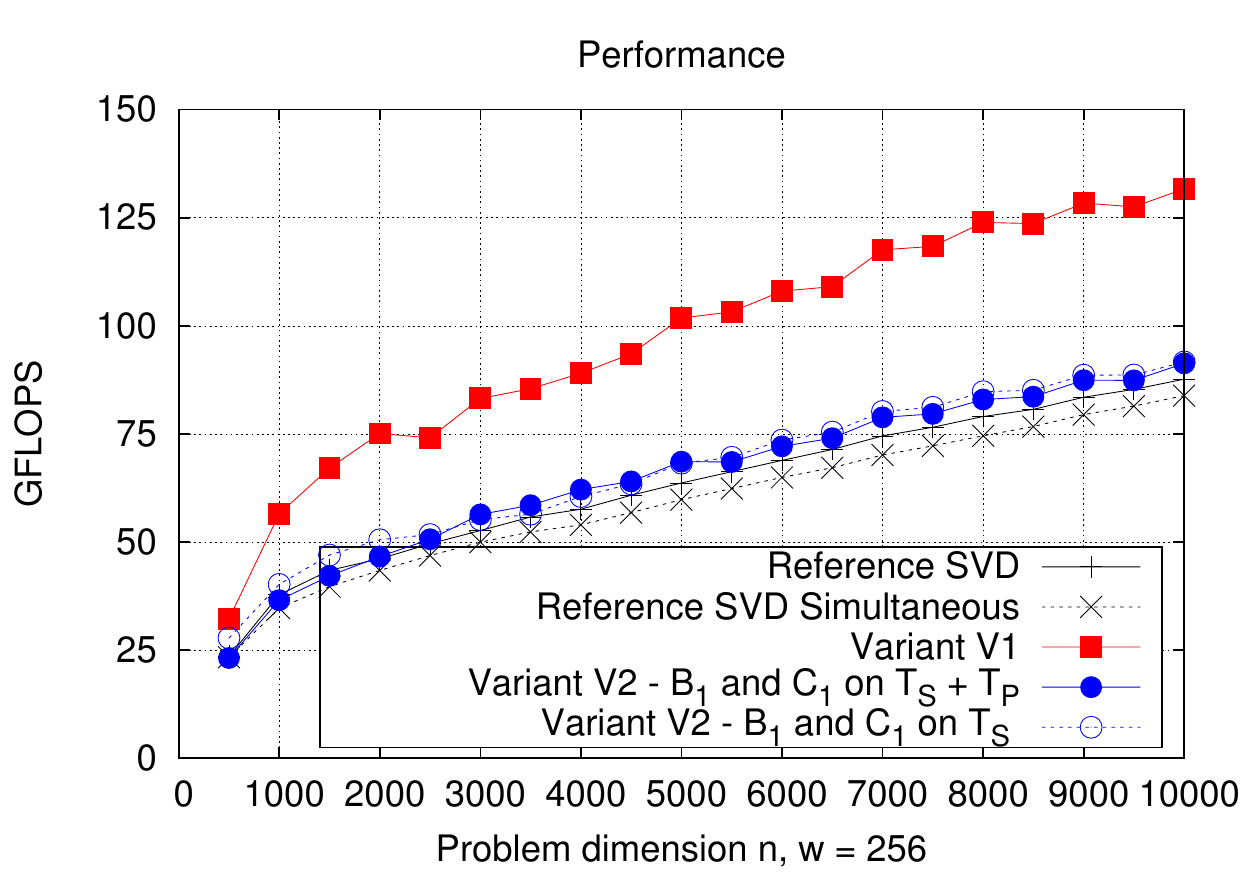}\\
\end{center}
\caption{Performance vs problem dimension of the SVD implementations with $w = 32, 64, 128, $ and $256$.
}
\label{fig:Banda_SVD}
\end{figure}

Figure~\ref{fig:Banda_SVD} reports the GFLOPS rates 
attained by the configurations for bandwidths ranging from \textcolor{black}{32 to 256 and square matrices}. 
For brevity, the analysis of the optimal block size is not presented as it revealed a similar behavior as that observed for SEVP.
Let us focus first on the implementations without look-ahead. From the plots, it is clear that
the ``Reference SVD'' implementation outperforms its ``Reference SVD Simultaneous'' counterpart.
At this point we would remark that the second implementation underlies variant~V2 of the SVD algorithm with look-ahead.

Focusing on Variant~V1, we detect the same drawback as that identified in Variant~V1 for SEVP in that, for
small bandwidths, the block size is strongly constrained. In contrast, 
large performance improvements are reported,
compared with all other implementations,
for medium and large bandwidths.

A less pleasant case is encountered for Variant~V2, which is not able to improve significantly the
performance results of the ``Reference SVD'' implementation for any bandwidth nor problem dimension though
it outperforms its baseline ``Reference SVD Simultaneous'' implementation.
The reason for this result is that, for this implementation, the update of $D_{22}$ cannot be fully overlapped with
the execution time of the next panel factorizations (both left and right). 
For Variant~V1, the execution of the panel factorizations is overlapped with the updates of $B_{1}^{R}$,
$C_{1}^{B}$, and $D$; but due to the data dependencies in Variant~V2, 
we can only overlap the execution of the panel factorizations
with the update of $D_{22}$, which exhibits a considerably more reduced number of flops.

\begin{figure}[t!]
\begin{center}
\includegraphics[width=0.49\columnwidth]{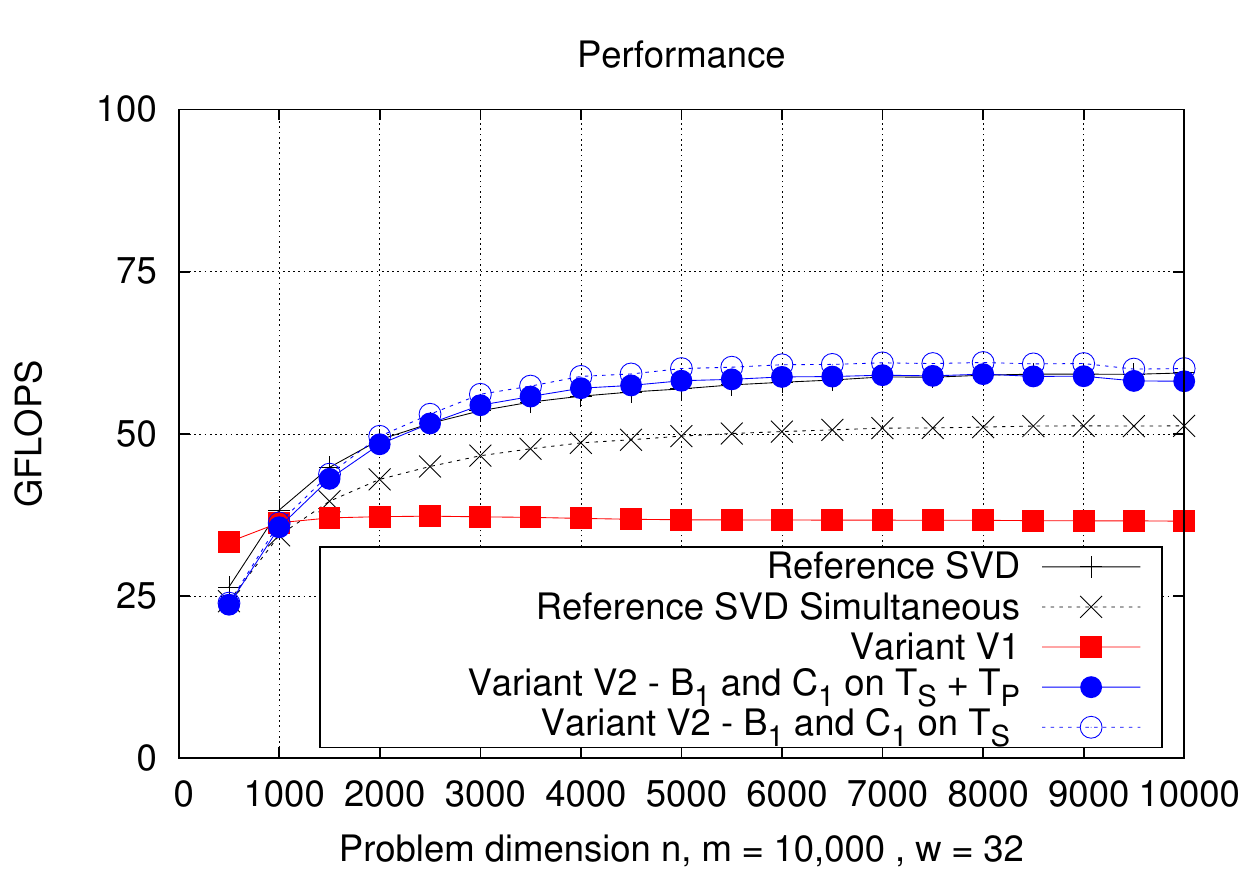}
\includegraphics[width=0.49\columnwidth]{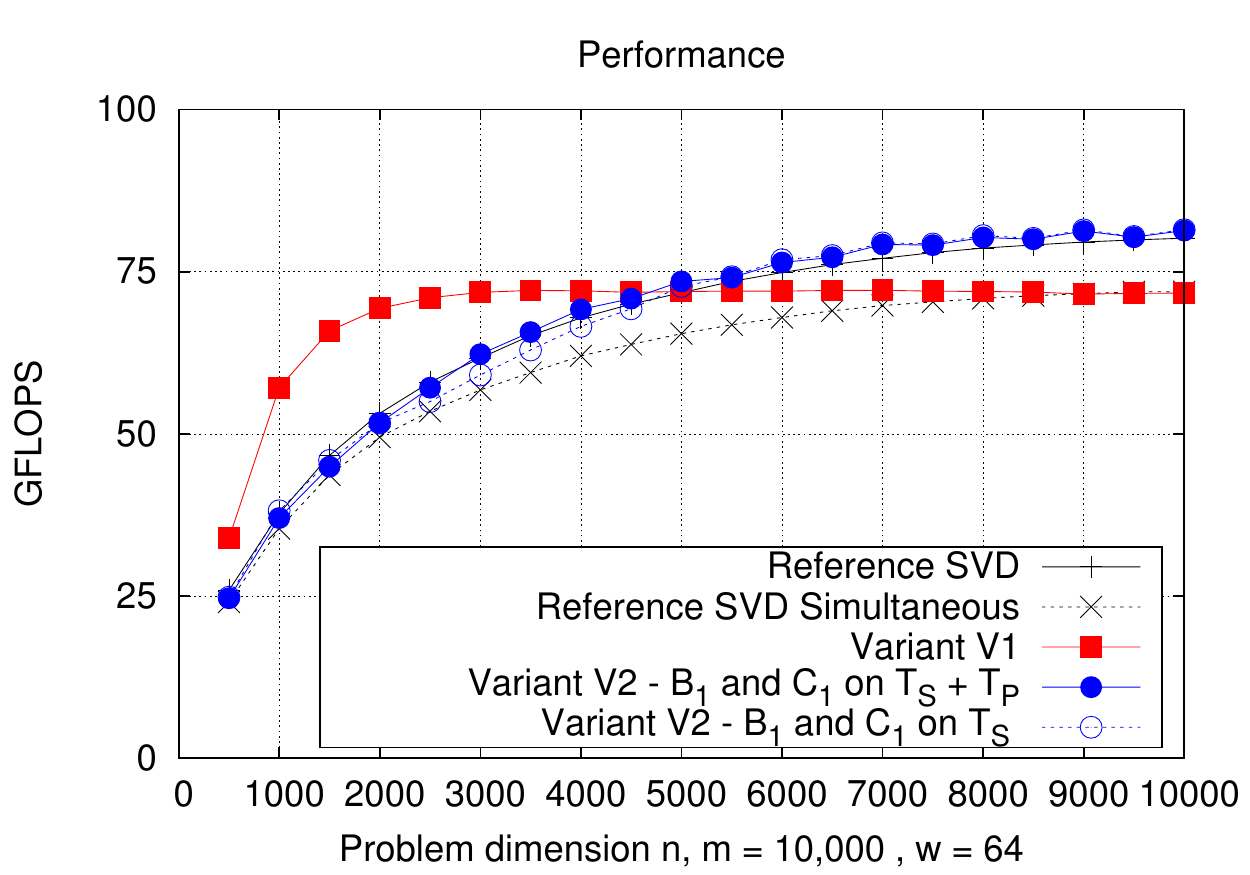}\\
\includegraphics[width=0.49\columnwidth]{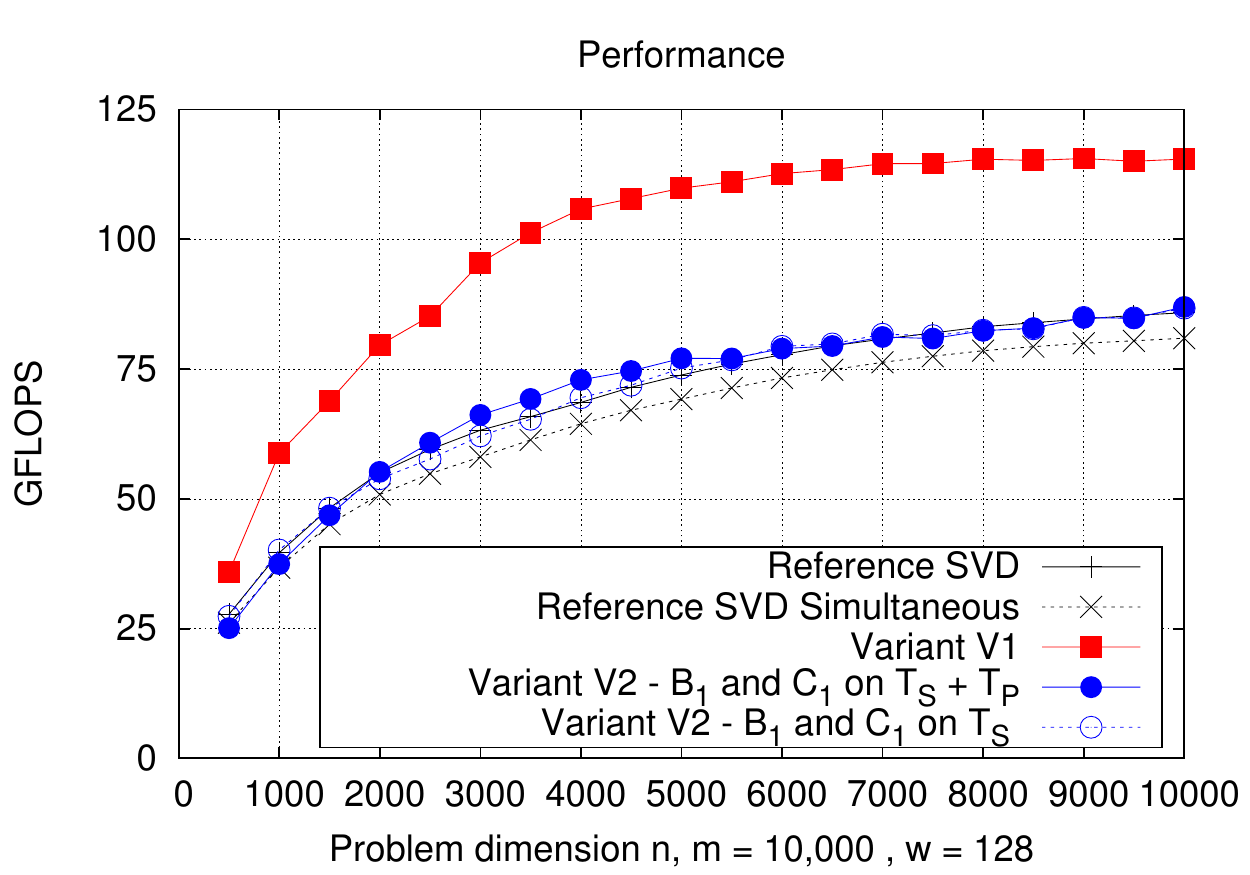}
\includegraphics[width=0.49\columnwidth]{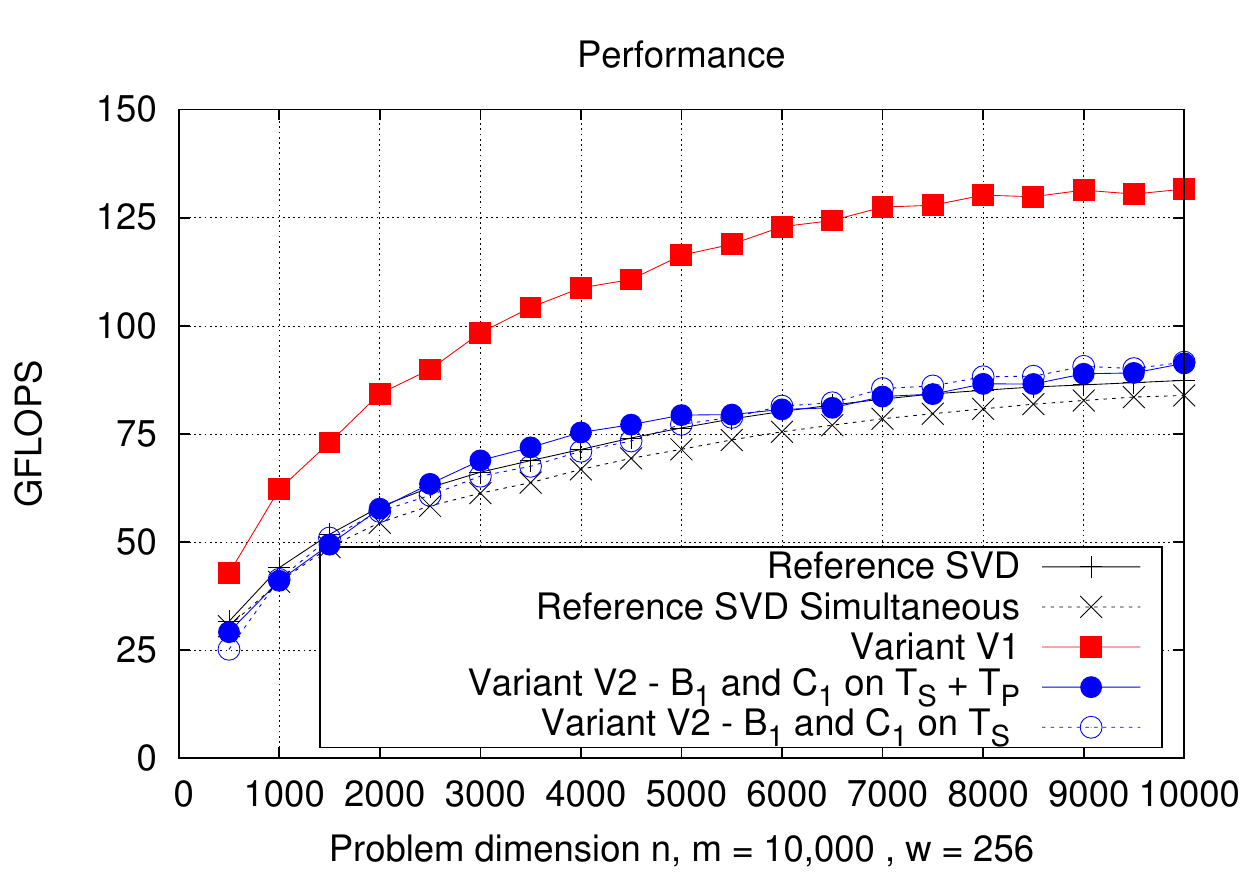}\\
\end{center}
\caption{Performance vs problem dimension of the SVD implementations for rectangular matrices with $w = 32, 64, 128, $ and $256$.
}
\label{fig:Banda_SVD_nsqu}
\end{figure}

\textcolor{black}{Figure~\ref{fig:Banda_SVD_nsqu} displays the GFLOPS rates 
observed for bandwidths ranging from 32 to 256 and non-square matrices. In the plots,
the $m$-dimension on the matrices is fixed to 10000, while the $n$-dimension is varied in the range 500--10000 
in steps of 500.  The plots reveal performance numbers that are very close to those observed for the reductions 
of square matrices, showing that
the new variants for TSR are not sensitive to the matrices shape.}

%% file: s5-remarks.tex
\section{Concluding Remarks}
\label{sec:remarks}

We have analyzed in detail the impact that static look-ahead exerts on the 
performance of two-sided routines that perform the reduction to compact band forms for SEVP and the SVD.
Our study shows that a correct selection of the look-ahead variant as well as
an appropriate mapping of tasks to cores are key to optimize performance.
Even more importantly, the block size plays a crucial role in the computational
throughput of these reduction routines. Decoupling this parameter from
the target bandwidth is a must, and therefore we have to depart from the solution 
adopted in the corresponding routines included in the current versions of LAPACK, PLASMA and MAGMA,
which simply set the block size to equal the bandwidth.

For the SVD, our analysis also advocates for an alternative option that reduces the original dense matrix to
a band form with the same upper and lower bandwidths, 
allowing an efficient exploitation of the look-ahead strategy.
This choice thus overcomes some of the difficulties of the traditional reduction
to band--triangular from that is adopted in LAPACK and MAGMA.

%% file: paper.bbl
\begin{thebibliography}{10}

\bibitem{ALIAGA2016314}
José~I. Aliaga, Pedro Alonso, José~M. Badía, Pablo Chacón, Davor
  Davidović, José~R. López-Blanco, and Enrique~S. Quintana-Ortí.
\newblock A fast band–krylov eigensolver for macromolecular functional motion
  simulation on multicore architectures and graphics processors.
\newblock {\em Journal of Computational Physics}, 309(Supplement C):314 -- 323,
  2016.

\bibitem{lapack}
Edward Anderson, Zhaojun Bai, L.~Susan Blackford, James Demmel, Jack~J.
  Dongarra, Jeremy~Du Croz, Sven Hammarling, Anne Greenbaum, Alan McKenney, and
  Danny~C. Sorensen.
\newblock {\em {LAPACK} Users' guide}.
\newblock SIAM, 3rd edition, 1999.

\bibitem{Ballard20153}
G.~Ballard, J.~Demmel, L.~Grigori, M.~Jacquelin, N.~Knight, and H.~D. Nguyen.
\newblock Reconstructing householder vectors from tall-skinny {QR}.
\newblock {\em J. Parallel \& Distributed Comp.}, 85:3--31, 2015.

\bibitem{CPE:CPE1680}
Paolo Bientinesi, Francisco~D. Igual, Daniel Kressner, Matthias Petschow, and
  Enrique~S. Quintana-Ort\'{\i}.
\newblock Condensed forms for the symmetric eigenvalue problem on
  multi-threaded architectures.
\newblock {\em Concurrency \& Comp.: Practice \& Exp.}, 23(7):694--707, 2011.

\bibitem{Bischof:2000:AST}
C.~H. Bischof, B.~Lang, and X.~Sun.
\newblock {Algorithm 807}: {The SBR Toolbox}---software for successive band
  reduction.
\newblock {\em ACM Trans. Math. Soft.}, 26(4):602--616, 2000.

\bibitem{Buttari200938}
Alfredo Buttari, Julien Langou, Jakub Kurzak, and Jack Dongarra.
\newblock A class of parallel tiled linear algebra algorithms for multicore
  architectures.
\newblock {\em Parallel Computing}, 35(1):38 -- 53, 2009.

\bibitem{Castaldo:2013:SLP:2491491.2491492}
Anthony~M. Castaldo, R.~Clint Whaley, and Siju Samuel.
\newblock Scaling {LAPACK} panel operations using parallel cache assignment.
\newblock {\em ACM Trans. Math. Soft.}, 39(4):22:1--22:30, July 2013.

\bibitem{catalan17}
Sandra Catal{\'{a}}n, Jos{\'{e}}~R. Herrero, Enrique~S. Quintana{-}Ort{\'{\i}},
  Rafael Rodr{\'{\i}}guez{-}S{\'{a}}nchez, and Robert~A. van~de Geijn.
\newblock A case for malleable thread-level linear algebra libraries: The {LU}
  factorization with partial pivoting.
\newblock {\em CoRR}, abs/1611.06365, 2016.

\bibitem{PDP:Davor2011}
D.~Davidovi\'c and E.~S. Quintana-Ort\'{\i}.
\newblock Applying {OOC} techniques in the reduction to condensed form for very
  large symmetric eigenproblems on {GPUs}.
\newblock In {\em Proceedings of the 20th Euromicro Conference on Parallel,
  Distributed and Network based Processing -- PDP 2012}, pages 442--449, 2012.

\bibitem{PIROBAND}
T.~A. Davis and S.~Rajamanickam.
\newblock Algorithm 8xx: {PIRO BAND}, pipelined plane rotations for band
  reduction.
\newblock {\em ACM Trans. Math. Soft.}
\newblock Submitted.

\bibitem{Dhillon:2006:DIM:1186785.1186788}
Inderjit~S. Dhillon, Beresford~N. Parlett, and Christof V\"{o}mel.
\newblock The design and implementation of the {MRRR} algorithm.
\newblock {\em ACM Trans. Math. Softw.}, 32(4):533--560, Dec. 2006.

\bibitem{blas3}
Jack~J. Dongarra, Jeremy Du~Croz, Sven Hammarling, and Iain Duff.
\newblock A set of level 3 basic linear algebra subprograms.
\newblock {\em ACM Trans. Math. Softw.}, 16(1):1--17, March 1990.

\bibitem{blas2}
Jack~J. Dongarra, Jeremy Du~Croz, Sven Hammarling, and Richard~J. Hanson.
\newblock An extended set of {FORTRAN} basic linear algebra subprograms.
\newblock {\em ACM Trans. Math. Softw.}, 14(1):1--17, March 1988.

\bibitem{Fernando1994}
K.~Vince Fernando and Beresford~N. Parlett.
\newblock Accurate singular values and differential {QD} algorithms.
\newblock {\em Numer. Mathematik}, 67(2):191--229, 1994.

\bibitem{GVL3}
Gene~H. Golub and Charles F.~Van Loan.
\newblock {\em Matrix Computations}.
\newblock The Johns Hopkins University Press, Baltimore, 3rd edition, 1996.

\bibitem{GROER1999969}
Benedikt Grosser and Bruno Lang.
\newblock Efficient parallel reduction to bidiagonal form.
\newblock {\em Parallel Computing}, 25(8):969 -- 986, 1999.

\bibitem{doi:10.1137/S0895479892242232}
Ming Gu and Stanley~C. Eisenstat.
\newblock A divide-and-conquer algorithm for the bidiagonal {SVD}.
\newblock {\em SIAM J. Matrix Analysis \& Appl.}, 16(1):79--92, 1995.

\bibitem{6114396}
A.~Haidar, H.~Ltaief, and J.~Dongarra.
\newblock Parallel reduction to condensed forms for symmetric eigenvalue
  problems using aggregated fine-grained and memory-aware kernels.
\newblock In {\em 2011 International Conference for High Performance Computing,
  Networking, Storage and Analysis (SC)}, pages 1--11, Nov 2011.

\bibitem{Haidar:2013:IPS:2503210.2503292}
Azzam Haidar, Jakub Kurzak, and Piotr Luszczek.
\newblock An improved parallel singular value algorithm and its implementation
  for multicore hardware.
\newblock In {\em Proceedings of the International Conference on High
  Performance Computing, Networking, Storage and Analysis}, SC '13, pages
  90:1--90:12, New York, NY, USA, 2013. ACM.

\bibitem{Moldaschl:2014:CES:2748152.2748600}
Michael Moldaschl and Wilfried~N. Gansterer.
\newblock Comparison of eigensolvers for symmetric band matrices.
\newblock {\em Sci. Comput. Program.}, 90(PA):55--66, Sept. 2014.

\bibitem{Petschow2013:208}
Matthias Petschow, Elmar Peise, and Paolo Bientinesi.
\newblock High-performance solvers for dense {H}ermitian eigenproblems.
\newblock {\em SIAM J. Scientific Comp.}, 35(1):C1--C22, Jan. 2013.

\bibitem{Quintana-Orti:2009:PMA:1527286.1527288}
Gregorio Quintana-Ort\'{\i}, Enrique~S. Quintana-Ort\'{\i}, Robert~A.
  {van~de~Geijn}, Field~G. {Van~Zee}, and Ernie Chan.
\newblock Programming matrix algorithms-by-blocks for thread-level parallelism.
\newblock {\em ACM Trans. Math. Softw.}, 36(3):14:1--14:26, 2009.

\bibitem{Str98}
Peter Strazdins.
\newblock A comparison of lookahead and algorithmic blocking techniques for
  parallel matrix factorization.
\newblock Technical Report TR-CS-98-07, Department of Computer Science, The
  Australian National University, {Canberra} 0200 {ACT}, {Australia}, 1998.

\bibitem{BLIS2}
Field G.~Van Zee, Tyler~M. Smith, Bryan Marker, Tze~Meng Low, Robert A. Van~De
  Geijn, Francisco~D. Igual, Mikhail Smelyanskiy, Xianyi Zhang, Michael
  Kistler, Vernon Austel, John~A. Gunnels, and Lee Killough.
\newblock The {BLIS} framework: Experiments in portability.
\newblock {\em ACM Trans. Math. Softw.}, 42(2):12:1--12:19, June 2016.

\end{thebibliography}
